\documentclass[12pt]{iopart}
\usepackage[english]{babel}
\usepackage{
amssymb,amsfonts,latexsym}
\usepackage[pdftex]{graphicx}
\usepackage{color}

\usepackage{times}
\usepackage[colorlinks,bookmarks=false,citecolor=blue,linkcolor=red,urlcolor=blue]{hyperref}

\usepackage{simplewick}
\usepackage{cancel}

\newcommand{\be}{\begin{equation}}
\newcommand{\ee}{\end{equation}}
\newcommand{\bea}{\begin{eqnarray}}
\newcommand{\eea}{\end{eqnarray}}
\def\HH{\mathcal H}

\begin{document}

\title{Quench dynamics of a Tonks-Girardeau gas released from a harmonic trap}

\author{Mario Collura, Spyros Sotiriadis, and Pasquale Calabrese}
\address{Dipartimento di Fisica dell'Universit\`a di Pisa and INFN, 56127 Pisa, Italy}
\date{\today}

\begin{abstract}
We consider the non-equilibrium dynamics of a gas of impenetrable bosons released from a harmonic trapping potential to a circle.
The many body dynamics is solved analytically and the time dependence of all the physically relevant correlations 
is described.  
We prove that, for large times and in the thermodynamic limit, the reduced density  matrix of any subsystem converges to a 
generalized Gibbs ensemble as a consequence of the integrability of the model. 
We discuss the approach to the stationary behavior at late times.
We also describe the time-dependence of the entanglement entropy which
attains a very simple form  in the stationary state. 

\end{abstract}

\section{Introduction}
Recent experiments on trapped ultra-cold atomic gases have shown that it is 
possible to follow and measure the unitary nonequilibrium evolution of an isolated quantum system
\cite{uc,kww-06,tc-07,hgm-09,tetal-11,cetal-12,getal-11,shr-12,rsb-13,mmk-13,fse-13}. 
A particular class of these nonequilibrium problems which is experiencing an enormous theoretical 
activity is that of a  sudden quench of a Hamiltonian parameter.
In a global quantum quench, the initial condition is the ground state of a translationally invariant Hamiltonian
which differs from the one governing the evolution by an experimentally tunable parameter such as a magnetic field.
In these experiments the two key questions are: i) how the correlations and entanglement spread into the 
system with time \cite{revq,cc-06,cc-06b,c-06,ir-00,lk-08,ic-09,fcc-09,ir-10,sc-10,bpk-12,CEF,CEFI,gcg-11,ri-11,ors-12,se-12}
and ii) whether the system relaxes (in some sense) to a stationary state, and if it does, 
how to characterize from first principles its physical properties at late times. 
For the latter question, it is widely believed that, depending on the integrability of the Hamiltonian governing the time evolution, 
the behavior of {\it local} observables can be   described either by an effective thermal 
distribution or by a generalized Gibbs ensemble (GGE), for non-integrable and integrable systems 
respectively (see e.g. \cite{revq} for a review). 
This scenario is corroborated by many investigations \cite{gg,gg2,cc-06b,c-06,msnm-07,wk-07,cdeo-08,bs-08,scc-09,r-09,CEF,dams-09,CEFII,f-13,eef-12,ccss-11,rs-12,bkl-10,bdkm-11,fm-10,can,cic-12,sfm-12,mc-12,ms-12,o-12,srgs-13,ce-13,fe-13,m-13,p-13,fe-13b}, 
but still a few studies \cite{kla-07,rf-11,bch-11,gme-11,gm-11,gp-08,g-13,cl-13} suggest that the 
behavior could be more complicated. 
Indeed, it has been argued that the initial state can affect this scenario, in particular if it breaks some  
symmetries of the Hamiltonian governing the subsequent evolution which tend to be recovered 
in a statistical ensemble such as thermal or GGE. 
The case that has been most largely studied is that of a non-translationally invariant initial state
generically referred to as inhomogeneous quenches \cite{chl-08,sc-08,eip-09,predra,dra,inh2,mpc-10,pv-13}.

Among the inhomogeneous quenches, a particularly relevant one which has been already 
experimentally realized (also in one dimension \cite{shr-12,rsb-13}) is the non-equilibrium dynamics of a gas released 
from a parabolic trapping potential. 
A very interesting experimental finding is that the spreading of correlations is 
ballistic for an integrable system and diffusive for a non-integrable one \cite{rsb-13}.
Both experimental \cite{shr-12,rsb-13} and theoretical \cite{gp-08,mg-05,dm-06,cro,dc-08,a-12,cv-10,v-12,hm-v,hm-v2,lbb-12,r-10,db-12} 
analyses concentrated on the expansion in the full one-dimensional space, which has the advantage 
to avoid unwanted finite-size effects.
However, if a repulsive gas expands on the full line, its density will decrease as time passes and 
for infinite time it goes to zero, making senseless to distinguish thermal and GGE states.

\begin{figure}[t]
\center\includegraphics[width=0.35\textwidth]{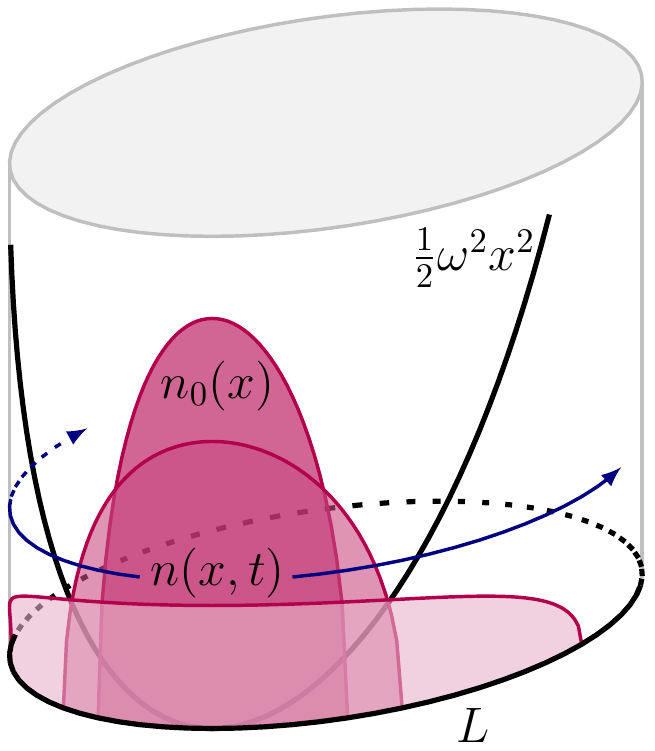}
\caption{Sketch of the trap release dynamic in a circle.}
\label{sketch}
\end{figure}

An alternative proposal by J.S Caux and R. Konik \cite{ck-12} is that of considering the release of a gas 
from a parabolic trap not in  free space but on a closed circle of length $L$ (as sketched in Fig. \ref{sketch}),
so that the gas has finite density even for infinite time. 
However, in a gas with a finite number of particles $N$ (or more generally in a system with a finite number of 
degrees of freedom) a stationary state cannot be approached because of revival and recurrence effects
(i.e. the system is always quasi-periodic). 
To circumvent this, the thermodynamic (TD) limit should be defined properly: 
for fixed final density $n=N/L$, when $N,L\to\infty$, at the same time the frequency of the initial 
confining potential $\omega$ should also vanish, i.e. $\omega\to 0$, but with $\omega N$ 
constant (i.e. fixed `initial density') \cite{ck-12,US}.
In order to tackle quite generally this problem, a new approach based on integrability has been developed and 
applied explicitly to the Lieb-Liniger Bose gas \cite{LiebPR130} (a preliminary analysis for non-integrable models has also 
been presented \cite{bck-13}). 
Using this approach, in Ref. \cite{ck-12}, it has been  shown numerically that for a Lieb-Liniger gas,
the time averaged correlation functions are well described by a GGE, apart from finite size effects
(the maximum number of particles considered in Ref.  \cite{ck-12} is $N=56$).
In principle this new approach allows also the study of the time evolution, but it is much more computationally demanding 
and it has not yet been done. 
As a consequence it has not yet been established whether (and in which sense) an infinite time limit exists
and, if yes, how it is approached.

In order to overcome these limitations, in a recent letter \cite{US} 
we presented  a full analytic solution of the nonequilibrium trap release 
 dynamics in the limit of strong coupling, i.e. in the Tonks-Girardeau regime \cite{TG}. 
This allowed us to understand that also the infinite time limit should be handled with care: 
in  the trap release dynamics, a stationary behavior is possible because of the interference of  the particles going around the 
circle $L$ many times (see Fig. \ref{sketch}), i.e. to observe a stationary value we must require $v t\gg L$ (with $v$ the
expansion velocity of the gas). 
This is very different from equilibration in standard global
quenches where, in order to avoid revival effects, the time should be such that the boundaries are never reached 
(i.e. one first considers the TD limit $L\to\infty$ and only after the infinite time limit $t\to\infty$, 
which, in finite systems, corresponds to the condition $vt\ll L$, see e.g. \cite{CEFII}). 
In the trap release problem the revival scale is $\tau_{\rm rev}\propto L^2$ (see also \cite{deg-13})
and so the infinite time limit in which a stationary behavior can be achieved  is 
$t/L\to \infty$ provided  $t/L^2\to 0$. 
In Ref. \cite{US}, we have showed that, in the TD limit, the reduced density matrix of any 
finite subsystem converges for long times (in the sense just explained) to the GGE one. 
This implies that any measurable local observable will converge to the GGE predictions.
In this manuscript, we extend the previous letter \cite{US} in several aspects.
First of all, we give complete derivation of all results in the GGE previously presented.
Secondly, for many observables we will characterize the full asymptotic time dependence 
and not restrict to the stationary results.
As particularly important new aspects absent in Ref. \cite{US}, we  study the time evolution of the entanglement 
entropies and we construct the GGE in terms of local integrals of motion.

The manuscript is organized as follows.
In Sec. \ref{sec2} we introduce the model under investigation and the quench protocol.
In Sec. \ref{sec3} we calculate the time evolution of the two-point correlation function and 
prove that for infinite time a stationary value is approached. 
We also discuss the approach to the stationary value. 
In Sec. \ref{sec4} we show that the stationary values of all local observables are described by a GGE 
both in fermionic momentum occupation numbers and in the local integrals of motion. 
In Sec. \ref{sec5} we compute the density-density correlation and in Sec. \ref{sec6} the bosonic 
one-particle density matrix (Fourier transform of the momentum distribution function).
In Sec. \ref{sec7} we move our attention to the entanglement entropies.
The trap release dynamics from a trapped gas to a larger trap is addressed in Sec. \ref{sec8}. 
Finally in Sec. \ref{concl} we draw our conclusions.

\section{The Model and quench protocol}
\label{sec2}

The Lieb-Liniger model describes a system of $N$ identical bosons in one dimension (1D) interacting 
via a pairwise Dirac-delta potential. 
In first quantization language, the Hamiltonian is given by \cite{LiebPR130}
\begin{equation}\label{HLL}
H_{\rm LL} = -\frac{1}{2}\sum_{j=1}^{N}\frac{\partial^{2}}{\partial x_{j}^{2}}  + c\sum_{i\neq j}\delta(x_{i}-x_{j}),
\end{equation}
where $c$ is the coupling constant and we set $\hbar=m=1$. 
For definiteness, we consider a system of length $L$ with periodic boundary conditions (PBC). 
In the repulsive regime, $c>0$, and in the TD limit, the equilibrium physics of the model depends on the 
single parameter $\gamma=c/n$  where $n = N/L$ is the particle density. 
Then in 1D, in stark contrast to higher dimensions, low densities lead one to the strong-coupling regime of 
impenetrable bosons $c\to\infty$, known as the Tonks-Girardeau limit \cite{TG}. 
In the attractive regime, $c<0$, the physics of the model is completely different (see e.g. \cite{mg-63,cc-07}) and will 
not be considered here. 

In second quantization language, the Hamiltonian (\ref{HLL}) can be rewritten as a quantum non-linear Schrodinger equation
\begin{equation}\label{HLL2}
H_{\rm LL} = \int_{-L/2}^{L/2} dx \left[\frac12 \partial_x \hat\Phi^{\dagger} (x) \partial_x \hat\Phi (x) 
+ c \hat\Phi^{\dagger} (x) \hat\Phi^{\dagger} (x) \hat\Phi (x) \hat\Phi(x) \right].
\end{equation}
where $\hat{\Phi}(x)$ and $\hat{\Phi}^{\dag}(x)$ are  the bosonic annihilation and creation field operators respectively.

The Lieb-Liniger model is Bethe ansatz integrable \cite{LiebPR130}, but the analytic calculation of the 
non-equilibrium dynamics in the TD limit is still a formidable task, despite  the numerous attempts in the 
literature \cite{cro,ck-12,a-12,deg-13,ksc-13,grd-10}.
For this reason, as already anticipated, we concentrate here in the impenetrable limit $c\to\infty$
in which the Hamiltonian (\ref{HLL2}) can be simply written as 
\begin{equation}\label{H_2}
H= \int_{-L/2}^{L/2} dx\, \hat{\Phi}^{\dag}(x)\left[-\frac{1}{2}\frac{\partial^{2} }{\partial x^2} \right]\hat{\Phi}(x),
\qquad {\rm with}\quad  \hat{\Phi}^2(x)=0, 
\end{equation}
and where the infinite coupling $c$ is encoded in the hard-core constraint $\hat{\Phi}^2(x)=0$,
i.e. the condition that two bosons cannot occupy the same position. 
At operator level, this constraint can be imposed by requiring that $\hat{\Phi}(x)^\dag$ and $\hat{\Phi}(x)$ 
commute at different spatial points and they anti-commute when evaluated at the same point.
In other words, they are similar to fermionic operators but they commute on different space positions. 
In order to restore a genuine Fermi algebra, fermionic field operators $\hat{\Psi}(x)$ and $\hat{\Psi}^{\dag}(x)$ are built
through a Jordan-Wigner transformation 
\begin{eqnarray}\label{JordanWigner}
\hat{\Psi}(x) & = & \exp\left\{i\pi\int_{0}^{x}dz\,\hat{\Psi}^{\dag}(z) \hat{\Psi}(z)\right\}\, \hat{\Phi}(x),\\
\hat{\Psi}^{\dag}(x) & = & \hat{\Phi}^{\dag}(x)\,\exp\left\{-i\pi\int_{0}^{x}dz\,\hat{\Psi}^{\dag}(z) \hat{\Psi}(z)\right\},\nonumber
\end{eqnarray}
which by construction satisfy $\{\hat{\Psi}(x),\hat{\Psi}^{\dag}(y)\}=\delta(x-y)$ and 
$\hat{\Psi}^{\dag}(x)\hat{\Psi}(x)=\hat{\Phi}^{\dag}(x)\hat{\Phi}(x)\equiv \hat{n}(x)$, with 
$\hat{n}(x)$ the density operator of both fermions and bosons.
This is the standard mapping between impenetrable bosons and free fermions \cite{TG} which ensures that 
all spectral and thermodynamical properties of the bosons can be simply obtained from free fermions. 
However, being the transformation (\ref{JordanWigner}) non-local, bosonic correlation functions are different 
from fermionic ones and they should be reconstructed with the help of Wick theorem, as explicitly done in the 
following. 

The Hamiltonian (\ref{H_2}) of $N$ impenetrable bosons is the one governing the time evolution in our problem and 
we have now to fix the initial many-body state. 

\subsection{The initial state}

The initial state we consider is the ground state of the Tonks-Giradeau gas in a harmonic confining potential,
i.e. the ground state of the Hamiltonian  
\begin{equation}\label{H_1}
H_0 = -\frac{1}{2}\sum_{j=1}^{N}\frac{\partial^{2}}{\partial x_{j}^{2}} + \sum_{j=1}^{N}V(x_{j}) + c\sum_{i\neq j}\delta(x_{i}-x_{j}),
\end{equation}
with $V(x)=\omega^{2}x^{2}/2$ and for $c\to\infty$.
The translationally invariant Lieb-Liniger model is recovered  for $\omega=0$. 
In the Tonks-Giradeau limit, the corresponding free fermionic Hamiltonian is 
\begin{equation}\label{HF_2}
H_0 = \int dx\, \hat{\Psi}^{\dag}(x)\left[-\frac{1}{2}\frac{\partial^{2} }{\partial x^2} + V(x)\right]\hat{\Psi}(x).
\end{equation}

The many body-ground state is the Slater determinant built with the lowest energy one-particle eigenfunctions.
This is easily worked out from the diagonalization  of the single-particle Hamiltonian
\begin{equation}\label{H_onebody}
\mathcal{H}_0=-\frac{1}{2}\frac{\partial^{2} }{\partial x^2} +V(x).
\end{equation}
Let us assume for the moment that $L\to\infty$ so that the  eigenfunctions of ${\cal H}$, 
for the parabolic potential $V(x)$, are the ones of the one-dimensional  harmonic oscillator
\begin{eqnarray}\label{chin}
\mathcal{H}_0\chi_{j}(x) & = &  \epsilon_{j}\chi_{j}(x),\qquad \epsilon_{j}  =  \omega(j+1/2),\\
 \chi_{j}(x) & = &  \frac{1}{\sqrt{2^{j}j!}}\left(\frac{\omega}{\pi}\right)^{1/4}\mathrm{e}^{-\omega x^2/2}H_{n}(x\sqrt{\omega}),\nonumber
\end{eqnarray}
with $H_{j}(x)$ the Hermite polynomials and $j=0,1,2,\dots$ a non-negative integer number.
Introducing now the fermionic operators $\hat{\xi}_{j}$ as 
\begin{equation}
\hat{\xi}_{j}  =  \int_{-\infty}^{\infty} dx\, \chi^{*}_{j}(x) \hat{\Psi}(x),\qquad
\hat{\Psi}(x)  =  \sum_{j=0}^{\infty} \chi_{j}(x)\hat{\xi}_{j},
\label{psi-xi}
\end{equation}
satisfying the canonical anti-commutation relations $\{ \hat{\xi}_{i} , \hat{\xi}^{\dag}_{j} \}=\delta_{ij}$,
the many-body Hamiltonian is diagonal in the $\hat{\xi}_{j},\, \hat{\xi}^{\dag}_{j}$ representation
\begin{equation}
H_0 = \sum_{j=0}^{\infty}\epsilon_{j} \hat{\xi}^{\dag}_{j} \hat{\xi}_{j}.
\end{equation}
Clearly, all previous results remain valid for any external potential $V(x)$ as long as one uses the corresponding 
eigenfunctions of the one-body Hamiltonian (\ref{H_onebody}).

In Fock space, the many-body ground state of $N$ impenetrable bosons in a parabolic trap  is
\begin{equation}\label{GS0}
|\Psi_{0}\rangle = \prod_{j=0}^{N-1}\hat{\xi}^{\dag}_{j}|\emptyset\rangle,
\qquad  {\rm and}\quad \langle\Psi_{0}|\hat{\xi}^{\dag}_{i}\hat{\xi}_{j}|\Psi_{0}\rangle = \delta_{ij}\theta(N-j),
\end{equation}
where $|\emptyset\rangle$ is the vacuum state annihilated by $\hat{\xi}_{j}$ for all $j$. 

Let us now consider the initial density profile 
\be
n_0(x)\equiv \langle \Psi_{0}| \hat{n}(x) |\Psi_{0}\rangle=\sum_{j=0}^{N-1} |\chi_j(x)|^2.
\ee
The sum over $j$ can be analytically carried out using the  Christoffel-Darboux formula for the Hermite polynomials $H_{j}(x)$
\begin{equation}
\sum_{j=0}^{N-1}\frac{H_{j}(x)H_{j}(y)}{2^j j!} = \frac{1}{2^{N}(N-1)!}\frac{H_{N}(x)H_{N-1}(y)-H_{N}(y)H_{N-1}(x)}{x-y},
\label{cdf}
\end{equation}
which in the limit $N\to\infty$ leads to 
\begin{equation}\label{CD_inf}
\frac{\mathrm{e}^{-x^2}}{\pi}\sum_{j=0}^{N-1}\frac{H_{j}(x)^2}{2^j j!}\simeq \frac{\sqrt{2N-\lambda^2}}{\pi}.
\end{equation}
Thus the TD initial density profile is 
\be  
n_{0}(x)= \frac{\sqrt{2N\omega -\omega^2 x^2}}{\pi} \theta(\ell-|x|),
\qquad \ell= \sqrt{2N/\omega}\,,
\label{n0}
\ee
which is the well-known Thomas-Fermi profile (straightforwardly obtained for free fermions also by  local 
density approximation).
Notice that for $x$ larger than the Thomas-Fermi radius $\ell$ the gas density is {\it exactly} zero in the TD limit. 
Also the trapped fermionic two-point correlation function is straightforwardly obtained from Christoffel-Darboux formula
\be
C(x,y)\equiv \langle \hat \Psi^\dag(x)\hat \Psi(y)\rangle=\sqrt\frac{N}{2\omega}
 \frac{\chi_{N}(x)\chi_{N-1}(y)-\chi_{N-1}(x)\chi_{N}(y)}{x-y},
\ee
with $\chi_j(x)$ the single particle wave-function in Eq. (\ref{chin}). When the two points $x,y$ are very close to the center of the trap,
i.e. $x,y\ll 1/\omega$ the above formula simplifies to
\be
C(x,y)\simeq \frac{\sin[\sqrt{2\omega N}(x-y)]}{\pi (x-y)},
\label{Ctrap1}
\ee
which is the translationally invariant result with $k_F^0=\sqrt{2\omega N}$.

The vanishing of the density and of the many-body wave-function, in the TD limit, for $|x|>\ell$  
is the fundamental property allowing us to treat  analytically also the time evolution  in a finite circle of length $L$.
Indeed, we now make the only crucial {\it physical assumption} of our treatment: 
we impose that the space initially occupied by the trapped gas as 
a whole is {\it within} the external box of length $L$, i.e. the PBC are irrelevant for the gas in the initial state 
which only ``sees'' the parabolic trap. 
This means that the extension of the gas in the trap, $2\ell$ in Eq. (\ref{n0}), must be smaller than the box size $L$:
\be
2\ell<L \Rightarrow  N< \omega L^2/8.
\label{crucass}
\ee
In terms of the number of particles $N$, this condition means that $N$ must be smaller 
than the first level of the parabolic potential that is affected by the PBC.
Furthermore, this is the hypothesis which allows us to talk about release of the gas,
because if the gas would feel the PBC before the quench, we would not have a trapped gas, but 
something more complicated.
Clearly, under the assumption (\ref{crucass}), the many-body ground state in infinite space (\ref{GS0}) is 
also the ground state for a finite circular geometry in the presence of the trap.

The trap-release condition can be also written in a maybe more transparent way in terms of the initial average 
density $n_0$ and final density $n$.
Indeed, by definition we have 
\be
n_0\equiv \frac{N}{2\ell}= \sqrt\frac{N\omega}8,
\label{n00}
\ee
and the trap-release condition becomes 
\be
n_0>n,
\ee
i.e. that the initial average density is larger than the final one signaling that the gas expands.

\subsection{The quench protocol}

In this section we describe the non-equilibrium dynamics which is the focus of the paper. 
The initial state is $|\Psi_0\rangle$ in Eq. (\ref{GS0}) and the Hamiltonian governing the evolution for $t>0$ is the 
Tonks-Girardeau in Eq. (\ref{H_2}) clearly with periodic boundary conditions.  
In practical terms, this protocol is a quench of the trapping potential from a given $\omega$ to $0$, i.e. 
a trap release at $t=0$.

The Hamiltonian (\ref{H_2}) in terms of the fermionic field operators is
\begin{equation}\label{HFq_2}\fl
H = \int_{-L/2}^{L/2} dx\, \hat{\Psi}^{\dag}(x)\left[-\frac{1}{2}\frac{\partial^{2} }{\partial x^2} \right]\hat{\Psi}(x),\qquad
{\rm with} \quad  \hat{\Psi}(x-L/2)=\hat{\Psi}(x+L/2),
\end{equation}
which is diagonalized by Fourier transform in terms of the free fermionic operators $\hat{\eta}_{k}$ and $\hat{\eta}^{\dag}_{k}$ 
(with $k=2\pi m/L$ and $m$ integer)
\begin{equation}
H = \sum_{k=-\infty}^{\infty} \frac{k^2}{2}\hat{\eta}^{\dag}_{k}\hat{\eta}_{k},\quad\hat{\eta}_{k} = \int_{-L/2}^{L/2}dx \, \varphi^{*}_{k}(x)\hat{\Psi}(x),\quad \varphi_{k}(x) = \frac{\mathrm{e}^{-ikx}}{\sqrt{L}}.
\end{equation}
The time evolution of an observable is obtained in a standard way:
\begin{enumerate}
\item
Write the desired observable in terms of the post-quench mode operators $\hat{\eta}_{k}$, 
whose time-evolution, in Heisenberg representation, is 
\begin{equation}
\hat{\eta}_{k}(t)  =  \mathrm{e}^{iH t} \hat{\eta}_{k} \mathrm{e}^{-iH t}  
=  \mathrm{e}^{itk^2 \hat{\eta}^{\dag}_{k} \hat{\eta}_{k}/2} \hat{\eta_{k}} \mathrm{e}^{-itk^2 \hat{\eta}^{\dag}_{k} \hat{\eta}_{k}/2}
=  \mathrm{e}^{-ik^2 t /2}\hat{\eta}_{k}.
\end{equation}
\item
Write the post-quench mode operators $\hat{\eta}_{k}$ as a function of the pre-quench operators $\hat{\xi}_{j}$, 
whose action on the initial state is trivial.
\end{enumerate}
The relation between pre-quench and post-quench mode-operators can be written as
\begin{equation}
\hat{\eta}_{k}  =  \int_{-L/2}^{L/2}dx \, \varphi^{*}_{k}(x)\sum_{j=0}^{\infty}\chi_{j}(x)\hat{\xi}_{j}
= \sum_{j=0}^{\infty}A_{k,j}\hat{\xi}_{j},
\end{equation}
where we introduced the overlap between the pre-quench and the post-quench one-particle eigenfunctions
\begin{equation}\label{Akj}
A_{k,j} \equiv \int_{-L/2}^{L/2}dx\,\varphi^{*}_{k}(x)\chi_{j}(x).
\end{equation}
The inverse relation is 
\be
\fl \hat{\xi}_{j}    =   \int_{-\infty}^{\infty} dx\, \chi^{*}_{j}(x) \sum_{k=-\infty}^{\infty}\varphi_{k}(x)\hat{\eta}_{k}
  =   \sum_{k=-\infty}^{\infty}A^{*}_{k,j}\hat{\eta}_{k} + \mathcal{O}\left(j^{-1/4}(\omega L)^{j-3/4}\mathrm{e}^{-\omega L^2/8}\right),
\label{mapeta}
\ee
where the term $\mathcal{O}(\cdots)$ is due to the different domain of integration between the pre-quench and 
the post-quench Hamiltonians.

All this derivation is completely general and it is the practical way we construct the exact time evolution for 
finite number of particles $N$, finite $L$ and $\omega$. 
However, the results greatly simplify in the TD limit if this is properly defined as follows. 
We should consider $N,L\rightarrow\infty$ at fixed density $n=N/L$ and, at the same time, $\omega\to 0$ with $\omega N$ 
constant (i.e. fixed `initial density'). This is exactly the same TD limit defined in Ref. \cite{ck-12}. 
In terms of these TD quantities the trap release condition $\ell < L/2$ reads $\sqrt{\omega N}> 2\sqrt{2}n$. 
Under this condition, the functions $\chi_{j}(x)$ entering in the definition of $|\Psi_{0}\rangle$ (i.e. with $j<N$)
are exponentially small outside 
the interval $[-L/2,L/2]$ and the mapping between the operators $\hat{\xi}_{j}$ and $\hat{\eta}_{k}$ in Eq. (\ref{mapeta})
is exact also with the integration domain in Eq. (\ref{Akj}) extended to $\pm\infty$. 
Thus, if the trap release condition is satisfied, the overlaps $A_{k,j}$ are simply  
the Fourier transforms of the eigenfunctions of the one-dimensional  harmonic oscillator, i.e. 
\begin{equation}\label{A_kj}
A_{k,j}  =  \frac{1}{\sqrt{L}}\int_{-\infty}^{\infty}dx\,\chi_{j}(x)\mathrm{e}^{ikx}
 =  i^{j}\sqrt{\frac{2\pi}{\omega L}}\chi_{j}(k/\omega).
\end{equation}

A very important quantity for the non-equilibrium dynamics is the expansion velocity of the gas in full space.
This is obtained straightforwardly from the analytic solution of the dynamics \cite{mg-05} which 
we will discuss later, but can be also simply  written down from elementary arguments.
Indeed this velocity is determined by the maximum energy single-particle occupied level in the 
initial state with energy $\epsilon_N=\omega(N-1/2)\simeq \omega N$. 
In terms of the post-quench Hamiltonian with single particle spectrum $\epsilon_k=k^2/2$,  $\epsilon_N$
corresponds to an initial Fermi-momentum $k^0_F=\sqrt{2\omega N}$. Since $v_k=d\epsilon_k/dk=k$, we have
for the Fermi velocity $v=v_{k_F^0}$
\be
v=\sqrt{2\omega N}\,.
\label{vexp}
\ee
Notice how the expansion velocity $v$ remains finite in the proper TD limit with $\omega N$
constant.


\section{The two-point fermionic correlation function}
\label{sec3}

The easiest observable that we can calculate is the two-point fermionic correlator 
\be
C(x,y;t)\equiv\langle\hat{\Psi}^{\dag}(x,t)\hat{\Psi}(y,t)\rangle.
\ee
Indeed, since the Hamiltonian is quadratic in the fermionic operators, the evolved state is a Slater determinant and 
Wick's theorem applies allowing to obtain (with some work as we shall see) all other observables.
In terms of one-particle wave functions the fermionic correlator is 
\begin{eqnarray}
\fl C(x,y;t) & = & \langle\Psi_{0}|\mathrm{e}^{iH_{0}t}\hat{\Psi}^{\dag}(x)\hat{\Psi}(y)\mathrm{e}^{-iH_{0}t}|\Psi_{0}\rangle
=\sum_{k,p=-\infty}^{\infty}\varphi^{*}_{k}(x)\varphi_{p}(y) \langle\Psi_{0}|\mathrm{e}^{iH_{0}t}\hat{\eta}^{\dag}_{k}\hat{\eta}_{p}\mathrm{e}^{-iH_{0}t}|\Psi_{0}\rangle \nonumber \\ \fl 
& = & \sum_{k,p=-\infty}^{\infty} \varphi^{*}_{k}(x)\varphi_{p}(y) \mathrm{e}^{i(k^2-p^2)t/2}
\sum_{i,j=0}^{\infty}A^{*}_{k,i} A_{p,j}\langle\Psi_{0}|\hat{\xi}^{\dag}_{i}\hat{\xi}_{j}|\Psi_{0}\rangle\nonumber\\ \fl
  & = & \sum_{j=0}^{N-1} \phi^{*}_{j}(x,t) \phi_{j}(y,t),\label{C_F}
\end{eqnarray}
which is a well-known result for Slater determinants.

The time evolved one-particle wave functions are the solutions to the Schr\"odinger equation 
\be
i\partial_{t}\phi_{j}(x,t) = \mathcal{H}\phi_{j}(x,t), \quad {\rm with} \quad  \phi_{j}(x,0)=\chi_{j}(x), 
\ee
where $\mathcal{H}=-\partial^{2}_{x}/2$ is the single particle Hamiltonian with PBC.
In terms of the overlaps in Eq. (\ref{A_kj}) the solutions to this equation read 
\begin{equation}\label{one_particle_0}
\phi_{j}(x,t) = \sum_{p=-\infty}^{\infty}A_{p,j}\varphi_{p}(x)\mathrm{e}^{-i p^2 t/2}.
\end{equation}

\subsection{The time average}
Let us first compute the time average of the fermionic correlation function since if a large time limit of Eq. (\ref{C_F}) exists, 
it should be equal to its time average. 
To this aim, it is convenient to split the double momentum sum in Eq. (\ref{C_F}) in a term with $p\neq\pm k$ and 
another with $p=\pm k$. After time-averaging only the latter terms survive to give
\begin{equation}
\overline{C(x,y;t)}  =   \underbrace{\frac{1}{L}\sum_{k=-\infty}^{\infty}\mathrm{e}^{ik(x-y)}B_{k,k}}_{C_{+}}
  +  \underbrace{\frac{1}{L}\sum_{k=-\infty}^{\infty}\mathrm{e}^{ik(x+y)}B_{k,-k}}_{C_{-}}
 - \frac{1}{L}B_{0,0},
\end{equation}   
where we defined
\begin{equation}\label{B_kp}
B_{k,p} \equiv \sum_{j=0}^{N-1}A^{*}_{k,j} A_{p,j} = \frac{2\pi}{\omega L}\sum_{j=0}^{N-1}\chi_{j}(k/\omega)\chi_{j}(p/\omega),
\end{equation}
in terms of the overlaps in Eq. (\ref{A_kj}).

We now calculate the TD limit of the three pieces $C_+$, $C_-$ and $B_{0,0}$ separately. 
The writing is simplified by the use of the Dirac notation for the one-particle states
\be
\varphi_{k}(x)=\langle x|k\rangle, \qquad {\rm and} \quad \chi_j(x)=\langle x|\chi_j\rangle,
\ee
(we use $|\chi_j\rangle$ instead of $|j\rangle$ to avoid confusion with the state $|k\rangle$).
Notice that, because of our normalization, the momentum operator $\hat{P}$ acts on free-waves as 
$\hat{P}|k\rangle = -k|k\rangle$.

\begin{figure}[t]
\center\includegraphics[width=\textwidth]{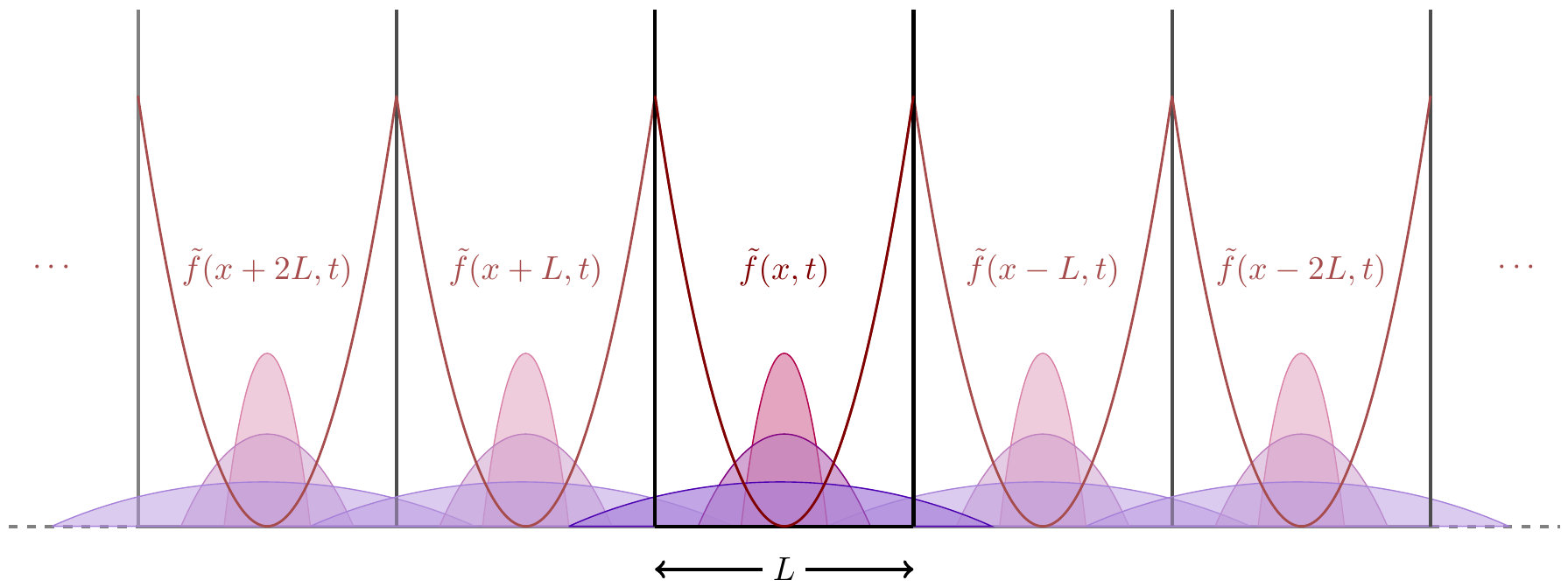}
\caption{Pictorial representation of the solution of the trap release dynamics in a ring as a superposition of replicas 
of the infinite-space time-evolved function $\tilde{f}(x,t)$ periodically shifted in space. 
See \ref{appA} for the mathematical derivation.}
\label{sketch_pbc}
\end{figure}

Let us first compute $C_{+}$: 
\begin{eqnarray}
\fl C_{+} & = & \frac{1}{L}\sum_{j=0}^{N-1}\sum_{k=-\infty}^{\infty}\mathrm{e}^{ik(x-y)}|A_{k,j}|^{2}
  =   \frac{1}{L}\sum_{j=0}^{N-1}\sum_{k=-\infty}^{\infty}\mathrm{e}^{ik(x-y)}|\langle k|\chi_{j}\rangle|^{2} 
  \nonumber\\ \fl & = &  
 \frac{1}{L}\sum_{j=0}^{N-1}\sum_{k=-\infty}^{\infty} \langle\chi_{j}|\mathrm{e}^{-i\hat{P}(x-y)}|k\rangle\langle k|\chi_{j}\rangle  
 =\frac{1}{L}\sum_{j=0}^{N-1} \langle\chi_{j}|\mathrm{e}^{-i\hat{P}(x-y)}|\chi_{j}\rangle.
\end{eqnarray}
These matrix elements can be calculated using $\hat{P}=(\hat{a}^{\dag}-\hat{a})\sqrt{\omega/2}$  
(with $\hat{a},\,\hat{a}^{\dag}$ the harmonic oscillator ladder operators) 
and the Baker-Campbell-Hausdorff formula:
\begin{eqnarray}
\fl \langle\chi_{j}|\mathrm{e}^{-i\hat{P}(x-y)}|\chi_{j}\rangle  &=&  
\langle\chi_{j}|\mathrm{e}^{(\hat{a}^{\dag}-\hat{a})(x-y)\sqrt{\omega/2}}|\chi_{j}\rangle   
=\mathrm{e}^{\omega(x-y)^{2}/4}\langle\chi_{j}|\mathrm{e}^{-\hat{a}(x-y)\sqrt{\omega/2}}\mathrm{e}^{\hat{a}^{\dag}(x-y)\sqrt{\omega/2}}|\chi_{j}\rangle  \nonumber \\
 &=&  \mathrm{e}^{\omega(x-y)^{2}/4} \sum_{m,n=0}^{\infty}\frac{(-1)^{m}}{m!n!}\left[\sqrt{\frac{\omega}{2}}(x-y)\right]^{m+n}
 \langle\chi_{j}|\hat{a}^{m}(\hat{a}^{\dag})^{n}|\chi_{j}\rangle  \nonumber\\
& =&   \mathrm{e}^{\omega(x-y)^{2}/4} \sum_{m,n=0}^{\infty}\frac{(-1)^{m}}{m!n!}\left[\sqrt{\frac{\omega}{2}}(x-y)\right]^{m+n}
\frac{\sqrt{(j+m)!(j+n)!}}{j!} \delta_{m,n}\nonumber\\
& =& \mathrm{e}^{\omega(x-y)^{2}/4} \frac{1}{j!}\sum_{m=0}^{\infty}\frac{(j+m)!}{m!^2}\left[-\omega(x-y)^{2}/2 \right]^{m} \nonumber\\
&=&  \mathrm{e}^{\omega(x-y)^{2}/4} {}_1F_{1}(j+1;1;-\omega(x-y)^{2}/2), 
\end{eqnarray}
where $_1F_1(a;b;z)$ is the hypergeometric function. 
To evaluate $C_{+}$ in the TD limit we need to calculate the coefficients of the powers 
of $\omega N (x-y)^2 / 2 $ with $\omega N = cst.$, $N,L\to\infty$ and $\omega\to 0$
\begin{eqnarray}\label{I_plus}
C_{+}  & =&   \frac{1}{L} \mathrm{e}^{\omega (x-y)^{2}/4} \sum_{j=0}^{N-1}\frac{1}{j!}\sum_{m=0}^{\infty}\frac{(j+m)!}{m!^2 N^m}\left[-\omega N(x-y)^{2}/2 \right]^{m}\nonumber \\
& = &  \frac{1}{L} \mathrm{e}^{\omega(x-y)^{2}/4} \sum_{m=0}^{\infty} \underbrace{\left( \sum_{j=0}^{N-1}\frac{(j+m)!}{j!} \right )}_{\frac{(m+N)!}{(1+m)(N-1)!}} \frac{\left[-\omega N(x-y)^{2}/2 \right]^{m}}{m!^2 N^m} \nonumber \\
 &= &   \frac{1}{L} \mathrm{e}^{\omega(x-y)^{2}/4} \sum_{m=0}^{\infty} \left(\frac{(m+N)!}{(N-1)! N^m} \right ) \frac{\left[-\omega N(x-y)^{2}/2 \right]^{m}}{m! (m+1)!}. 
\end{eqnarray}
The large $N$ limit of the expression in the last line in the round parenthesis is just $N$ and therefore we obtain 
\begin{equation}\fl
C_{+}  \simeq  \frac{N}{L}\sum_{m=0}^{\infty}\frac{1}{m!(m+1)!} \left[-\omega N(x-y)^{2}/2 \right]^{m}
 =  2n\frac{J_{1}[ \sqrt{2\omega N}(x-y)]}{\sqrt{2\omega N}(x-y)},
\end{equation}
with $J_{1}(z)$ the Bessel function.
Notice $\sqrt{2\omega N}=k_F^0$, the initial Fermi momentum.

In order to evaluate $C_{-}$ we notice from Eq. (\ref{A_kj}) that $A_{-k,j} = (-1)^{j}A_{k,j}$. 
Thus the only differences compared to $C_{+}$ are (i) an extra sign $(-1)^{j}$ and (ii) the replacement  of $y$ with $-y$. 
Therefore the calculation of $C_{-}$ follows the same steps as for $C_{+}$ up to the middle line in Eq. (\ref{I_plus}), 
which now becomes
\begin{equation}
C_{-} = \frac{1}{L} \mathrm{e}^{\omega(x+y)^{2}/4} \sum_{m=0}^{\infty}\left( \sum_{j=0}^{N-1}(-1)^{j}\frac{(j+m)!}{j!} \right )
\frac{\left[-\omega N(x+y)^{2}/2 \right]^{m}}{m!^2 N^m},
\end{equation}
but unlike $C_+$, the piece in the round parenthesis does not grow like $N$ for large $N$, but it has a finite limit because 
$\sum_{j=0}^{\infty}(-1)^{j}\frac{(j+m)!}{j!} = m!/2^{m+1}$. Therefore, for $N\to\infty$, we have
\begin{equation}
C_{-}  \simeq  \frac{1}{2L} \mathrm{e}^{\omega(x+y)^{2}/4} \sum_{m=0}^{\infty} \frac{\left[-\omega(x+y)^{2}/4 \right]^{m}}{m! }
=  \frac{1}{2L}.
\end{equation}
Thus, unlike $C_{+}$ which is finite in the TD limit, $C_{-}$ decays to zero as $1/L$ giving just a finite-size correction to the 
correlation function.
 
The calculation of the last term $-B_{0,0}/L$ is straightforward and in the TD limit we have
\be
-\frac{B_{0,0}}L= -\frac{(2\pi n)^{3/2}}{\sqrt{2\omega N}L^{1/2}},
\ee
which also vanishes for large $L$.

Summing up, the time average of the fermionic correlation function in the thermodynamic limit gets a non vanishing
contribution only from $C_+$ and so it is
\be
\overline{C_{F}(x,y;t)} = 2n\frac{J_{1}[ \sqrt{2\omega N}(x-y)]}{\sqrt{2\omega N}(x-y)}=
2n\frac{J_{1}[ k_F^0 (x-y)]}{k_F^0(x-y)}.
\label{Caver}
\ee

\subsection{The time dependent one-particle problem}
As detailed in \ref{appA}, Fourier analysis allows us to rewrite the one-particle evolution in Eq. (\ref{one_particle_0}) 
in terms of the time evolved wave function in the infinite-space $\phi_{j}(x,t)$
\begin{equation}\label{phi_L_t}
\phi_{j}(x,t) = \sum_{p=-\infty}^{\infty}\phi^{\infty}_{j}(x+pL,t).
\end{equation}
This formula is valid for any one-particle time-dependent problem with PBC and its physical 
meaning is very simple: in a circle of length $L$ with PBC, the time evolution is equivalent to 
the sum (superposition) of the time evolution of infinite copies (replicas labeled by $p$ in Eq. (\ref{phi_L_t})) 
of the initial state in infinite space but shifted by integer multiples of $L$
(see Fig. \ref{sketch_pbc} for a pictorial representation). 
For the particular case at hand we have
\begin{eqnarray}
\phi^{\infty}_{j}(x,t) & = & \frac{i^{j}}{\sqrt{2\pi\omega}}\int_{-\infty}^{\infty}dk\,\chi_{j}(k/\omega)\mathrm{e}^{-ik^2 t/2}\mathrm{e}^{-ikx} 
\\ & = & 
\frac{i^{j}(\omega/\pi)^{1/4}}{\sqrt{2^{j+1}j!\pi\omega}}
\int_{-\infty}^{\infty}dk\, \mathrm{e}^{-ikx -\frac{1+i\omega t}{2\omega}k^2}H_{j}(k/\sqrt{\omega})\nonumber,
\end{eqnarray}
and using the following property of the Hermite polynomials
\begin{displaymath}
\int_{-\infty}^{\infty}dx\, \mathrm{e}^{-(x-y)^2}H_{j}(ax) = \sqrt{\pi}(1-a^2)^{j/2}H_{j}\left( \frac{ay}{\sqrt{1-a^2}}\right),
\end{displaymath}
we get
\begin{eqnarray}
\int_{-\infty}^{\infty}dk\, \mathrm{e}^{-ikx -\frac{1+i\omega t}{2\omega}k^2}
H_{j}\Big(\frac{k}{\sqrt{\omega}}\Big)& = & (-i)^{j}\sqrt{\frac{2\pi\omega}{1+i\omega t}}\Big( \frac{1-i\omega t}{1+i\omega t} \Big)^{j/2}
\\&& \qquad\times
\mathrm{e}^{-\frac{\omega x^2}{2(1+i\omega t)}} H_{j}\Big(x\sqrt{\frac{\omega}{1+\omega^2 t^2}}\Big),\nonumber
\end{eqnarray}
and therefore
\begin{eqnarray}\label{phi_inf_t}
\fl \phi^{\infty}_{j}(x,t) & = & \frac{1}{\sqrt{2^j j!}}\left( \frac{\omega}{\pi} \right)^{1/4}\frac{1}{\sqrt{1+i\omega t}} \left( \frac{1-i\omega t}{1+i\omega t} \right)^{j/2}
\mathrm{e}^{-\frac{\omega x^2}{2(1+i\omega t)}} H_{j}\left(x\sqrt{\frac{\omega}{1+\omega^2 t^2}}\right)\nonumber \\ \fl
 & = & \frac{1}{\sqrt{1+i\omega t}} \left( \frac{1-i\omega t}{1+i\omega t} \right)^{j/2} \mathrm{e}^{-i\frac{t \omega^2 x^2}{2(1+\omega^2 t^2)}} \chi_{j}\left(\frac{x}{\sqrt{1+\omega^2 t^2}}\right),
\end{eqnarray}
which coincides with the result in Ref. \cite{mg-05}.
The full time dependence in the ring is obtained  by plugging the above equation (\ref{phi_inf_t}) in Eq. (\ref{phi_L_t}).

\begin{figure}[t]
\center\includegraphics[width=0.65\textwidth]{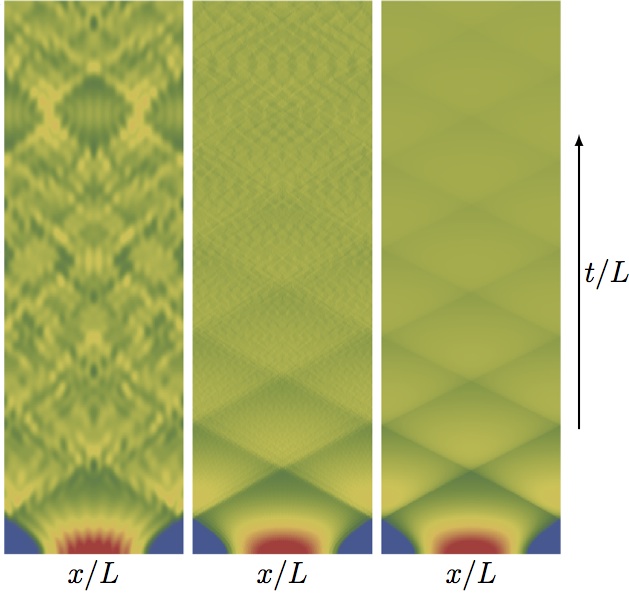}
\caption{Color plot of the numerically calculated density evolution $n(x,t)$ for $N=10,100,\infty$ (from left to right) 
at $N/L=1/2$ and $\omega N=5$ as a function of the rescaled space variable $x/L$ and for rescaled times $t/L\in[0,2].$}
\label{density_color}
\end{figure}

\begin{figure}[t]
\center\includegraphics[width=0.5\textwidth]{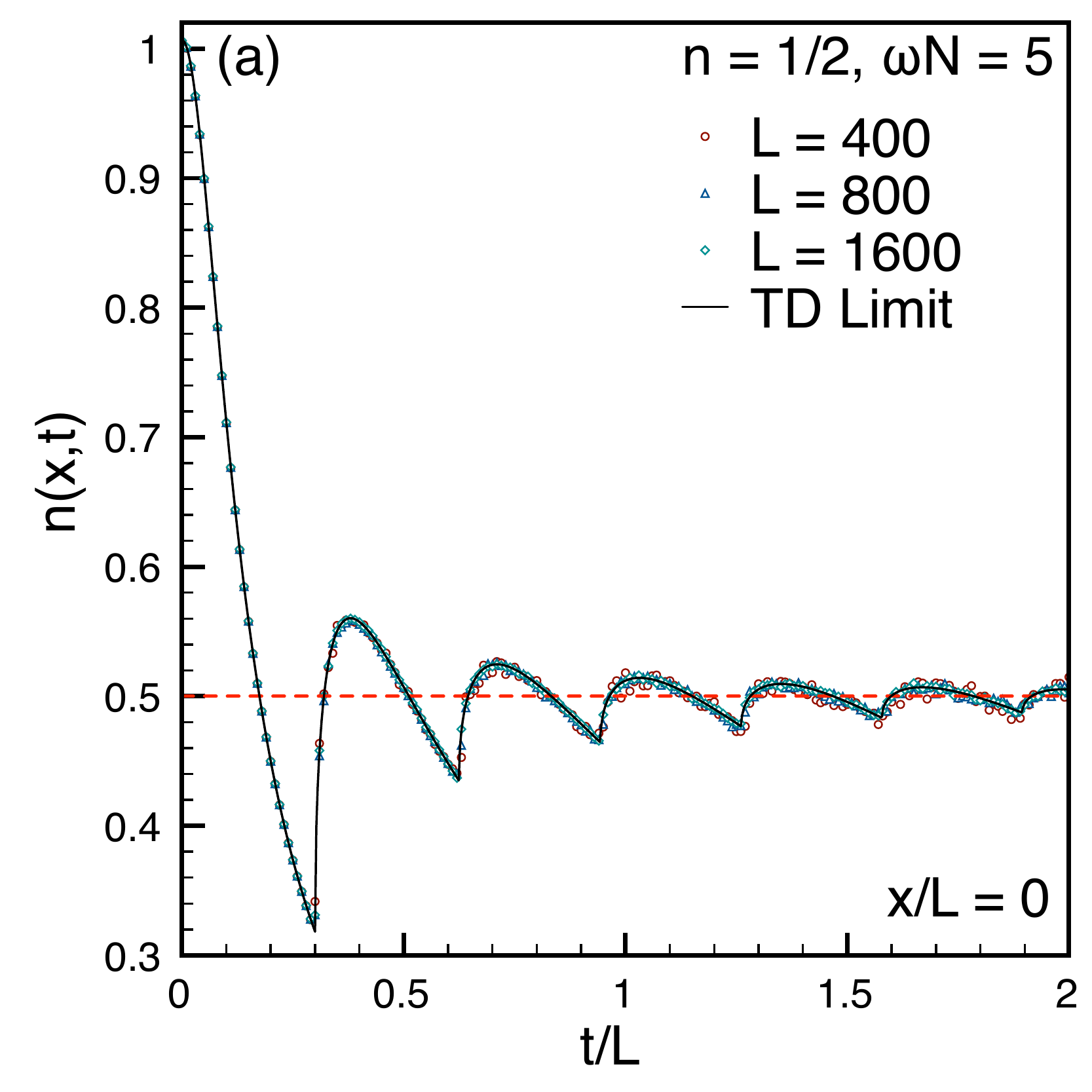}\includegraphics[width=0.5\textwidth]{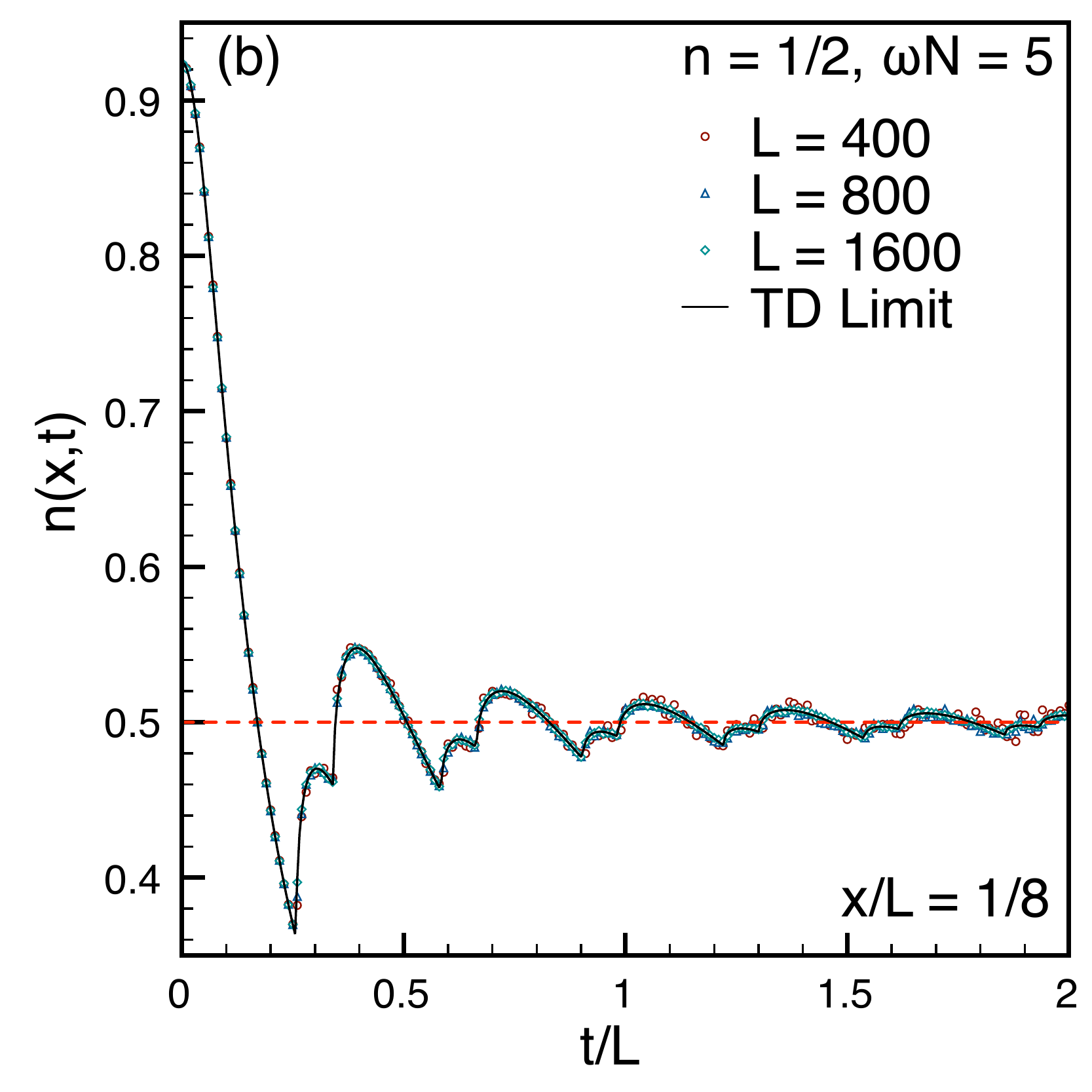}\\\includegraphics[width=0.5\textwidth]{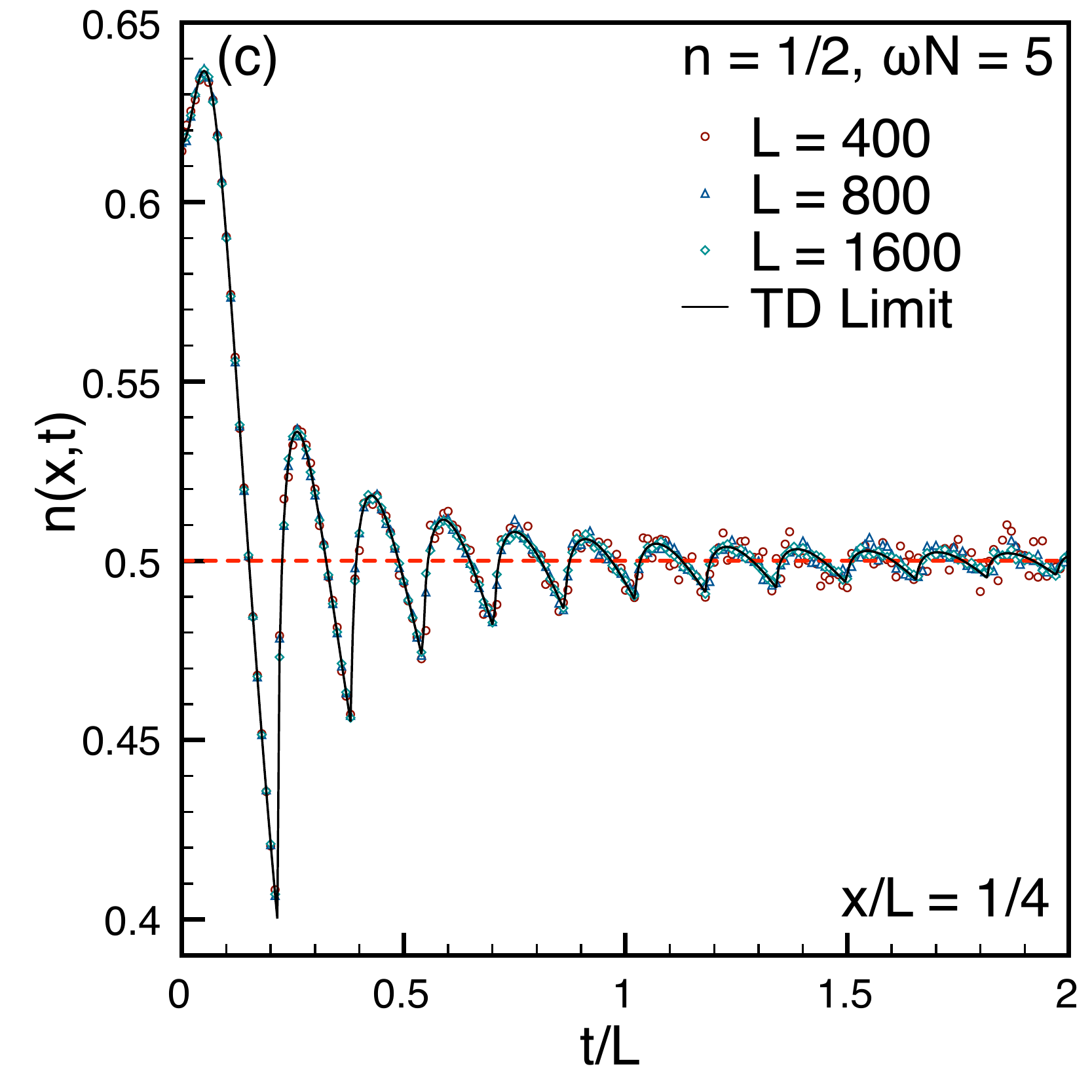}\includegraphics[width=0.5\textwidth]{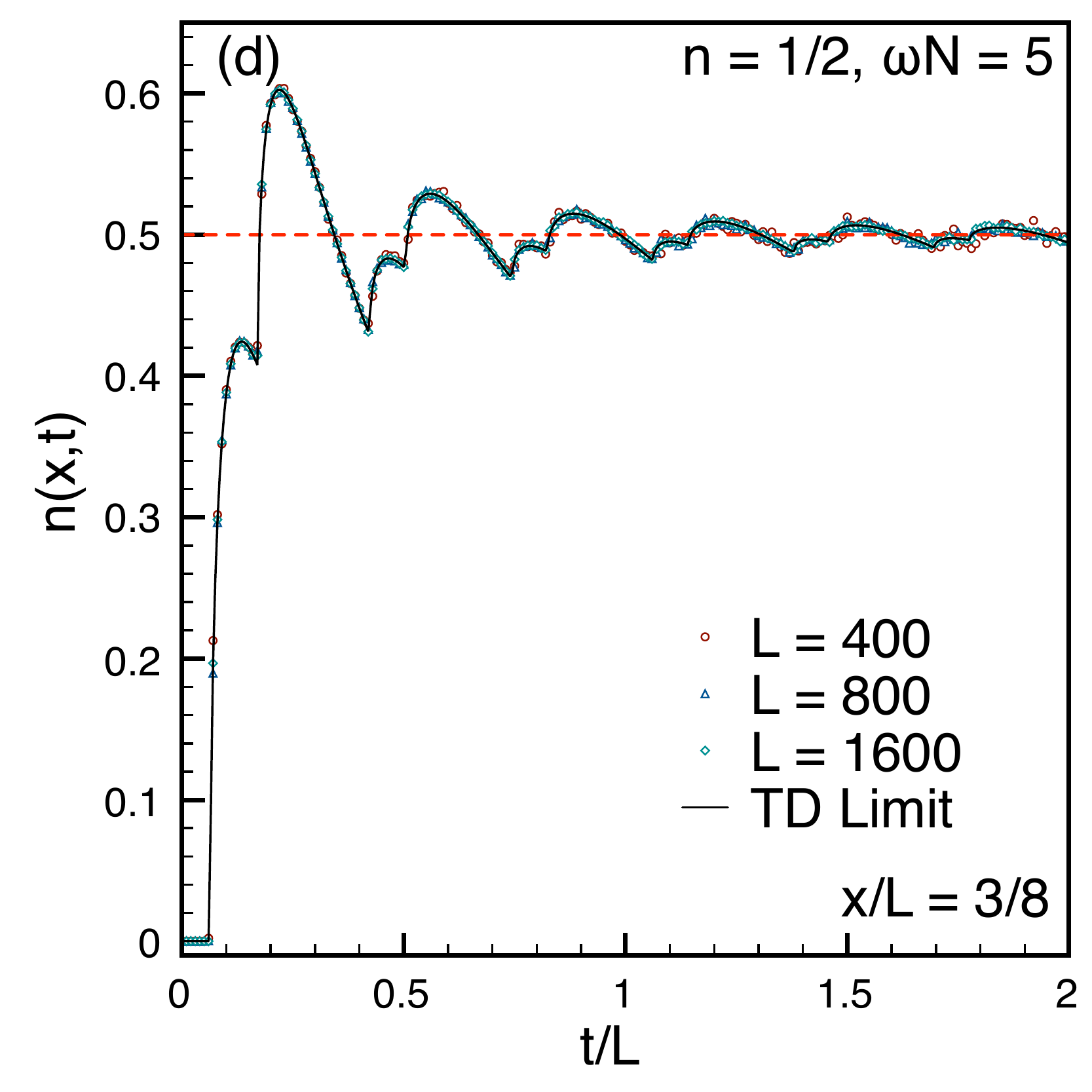}
\caption{In each panel we plot the time evolution of the density $n(x,t)$ as function of the rescaled time $t/L$
at fixed $x/L$: for different (large enough) sizes the curves collapse on top of each other. 
Dashed red lines indicate the equilibration value $n=N/L$ reached at infinite time. 
The symbols are the exact dynamics [cf. Eq. (\ref{nxt_1})]
for finite $N$, while full black lines are the TD limit in Eq. (\ref{nxt_TD}).} 
\label{figden}
\end{figure}

\begin{figure}[t]
\center\includegraphics[width=0.5\textwidth]{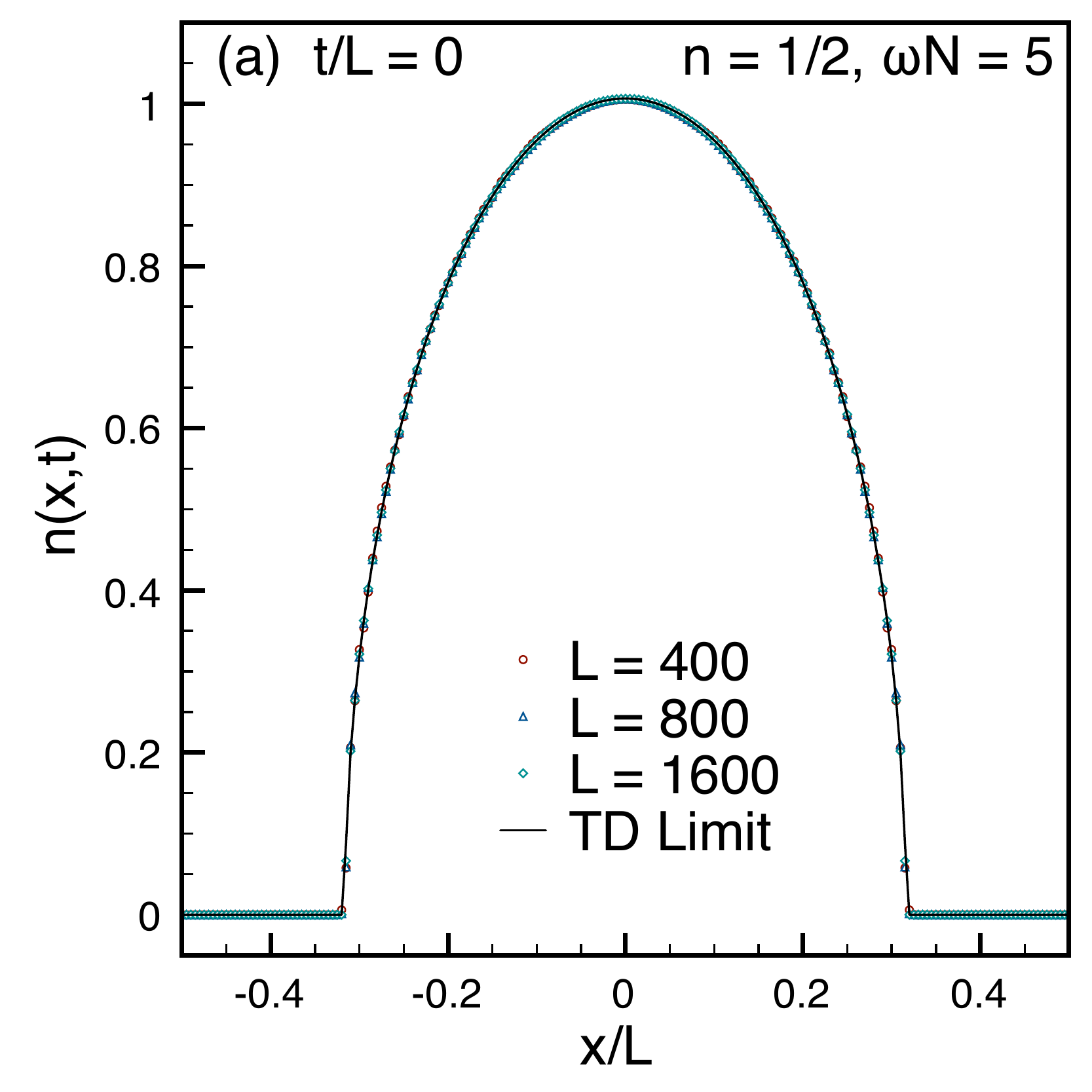}\includegraphics[width=0.5\textwidth]{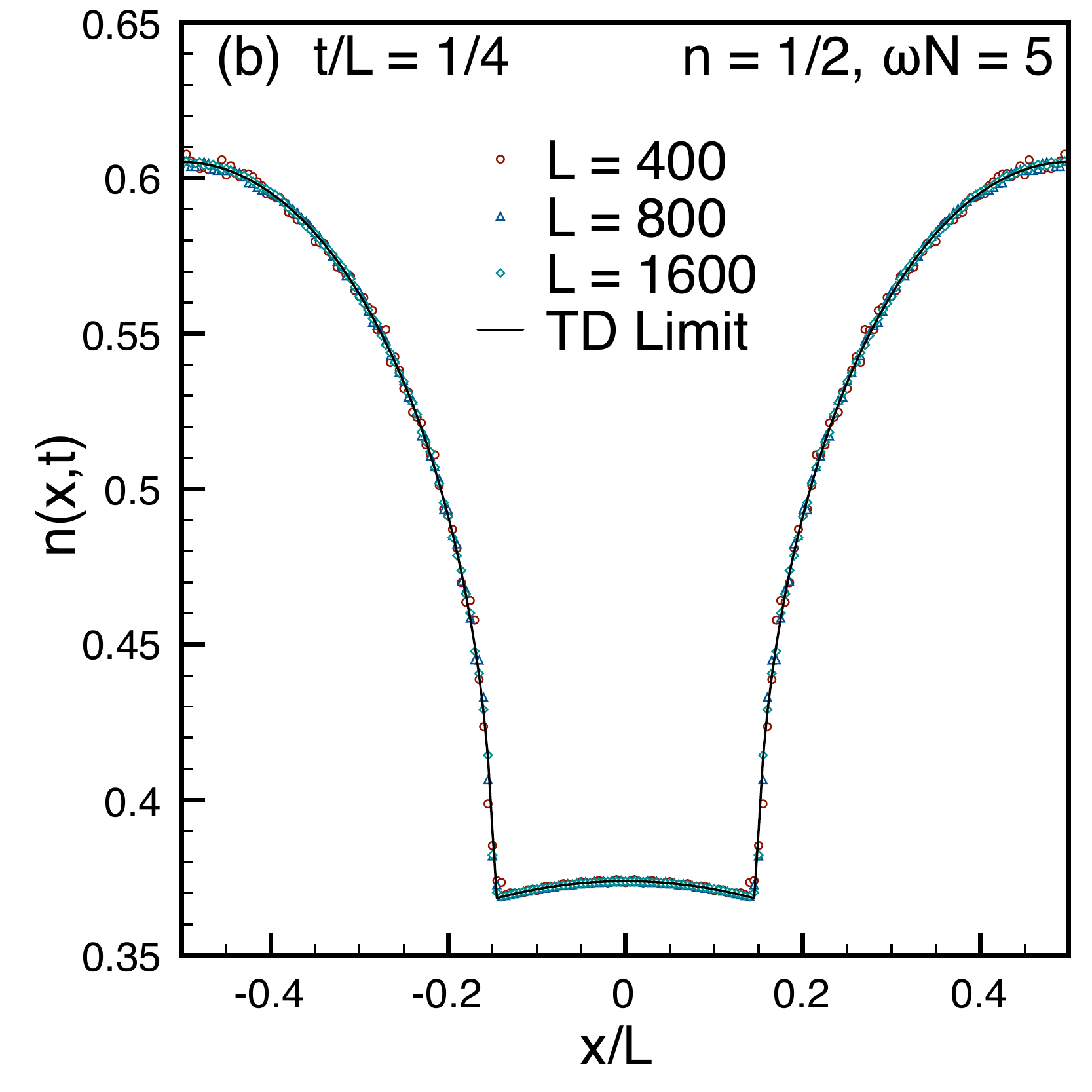}\\
\includegraphics[width=0.5\textwidth]{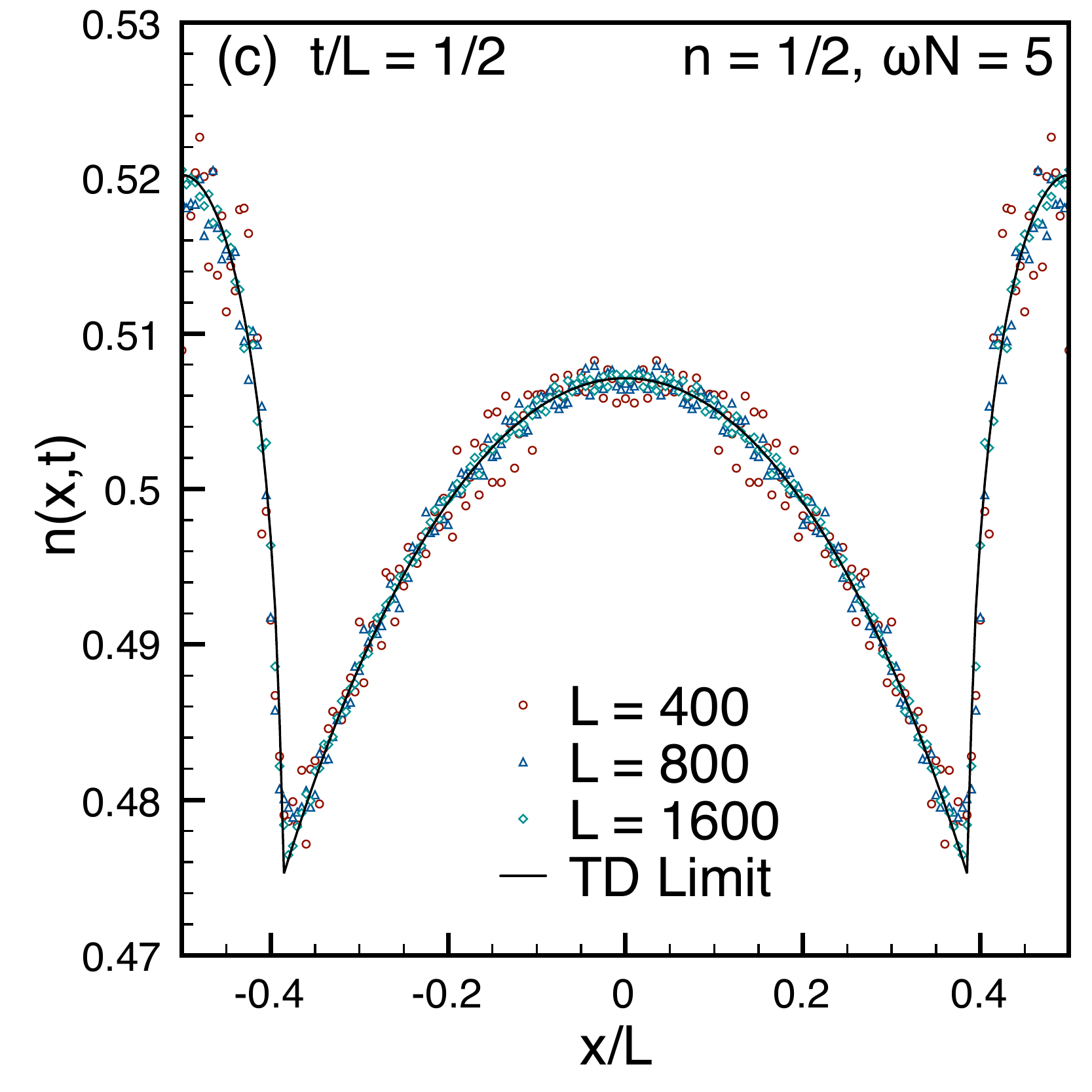}\includegraphics[width=0.5\textwidth]{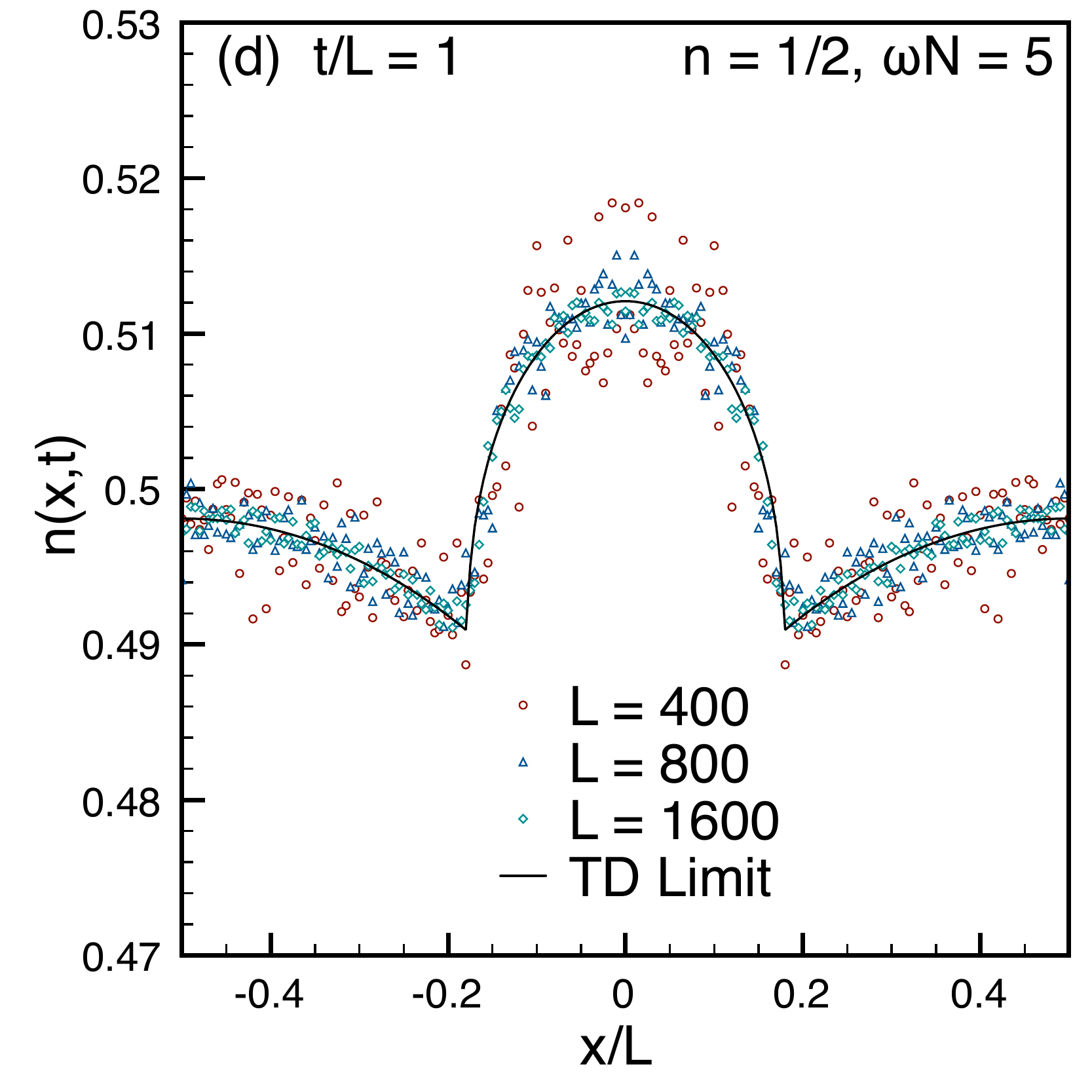}
\caption{In each panel we report the density profile $n(x,t)$ as function of $x/L$ for fixed $t/L$: again 
the results for various sizes collapse on a single curve when $L$ increases. 
Symbols are the exact dynamics for finite $N$ [cf. Eq. (\ref{nxt_1})], while full black lines are the TD limit in Eq. (\ref{nxt_TD}). 
As the time increases, the profiles tighten close to the equilibration value $n=N/L$ (note the vey different vertical scale 
in the four panels). 
} 
\label{figden2}
\end{figure}

\subsection{The time evolution of the density profile}
We start the time-dependent analysis of the many-body problem from the diagonal part of the fermionic correlation function, 
i.e. the particle density (both for fermions and bosons). 
Plugging the one-particle wave-functions (\ref{phi_inf_t}) and (\ref{phi_L_t}) into Eq. (\ref{C_F}) we obtain
\begin{eqnarray}\label{nxt_1}
n(x,t) &=& \frac{1}{\sqrt{1+\omega^{2}t^2}}\sum_{p,q=-\infty}^{\infty}
\exp\left\{i \frac{\omega^2 t}{2(1+\omega^2 t^2)} [(x+pL)^2-(x+qL)^2] \right\}\nonumber\\
&\times& \sum_{j=0}^{N-1}\chi_{j}\left(\frac{x+pL}{\sqrt{1+\omega^{2}t^2}}\right)\chi_{j}\left(\frac{x+qL}{\sqrt{1+\omega^{2}t^2}}\right),
\end{eqnarray}
which is an exact formula for any $N,L,\omega$.

The physical interpretation of Eq. (\ref{nxt_1}) is again the one suggested by Fig. \ref{sketch_pbc}.
The wave-functions of periodically placed replicas expand in infinite space and eventually overlap with each other when they 
reach a boundary between two replicas, that happens (approximately) at times which are integer multiples 
of the characteristic time $\tau\sim L/v$.
Therefore, if $t<\tau$ the boundaries are not reached and the system does not feel the PBC. 
When the time becomes much larger than $\tau$, the overlap between replicas (or the turning of the particles 
around the circle) leads, as we shall prove, to equilibration manifested as a uniform distribution. 
However, as already stressed in the introduction, the infinite time limit should be handled with care because revivals will 
take place for larger time scales of the order of $L^2$ (because the fundamental frequency of the double momentum sum 
is $2\pi/L^2$, i.e. the revival period is $L^2/2\pi$, see also \cite{deg-13} for a general treatment). 
Therefore in the TD limit, the physically relevant scaling regime is obtained by taking $x/L=cst.$ and $t/L=cst.$ and 
consequently $L^2/t\to\infty$, so that the revivals are eliminated.
In this regime the leading behavior of Eq. (\ref{nxt_1}) is extracted by stationary phase arguments and  it comes from 
the `diagonal replicas' , i.e. the terms with $p=q$. 
Indeed, in the TD limit with $\omega N= \omega L n = cst.$, the phase of the exponent in (\ref{nxt_1}) is stationary only for $p= \pm q$,
but the terms with $p=-q$ give a finite-size correction going like $L^{-1}$. 
Thus, in the TD limit, the leading behavior of the time-dependent density profile is given by
\begin{equation}
n(x,t) = \frac{1}{\sqrt{1+\omega^{2}t^2}}\sum_{p=-\infty}^{\infty}\sum_{j=0}^{N-1}
\left| \chi_{j}\left(\frac{x+qL}{\sqrt{1+\omega^{2}t^2}}\right) \right|^{2}.
\label{nxt2}
\end{equation}
To perform the sum over $j$, we use the Christoffel-Darboux formula for the Hermite polynomials $H_{j}(x)$ in Eq. (\ref{cdf})
which in the limit $N\to\infty$ leads to Eq. (\ref{CD_inf}).
Thus Eq. (\ref{nxt2}) can be written in terms of the particle density at initial time $n_0(x)$ in Eq. (\ref{n0}) as 
\begin{equation}\label{nxt_TD}
n(x,t) = \frac{1}{\sqrt{1+\omega^{2}t^2}}\sum_{p=-\infty}^{\infty}n_{0}\left(\frac{x+qL}{\sqrt{1+\omega^{2}t^2}}\right),
\end{equation}
showing that the density profile in the TD limit is simply given by the sum of the replicated densities and all 
the interference effects are subleading in $L$, as probably expected.

In Figs. \ref{density_color}, \ref{figden} and \ref{figden2} we show the numerically calculated exact time dependent density for finite 
but large $N$. 
For large enough systems, the numerical data perfectly agree with the above TD prediction for any time. 
The infinite-time limit $t/L\to\infty$ of Eq. (\ref{nxt_TD}) is straightforward and gives the expected result $n(x,\infty)=n$.

\begin{figure}[t]
\center\includegraphics[width=\textwidth]{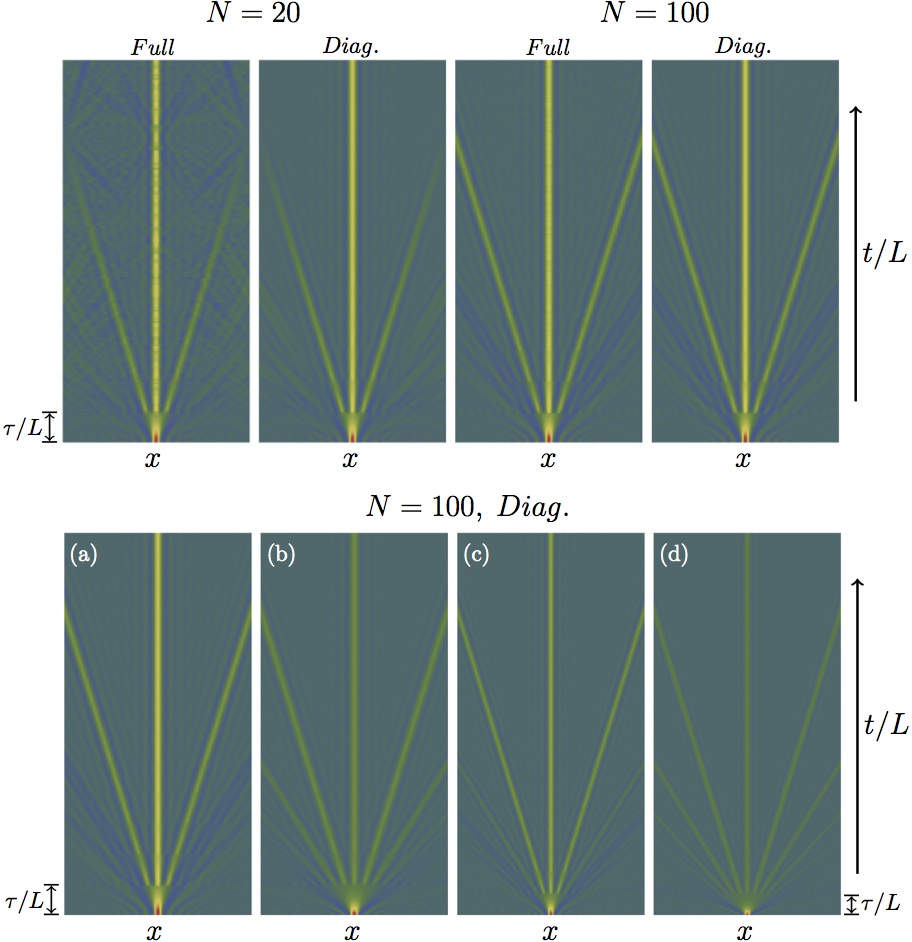}
\caption{Top: Color plot of the numerically calculated fermionic correlation function ${\rm Re}[C(x,0;t)]$ for $N=20,100$ 
considering both the full replica sum on $p,q$ (\textit{Full}) in Eq. (\ref{Cxyt_1}) and only the diagonal part $p=q$ (\textit{Diag.})
in Eq. (\ref{Cxyt_2}). We fix $x\in[-20,20]$ and $t/L\in[0,4]$. 
Notice how the differences between the full and the diagonal calculation, visible for $N=20$, disappear already for $N=100$. 
Bottom: For $N=100$ we report the correlation function ${\rm Re}[C(x,0;t)]$ (calculated as sum over only the diagonal 
term $p=q$ in Eq. (\ref{Cxyt_2})) for several values of TD parameters. 
From left to right: (a) $n=1/2,\,\omega N=5$; (b) $n=1/4,\; \omega N = 5$; (c) $n=1/2,\; \omega N = 10$; $n=1/4,\; \omega N = 10$. 
Notice (i) the slope of the signal lines (yellow) is equal to integer multiples of $2\pi/L$ 
(ii) the difference in the time-scale $\tau/L$ (at the bottom of the figure) which depends on $\omega N$.}
\label{corr_color}
\end{figure}

\begin{figure}[t]
\includegraphics[width=0.49\textwidth]{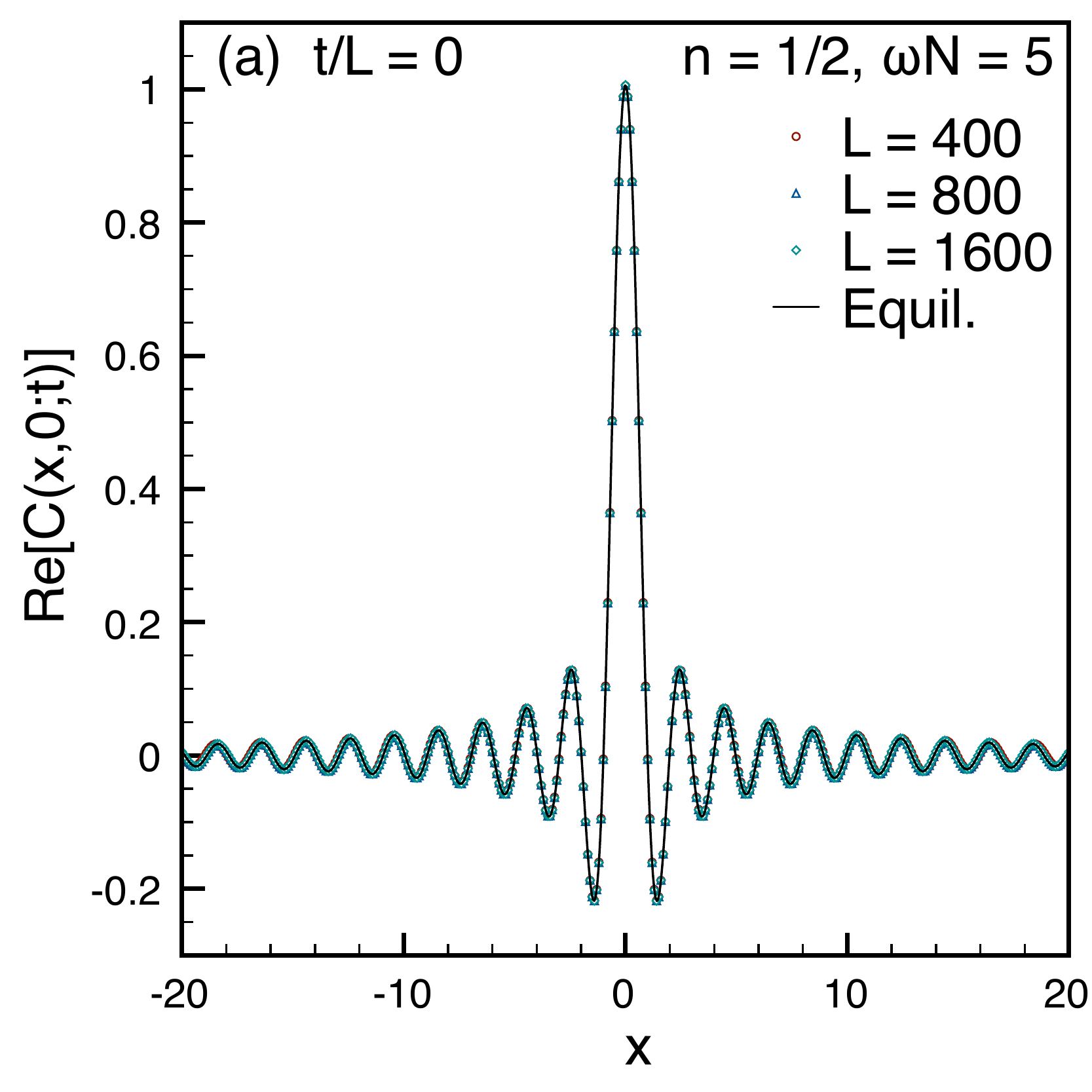}\includegraphics[width=0.49\textwidth]{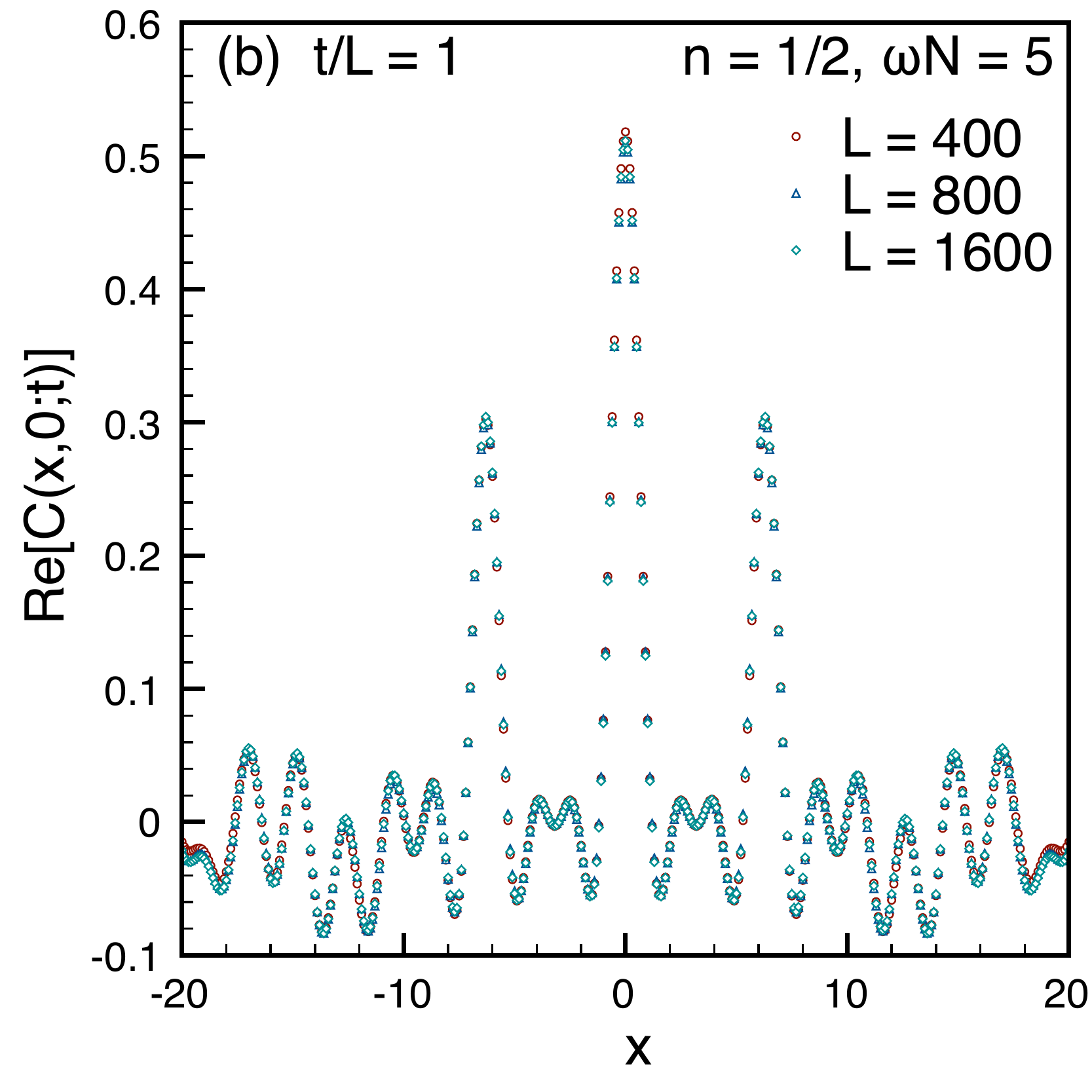}\\
\includegraphics[width=0.49\textwidth]{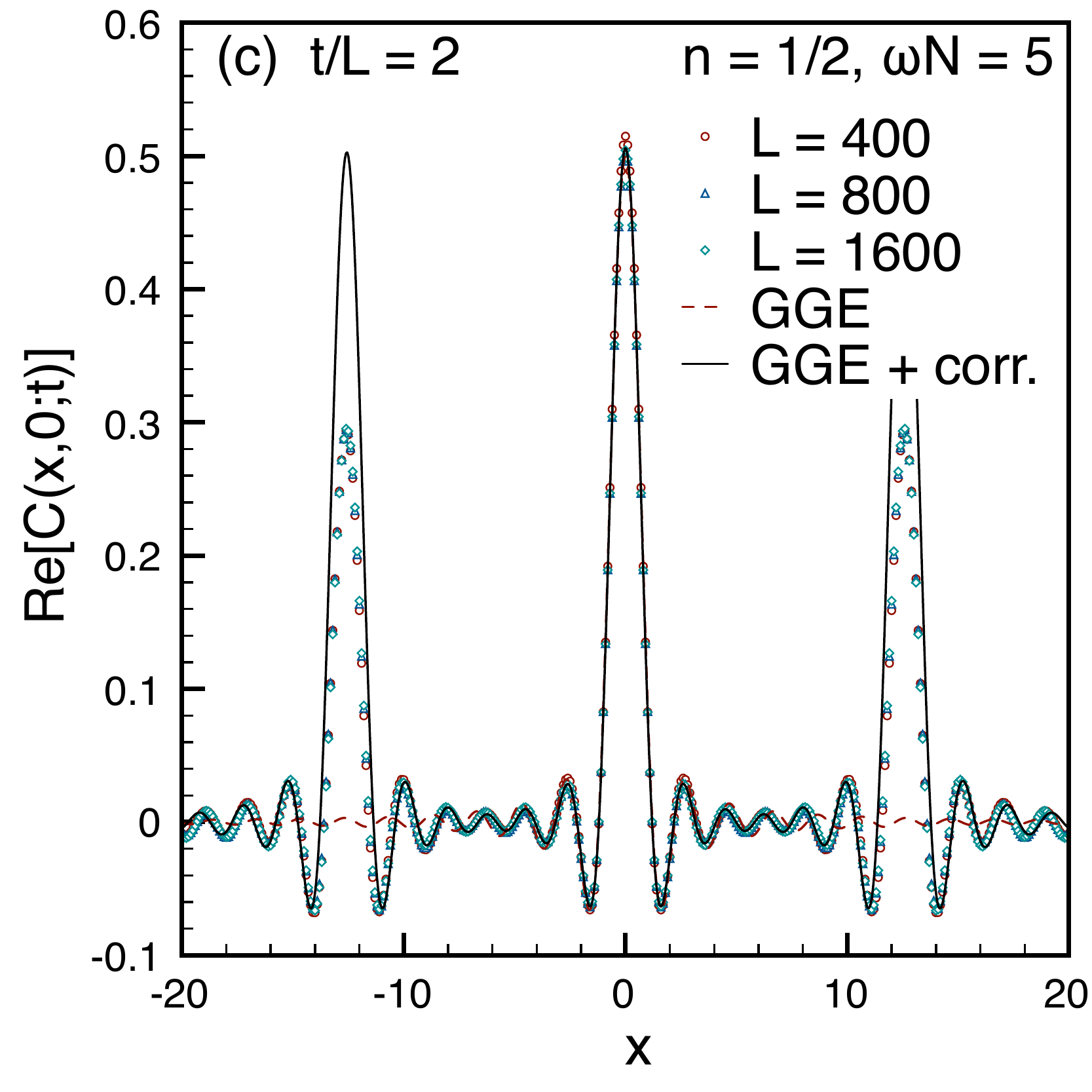}\includegraphics[width=0.49\textwidth]{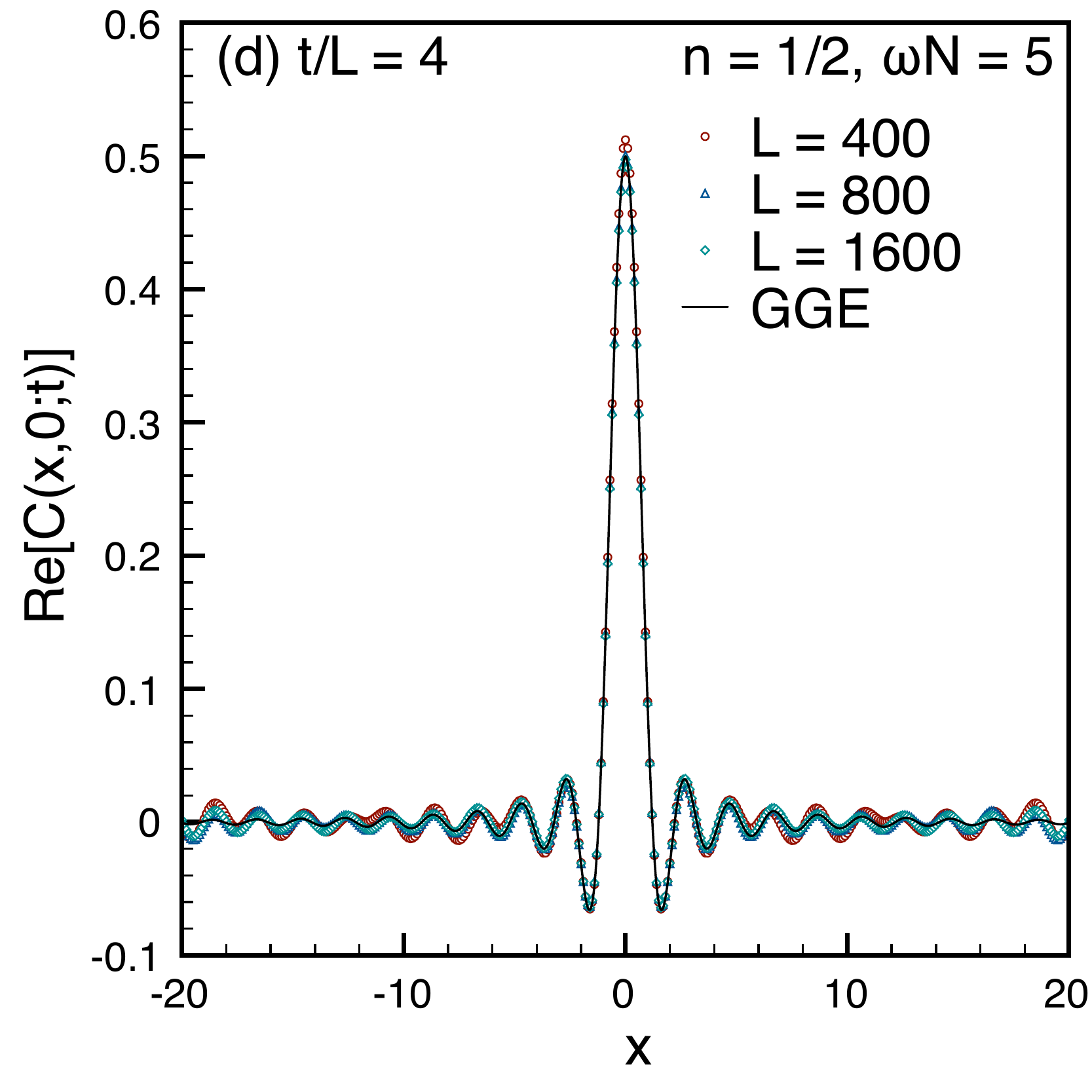}\\
\caption{Snapshots of the correlation ${\rm Re}[C(x,0;t)]$ at different rescaled times $t/L$ and sizes. 
The data for different large enough sizes nicely collapse on top of each other. 
In panel (a) for $t/L=0$, the full line is the initial correlation in the TD limit (cf. Eq. (\ref{Ctrap1})). 
Panels (b) and (c) show that, as time increases, two symmetric peaks are expelled from the central region.
In the panel (c) for $t/L=2$, the full line is the stationary value plus the first-order correction in Eq. (\ref{Cxyt_larget_corr}) 
which correctly describes the position of the two moving peaks, but not their amplitudes. 
For comparison,  in panel (c)  we report also the leading contribution for infinite time (red dashed line). 
In panel (d) with $t/L=4$, in the considered spatial region $x\in[-20,20]$, the time evolved data are 
almost indistinguishable from the stationary values (full line). 
} 
\label{figCxy}
\end{figure}

\subsection{The time evolution of the two-point fermionic correlation and its large-time limit}

The calculation of the time evolution of the two-point fermionic correlator is similar to the one just reported for the density. 
For finite $L,N,\omega$, plugging Eq. (\ref{phi_L_t}) into Eq. (\ref{C_F}) we have an exact starting point 
\begin{eqnarray}\label{Cxyt_1}\fl
C(x,y;t) &=& \frac{1}{\sqrt{1+\omega^{2}t^2}}
\sum_{p,q=-\infty}^{\infty}\exp\left\{i \frac{\omega^2 t}{2(1+\omega^2 t^2)} [(x+pL)^2-(y+qL)^2] \right\}\nonumber\\ \fl
&& \qquad\times
\sum_{j=0}^{N-1}\chi_{j}\left(\frac{x+pL}{\sqrt{1+\omega^{2}t^2}}\right)\chi_{j}\left(\frac{y+qL}{\sqrt{1+\omega^{2}t^2}}\right),
\end{eqnarray}
which once again can be easily interpreted in terms of replicas.

\begin{figure}[t]
\includegraphics[width=0.5\textwidth]{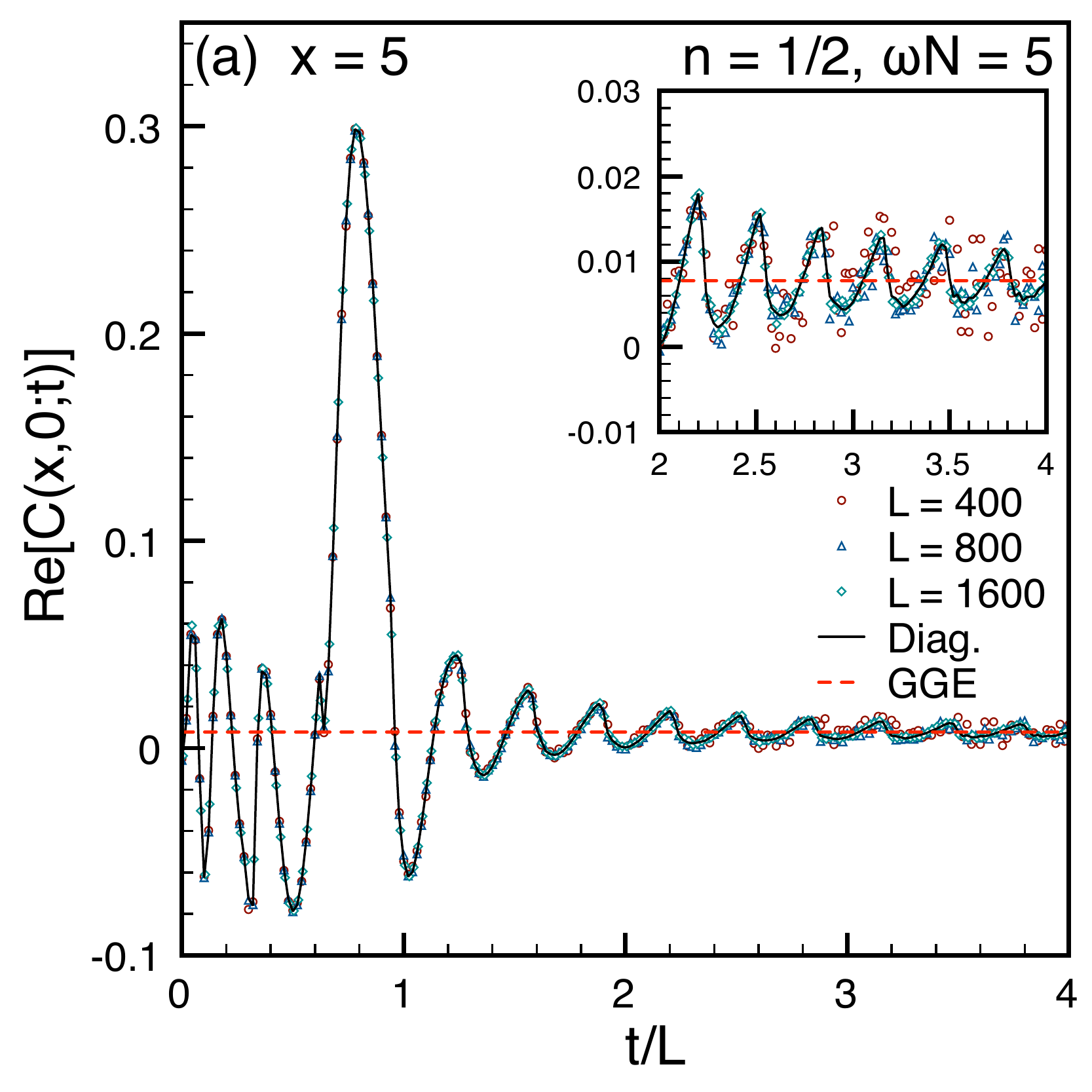}\includegraphics[width=0.5\textwidth]{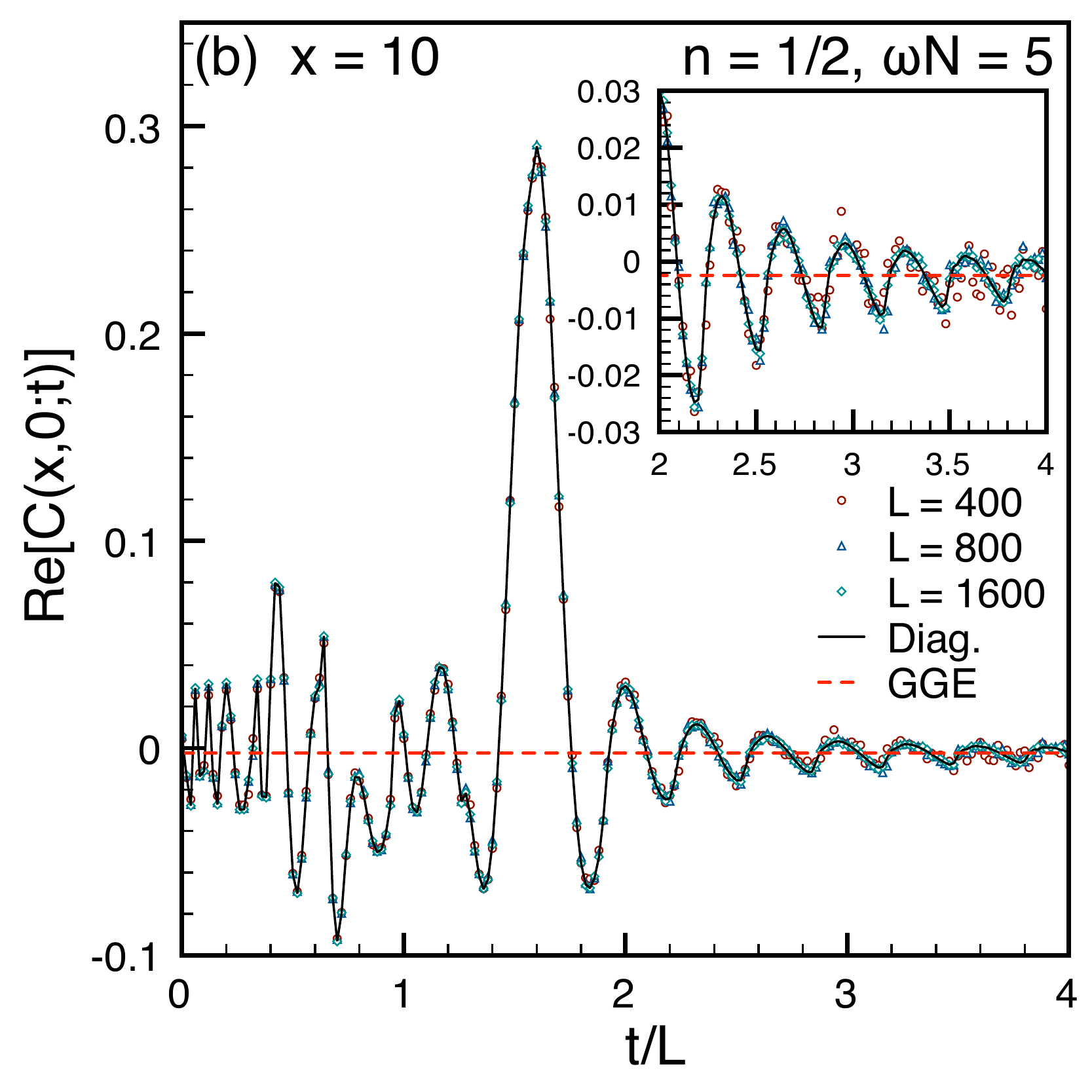}\\
\includegraphics[width=0.5\textwidth]{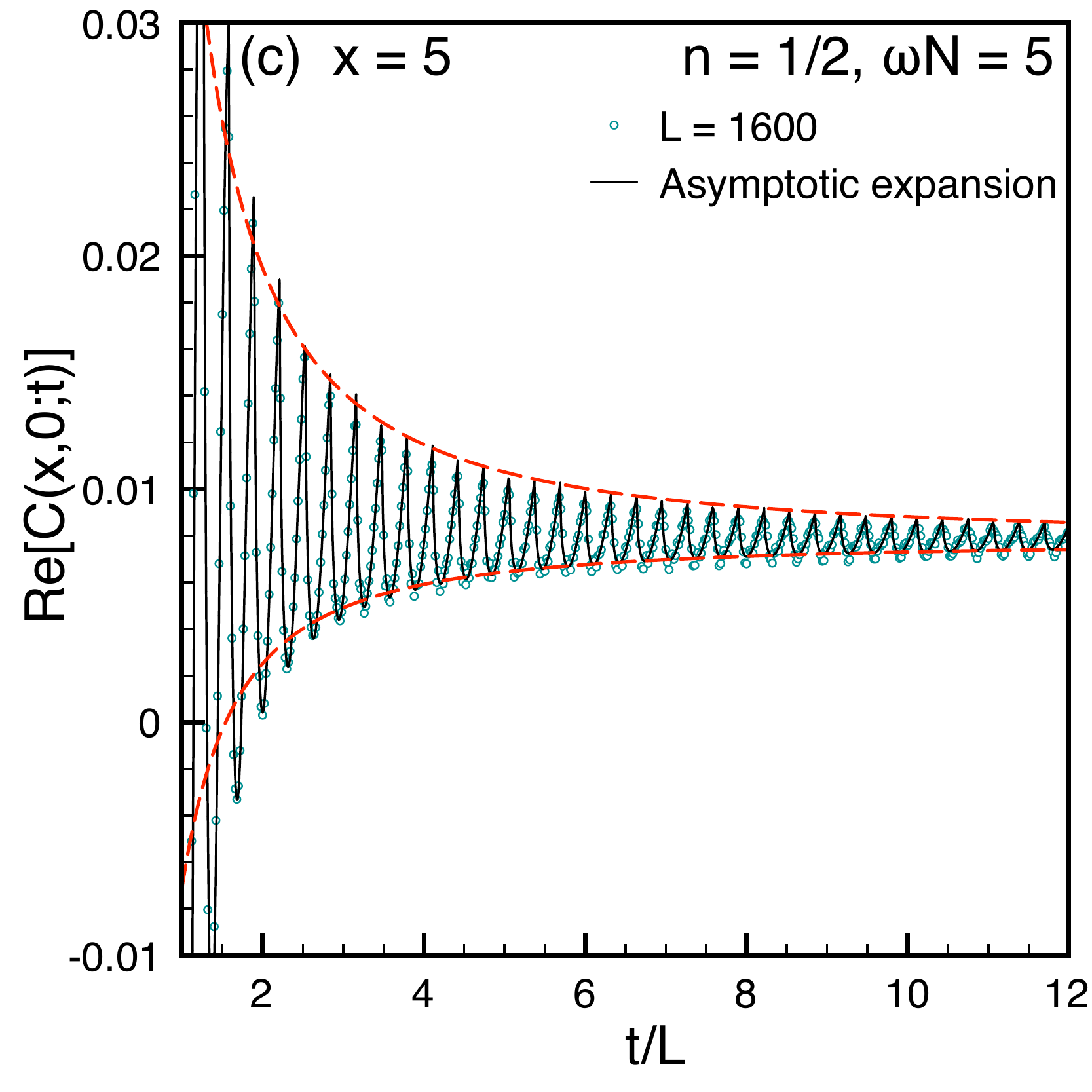}\includegraphics[width=0.5\textwidth]{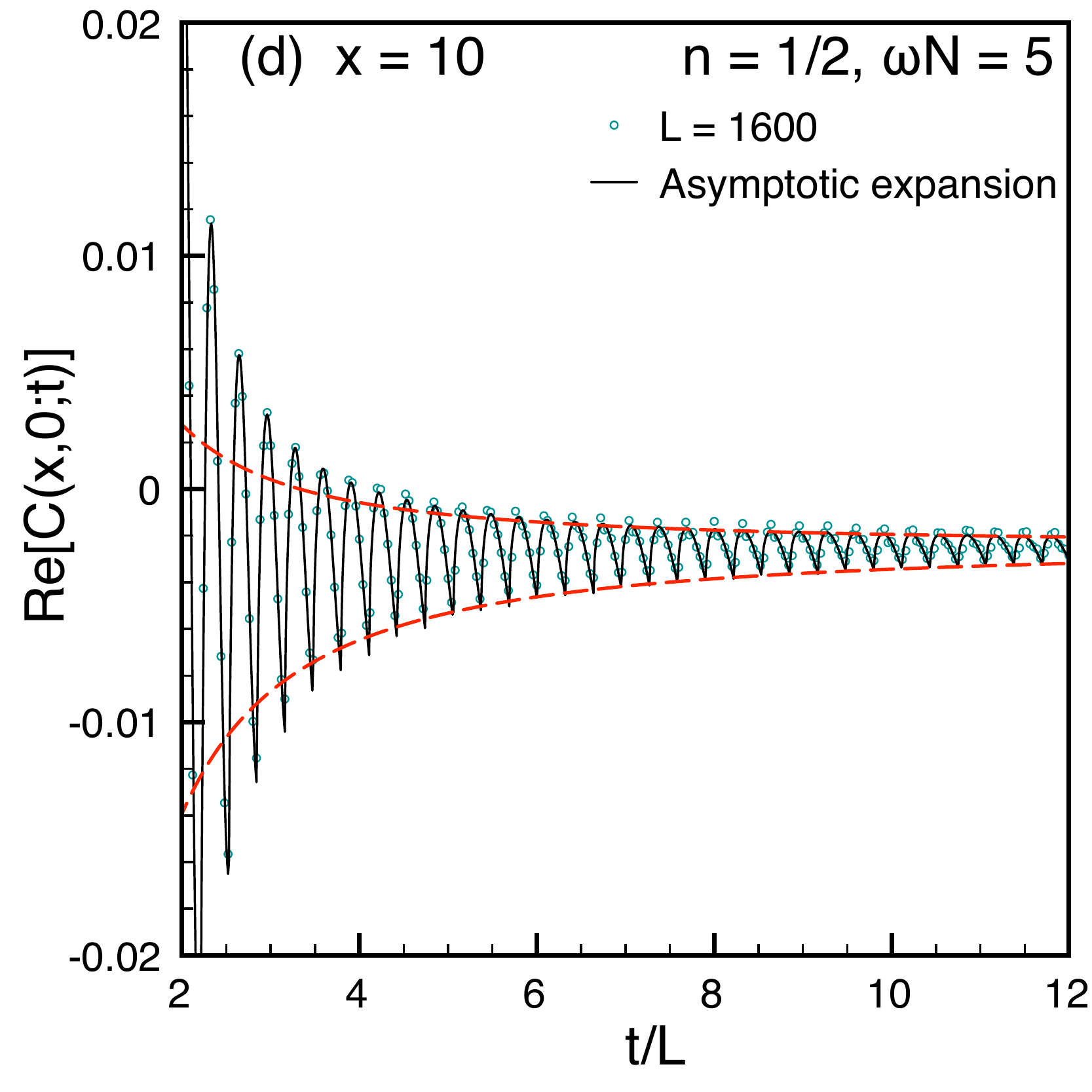}
\caption{Time dependence of the correlation ${\rm Re}[C(x,0;t)]$ for $x=5$ (a,c) and $x=10$ (b,d). 
The dynamics is rather irregular before the propagating peak travels the distance $x$ (in a time $x/v_{\rm peak}$).
For later times instead there is a simple damped oscillatory behavior around the stationary value (dashed line in top panels)
which is zoomed in the insets for clarity.
In (a,b) the points represent the full correlation function in Eq. (\ref{Cxyt_1}) while the full lines are the diagonal sum 
in Eq. (\ref{Cxyt_2}).
In the two bottom panels (c,d), the points represent ${\rm Re}[C(x,0;t)]$ from the diagonal sum in Eq. (\ref{Cxyt_2}) 
for larger time scales. 
They are compared with the asymptotic expansion in Eq. (\ref{CF_corr})
(full lines) which is almost  indistinguishable from the data soon after the moving peak passed through. 
Also the asymptotic $t^{-3/2}$ behavior for the envelopes of maxima and minima  is reported
(dashed lines cf. Eq. (\ref{Ct32})).
}
\label{figcorrtime}
\end{figure}

As for the density, a stationary phase argument allows us to conclude that only diagonal terms $p=q$ 
contribute to the TD limit. (To quantitatively support this statement, in Fig. \ref{corr_color} we compare the full sum with the one 
restricted over the diagonal terms: for $N=20$ small differences are visible, but they are negligible already for $N=100$.)
This leads to
\begin{eqnarray}\label{Cxyt_2}\fl
C(x,y;t) &=& \frac{\mathrm{e}^{i \frac{\omega^2 t(x^2 - y^2)}{2(1+\omega^2 t^2)}}}{\sqrt{1+\omega^{2}t^2}}
\sum_{p=-\infty}^{\infty} 
e^{i \frac{\omega^2 t(x-y)pL}{1+\omega^2 t^2}}
\sum_{j=0}^{N-1}\chi_{j}\left(\frac{x+pL}{\sqrt{1+\omega^{2}t^2}}\right)\chi_{j}\left(\frac{y+pL}{\sqrt{1+\omega^{2}t^2}}\right).
\end{eqnarray}
As for the density, we can use the Christoffel-Darboux formula (\ref{cdf}) to evaluate the sum over the number of particles, 
but the analytic progress is not enough to get the full time dependence of the correlation function.

In Figs. \ref{corr_color}, \ref{figCxy} and \ref{figcorrtime} we report the numerically calculated correlation function for finite $N$ 
in order to understand how the infinite time limit is approached. 
Fig. \ref{corr_color} is a density plot for ${\rm Re}[C(x,0;t)]$ revealing clearly the various velocities entering in the dynamics. 
Fig.  \ref{figCxy} reports the $x$ dependence of the same correlation for four different times, while 
Fig. \ref{figcorrtime} shows the time dependence for two different values of $x$. 
All these figures show that the stationary value of $C(x,x_{0},t)$ as a function of $x$ and $t$ is approached starting 
from a neighborhood of $x_{0}$ and growing inside a {\it cone} bounded by two moving peaks with velocity $v_{\rm peak} = 2\pi/L$. 
Interestingly, this velocity depends only on the final geometry (very differently from the expansion velocity $v=\sqrt{2\omega N}$
depending only on the initial state). 
Conversely, the amplitude of the peaks depends on the initial condition through $\omega N$. 
Looking at Fig. \ref{corr_color} more carefully, one can notice the presence of 
secondary peaks of two different types. 
First, there are peaks generated at times which are integer multiples of $\tau/L \sim 1/v$, where 
$v=\sqrt{2\omega N}$ is the expansion velocity of the gas. 
The interference of this series of peaks produces finally a stationary correlation. 
Second (as clear  in Fig. \ref{corr_color} from the panels with different $\omega N$, but same $L$)
there are sub-leading moving peaks with velocities equal to integer multiplies of $v_{\rm peak}$. 
We will show in the following that this second family of  peaks can be explained in the large-time limit as a 
series of subleading corrections to the stationary behavior.

\begin{figure}[t]
\center\includegraphics[width=\textwidth]{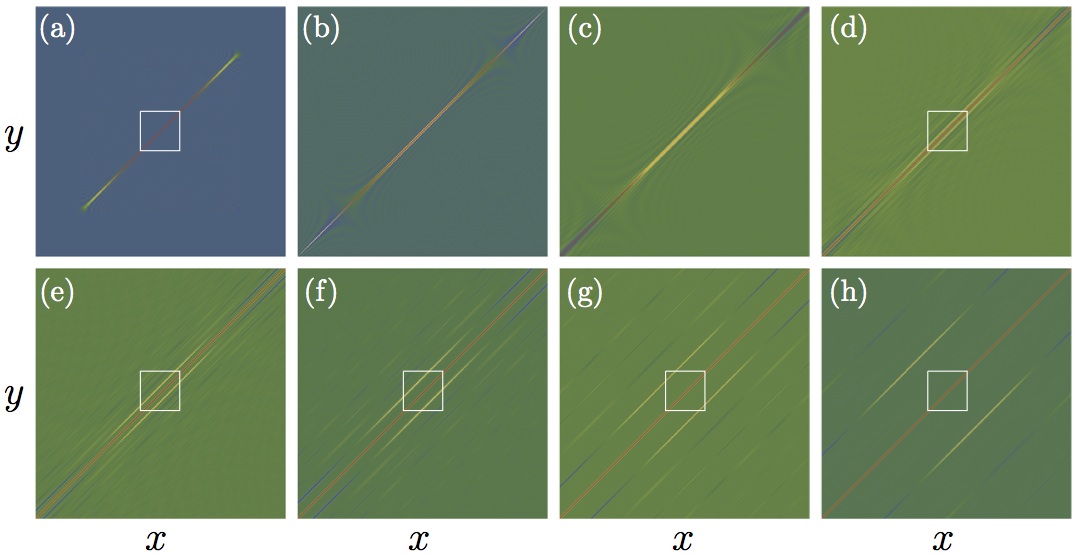}
\caption{Color snapshots of the fermionic correlation  ${\rm Re}[C(x,y;t)]$ for $N=100$, $L=200$ ($n=1/2$) and $\omega N = 5$  
with $x,y\in [-L/2,L/2]$. 
(a,b,c,d) From left to right correlations are calculated at $t/L=0,1/8,1/4,1/2$. 
(e,f,g,h) From left to right correlations are calculated at $t/L=1,2,4,8$. 
The numerical evaluation is done using only the diagonal part of the replica sum, which is exact in the TD limit. 
The small white squares are the regions zoomed  in Fig. \ref{corr_snapshot_zoom}.}
\label{corr_snapshot}
\end{figure}

\begin{figure}[t]
\center\includegraphics[width=\textwidth]{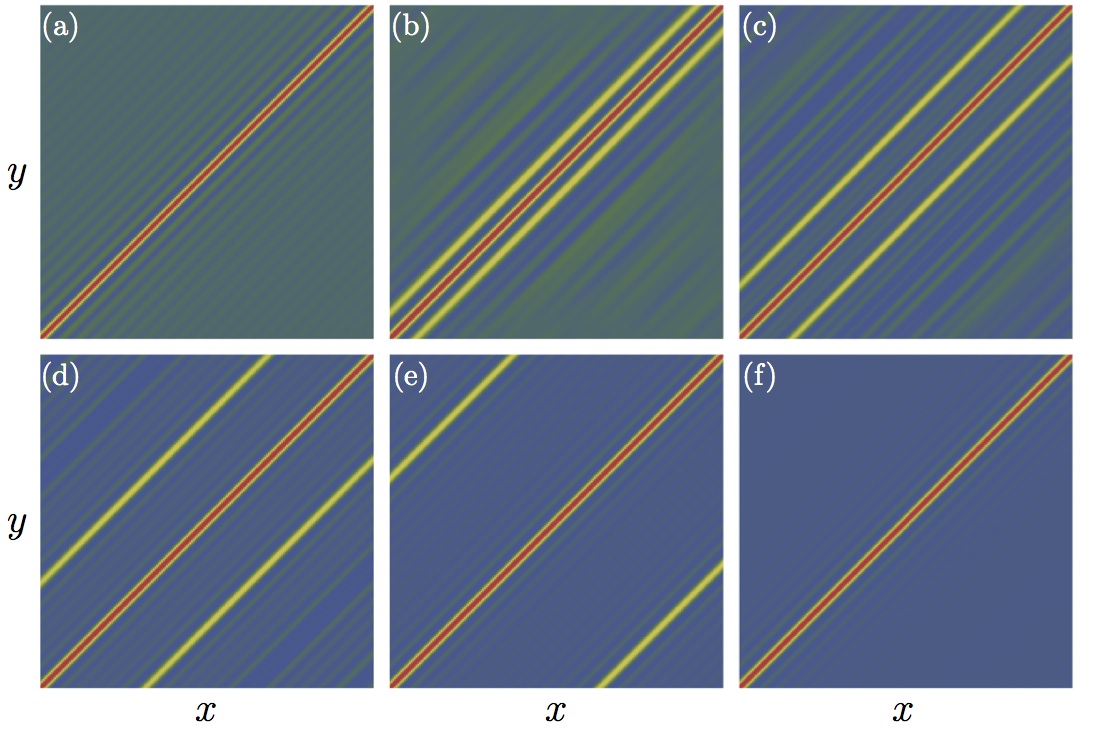}
\caption{Same data as in Figure \ref{corr_snapshot} 
zoomed in the region $x,y\in[-20,20]$. 
(a,b,c) From left to right correlations are calculated at $t/L=0,1/2,1$. 
(d,e,f) From left to right correlations are calculated at $t/L=2,4,8$.}
\label{corr_snapshot_zoom}
\end{figure}

Let us now take the TD and large time limit.
In the TD limit, unlike the density which depends on the scaling variables $x/L$ and $t/L$,  
the presence of the phase factor in Eq. (\ref{Cxyt_2}) breaks the spatial scaling: 
the two-point fermionic correlation function depends on $x$, $y$ and $t/L$. 
This is clearer in the large time limit $t\gg \omega^{-1}$, when we have 
(in terms of the TD quantities $\omega N,\,N/L,\,\ldots$)
\begin{equation}\label{Cxyt_larget}
C(x,y;t) \simeq \frac{1}{\omega t}\sum_{p=-\infty}^{\infty}\mathrm{e}^{i (x-y)pL/t}
\sum_{j=0}^{N-1}\chi_{j}\left(\frac{x+pL}{\omega t}\right)\chi_{j}\left(\frac{y+pL}{\omega t}\right).
\end{equation}
To explicitly take the TD limits it is convenient  to introduce $\tilde{\chi}_{j}(x)$ as the eigenfunctions of an harmonic oscillator 
with $\omega=1$ because in the eigenfunctions $\chi_{j}(x)$ (cf. Eq. (\ref{chin})) the $x$ variable is multiplied by $\sqrt{\omega}$. 
Eq. (\ref{Cxyt_larget}) is then rewritten as 
\begin{equation}\label{Cxyt_larget2}
C(x,y;t) \simeq \frac{1}{\sqrt{\omega} t}\sum_{p=-\infty}^{\infty}\mathrm{e}^{i (x-y)pL/t}
\sum_{j=0}^{N-1}\tilde{\chi}_{j}\left(\frac{x+pL}{\sqrt{\omega} t}\right)\tilde{\chi}_{j}\left(\frac{y+pL}{\sqrt{\omega} t}\right).
\end{equation}
Therefore, taking both TD and large-time limits $t/L\rightarrow \infty$ (with $t/L^{2}\rightarrow 0$), 
the factor $L/t\sqrt{\omega}$ goes to zero and, using Eq. (\ref{sumtoint}), we can recast the sum over $p$ as an integral
\be\label{Cxyt_larget3}\fl
C(x,y;t\rightarrow\infty) \simeq\frac{1}{L}\int_{-\infty}^{\infty} dz\, \mathrm{e}^{i\sqrt{\omega}\,(x-y)z}\, 
\sum_{j=0}^{N-1}\tilde{\chi}_{j}\left(\frac{x}{\sqrt{\omega} t}+z\right)\tilde{\chi}_{j}\left(\frac{y}{\sqrt{\omega} t}+z\right).
\ee
In the TD and infinite time limit, the terms $x/t\sqrt{\omega}$ and $y/t\sqrt{\omega}$ can be neglected and 
we can further simplify Eq. (\ref{Cxyt_larget3}) obtaining
\begin{equation}\label{Cxyt_larget4}
C(x,y;t\to\infty) =\frac{1}{L}\int_{-\infty}^{\infty} dz\, \mathrm{e}^{i\sqrt{\omega}\,(x-y)z}\,
\sum_{j=0}^{N-1}|\tilde{\chi}_{j}(z)|^2,
\end{equation}
which in the large $N$ limit becomes
\be\label{Cxyt_larget_leading}
C(x,y;t\rightarrow\infty)  =  \frac{1}{L}\int_{-\infty}^{\infty}dz\,\mathrm{e}^{i\sqrt{\omega}\,(x-y)z}\, \tilde{n}_{0}(z)
 =  2n\frac{J_{1}[ \sqrt{2\omega N}(x-y)]}{\sqrt{2\omega N}(x-y)},
\ee
where $n_{0}(x) = \sqrt{\omega}\,\tilde{n}_{0}(x\sqrt{\omega})$, i.e. $\tilde{n}_{0}(x) = \sqrt{2N-x^2}/\pi$. 

Eq. (\ref{Cxyt_larget_leading}) for the infinite time limit of the correlation function coincides with the 
time average $\overline{C(x,y;t)}$ obtained in Eq. (\ref{Caver})
showing explicitly that, in the TD and large-time limits, a stationary correlation function is approached without time average. 
Furthermore, this result perfectly matches the numerical calculation for large enough value of $t/L$ 
(as long as we observe the correlations inside the cone bounded by two propagating peaks with velocity $v_{\rm peak}=2\pi/L$, 
see Fig. \ref{corr_color}). 
Notice that the analytic large-time calculation is much simpler than the full time dependence which is 
accessible only numerically.

The peculiar approach to the infinite time limit and the presence of the moving peaks
 can be explained taking into account the  corrections to the integral 
in Eq. (\ref{Cxyt_larget_leading}). 
From Eq. (\ref{sumtoint}), starting from Eq. (\ref{Cxyt_larget2}), 
the leading  correction to  Eq. (\ref{Cxyt_larget_leading}) is
\be\label{Cxyt_larget_corr0}
\frac{1}{L}\int_{-\infty}^{\infty} dz\, \mathrm{e}^{i\sqrt{\omega}\,[(x-y)\pm 2\pi t/L ]z}\,
\sum_{j=0}^{N-1}\tilde{\chi}_{j}\left(\frac{x}{\sqrt{\omega} t}+z\right)\tilde{\chi}_{j}\left(\frac{y}{\sqrt{\omega} t}+z\right),
\ee
which, in the TD limit and for $\sqrt{\omega}t\to\infty$, becomes
\begin{equation}\label{Cxyt_larget_corr}
\frac{1}{L}\int_{-\infty}^{\infty}dz\,\mathrm{e}^{i\sqrt{\omega}\,[(x-y)\pm 2\pi t/L ]z}\, \tilde{n}_{0}(z) = 
2n\frac{J_{1}[ \sqrt{2\omega N}(x-y \pm 2\pi t/L) ]}{\sqrt{2\omega N}(x-y \pm 2\pi t/L)}
\end{equation}
This correction correctly identifies the location of the first two propagating peaks at $x-y\sim 2\pi t/L$ justifying the 
result we anticipated for $v_{\rm peak}= 2\pi/L$.
However, the amplitude of the moving peaks is lower than what is predicted by Eq. (\ref{Cxyt_larget_corr}) 
because it gets contributions from the interference between the eigenfunctions $\chi_{j}(x)$ 
evaluated at different points which have been neglected going from Eq. (\ref{Cxyt_larget2}) to Eq. (\ref{Cxyt_larget_corr0}). 
Interestingly, Eq. (\ref{Cxyt_larget_leading}) and Eq. (\ref{Cxyt_larget_corr}) are translationally invariant 
symmetric integrals over even functions, and therefore they are real. 
This is not true for the full correlation in Eq. (\ref{Cxyt_larget2}) which is, in general, complex and not translationally invariant. 
As expected, only in the large-time limit one recovers the translational invariance. 
In the very same way, all secondary peaks with velocities $m v_{\rm peak}$ (with $m$ integer) are qualitatively described 
by the sum
\be\label{CF_corr}
C(x,y;t) =2n \sum_{m=-\infty}^{\infty}\frac{J_{1}[ \sqrt{2\omega N}(x-y + 2\pi m t /L) ]}{\sqrt{2\omega N}(x-y + 2\pi m t /L) },
\ee
but again the amplitudes of these peaks are reduced by interference effects which are encoded in Eq. (\ref{Cxyt_larget2}).
For any $x,y$ and for times slightly larger than the expulsion of the moving peak, i.e. $v_{\rm peak}t\gtrsim x-y$, 
Eq. (\ref{CF_corr}) reproduces very precisely the data as shown in the two bottom panels in Fig. \ref{figcorrtime}.
Furthermore, one can expand Eq. (\ref{CF_corr}) for very large time and obtain 
\be\fl 
C(x,0;t) \simeq C(x,0;t\to\infty) + \frac{n}{\pi^2 (\sqrt{2\omega N} t/L)^{3/2}} \cos(\sqrt{2\omega N} x)F(2 \pi \sqrt{2\omega N} t / L),
\label{Ct32}
\ee
where we defined the real function
\be
F(z) = \frac{i-1}{\sqrt{2}} [\mathrm{Li}_{3/2}(\mathrm{e}^{-i z}) + i \mathrm{Li}_{3/2}(\mathrm{e}^{iz})],
\ee
with $F_{\rm min}  =  -\sqrt{2} \zeta(3/2)$ and $F_{\rm max}  =  1.6592637\dots$ being respectively 
the minimum and the maximum of $F(z)$.
The maxima and minima of $C(x,0;t)$ depends on $x$ because of the oscillating cosine.
This shows that the approach to GGE is power-law like with an exponent $3/2$, similarly to what found for the 
Ising model \cite{CEFII}.
This power-law behavior is compared to the numerical data in Fig. \ref{figcorrtime}.

All the figures we reported up to now are just for $C(x,0;t)$, but the general conclusions we outlined are 
true for arbitrary $x$ and $y$ as 
Figs. \ref{corr_snapshot} and \ref{corr_snapshot_zoom}  show.
These two figures are density plots for the real part of $C(x,y;t)$ as a function of $x$ and $y$ for different times $t/L$ 
which can be explained as follows.
The system starts from an inhomogeneous initial condition which determines the time evolution, especially during the early 
stages where the breakdown of the translational invariance is evident. 
However, as time goes on, a strip opens around the diagonal and in that region translational invariance is (approximately) restored. 
If one zooms in a small region as in Fig. \ref{corr_snapshot_zoom}, the system  appears almost homogeneous for all times.

\section{The reduced density matrix and the GGE.}
\label{sec4}

In the previous section, we have proved analytically that the fermionic correlation function for large 
time approaches a stationary value.
We are going to show in this section that this is true for arbitrary correlations of local observables 
and that their value can be inferred a priori without solving the non-equilibrium evolution by the so called 
generalized Gibbs ensemble (GGE).
A GGE can be written generically as 
\be
\rho_{GGE}= \frac1{Z_{\rm GGE}} \exp\left({-\sum \lambda_i \hat{I}_i}\right), 
\label{GGEdef}
\ee  
where $I_j$ are some integrals of motion and $Z_{\rm GGE}$ ensures the normalization condition $\Tr \rho_{GGE}=1$ (this generalizes
the canonical Gibbs ensemble where we only have $I_1=H$ and $\lambda_1=\beta=1/kT$).
In Ref. \cite{gg}, it has been proposed that an integrable system after a quantum quench in the infinite time limit is described 
by a GGE where the $I_j$'s represent a complete set of independent integrals of motion and the Lagrange multipliers $\lambda_j$ 
are fully determined by the initial state $|\Psi_0\rangle$  through the conditions 
\be 
\langle \Psi_0| \hat{I}_i |\Psi_0\rangle={\rm Tr} [\rho_{GGE}\hat{I}_i].
\ee
  
However, for a closed system evolving under Hamiltonian dynamics, the existence of a stationary state described by $\rho_{GGE}$
may seem paradoxical because the whole  system is always in a pure 
state and cannot be described by a mixed state at infinite time.
This apparent paradox is solved with the help of the reduced density matrix \cite{cdeo-08,bs-08,CEFII}.
Let us consider a spatial interval $A$, and its reduced density matrix 
\be
\rho_A(t)\equiv{\rm Tr}_B [\rho(t)],
\ee 
where $B$ is the complement of $A$ and $\rho(t)=|\Psi(t)\rangle\langle \Psi(t)|$ is the time dependent density matrix of the 
whole pure system. 
With some abuse of language, we can say that a system is stationary if, after the TD limit is properly taken for the whole system, 
the limit 
\be 
\displaystyle\rho_{A,\infty}\equiv \lim_{t\to\infty} \rho_A(t),
\ee 
exists for any finite $A$  \cite{CEFII}.
Furthermore we say that a system is described by a statistical ensemble, e.g. Gibbs or GGE, with density matrix $\rho_E$
if the reduced density matrix $\rho_{A,E}\equiv {\rm Tr}_B [\rho_E]$ equals $\rho_{A,\infty}$.

For a gas of free fermions, by means of Wick theorem, any observable can be obtained from the two-point correlator and so 
also the reduced density matrix. 
The construction of $\rho_A$ in terms of the fermionic correlation $C(x,y)$ in continuous space 
has been detailed in \cite{cmv-11,cmv-11b} 
(generalizing the lattice approach \cite{pes}).
Indeed in the fermionic basis, the reduced density matrix of a subsystem $A$ extending from $x_1$ to $x_2$ can be written as
\be
\rho_A \propto \exp \Big(- \int_{x_1}^{x_2} d y_1 dy_2 \hat\Psi^\dagger(y_1) {\mathbb H}(y_1,y_2) \hat\Psi(y_2)\Big)\,,
\label{rhoa}
\ee
where ${\mathbb H}=\ln [(1-C)/C]$ and the normalization constant is fixed requiring ${\rm Tr}\rho_A=1$.
The easiest way to understand this equation is the continuum limit of the formula for lattice free fermions \cite{pes}, but 
can also be  obtained following the standard derivation in Ref. \cite{pes} in path integral formalism. 
As a fundamental point, the non-local Jordan-Wigner transformation (\ref{JordanWigner}) mapping the Tonks-Girardeau gas to 
free fermions is local within any given {\it compact} subspace in systems with PBC, i.e. 
the bosonic degrees of freedom within $A$ can be written only in terms of fermions in $A$, 
as clear from Eq. (\ref{JordanWigner}).
This is analogous to lattice models such as the Ising chain \cite{CEF,CEFII,f-13}. 
Thus, if  for finite $x$ and $y$, $C(x,y;t\to\infty)$ is described by a statistical ensemble, also $\rho_A$ will be
and consequently  the expectation value of any bosonic of fermionic observable local within $A$. 

Another important point concerns the issue of which integrals of motion $I_j$ must 
be included in the definition (\ref{GGEdef}) of the GGE density matrix.
Indeed, any quantum system, integrable or not, has too many integrals of 
motion. For example the projectors 
on Hamiltonian eigenstates 
are integrals of motion,  
but such conservation laws cannot be always important for the late time behavior after a quantum quench, 
otherwise no system would ever thermalize.  
Following Refs. \cite{CEFII,fe-13}, in Eq. (\ref{GGEdef}) 
we include only the \emph{local} integrals of motion: 
these are characterized by arising from an integral of a local current density $J_j(x)$ as $I_j=\int dx J_j(x)$.

At this point, we have all ingredients to construct the GGE which is expected to describe the stationary value of correlations 
of local observables in the trap release dynamics of a Tonks-Girardeau gas.
For free fermionic models, instead of using the local integrals of motion, it is simpler to work  
with the momentum occupation modes $\hat n_k=\eta^\dagger_k \eta_k$ 
which are non-local integrals of motion, but can be written as linear combinations of local integrals of motion \cite{fe-13} 
(however we will describe also the GGE with local charges in the following subsection and also describe the linear mapping 
between the two). 
The initial values of $\hat n_k$ are 
\bea\label{nkGGE}\fl
\langle \Psi_0| \hat n_k |\Psi_0\rangle & =  & \sum_{i,j} A^{*}_{k,i} A_{k,j} \langle \Psi_0| \hat{\xi}^{\dag}_{i}\hat{\xi}_{j} |\Psi_0\rangle
= 
\sum_{j=0}^{N-1} |A_{k,j}|^2 \simeq \frac{2}{L} \sqrt{\frac{2N}{\omega}} \sqrt{1-\frac{k^2}{2\omega N}},
\eea
and zero if the argument of the square root is negative (with the last equality above valid only in the TD limit). 
In the GGE we have 
\be
n_{GGE}(k)\equiv{\rm Tr} [\rho_{GGE} \hat n_k]= \frac1{e^{\lambda_k}+1},
\ee
and equating the last two equations we have 
\be
\lambda_k=\ln \left[\frac1{\langle \Psi_0| \hat n_k |\Psi_0\rangle}-1\right]= 
\ln \left[\frac{L\omega}{2} \frac1{\sqrt{2\omega N-k^2}}- 1\right]
\label{lamkgge}
\ee
The real space fermionic correlation $C(x,y)$ in the GGE is just the Fourier transform of $n_{\rm GGE}(k)$, but 
Eq. (\ref{nkGGE}) is the momentum distribution appearing in the integral definition $C(x,y;t\to\infty)$ in 
Eq. (\ref{Cxyt_larget_leading}). Thus $C(x,y)$ in the GGE trivially equals the infinite time limit after the trap release.
Since via Eq. (\ref{rhoa}), the two-point fermion correlation determines the full reduced density matrix, 
this equality shows that all stationary quantities of the released gas are described by a GGE.
Very interestingly, in Ref. \cite{eef-12} it has been shown that all 
non-equal time stationary properties are always determined by the same ensemble describing the static quantities,
and so, even in our case, they are encoded solely in the GGE.

\subsection{Local integrals of motion and GGE}
The local integrals of motion are linear combinations of the fermionic occupation modes $\hat{n}_{k}$.
Indeed, the linear combinations \cite{ck-12}
\begin{equation}
\hat{I}_{j} = \sum_{k} k^{j} \hat{n}_{k},
\end{equation}
satisfy the commutation relations $[\hat{I}_{i},\hat{I}_{j}]=0$ and they are local, 
in the sense that they can be written as integrals of one-point differential operators \cite{dk-90-11}:
\begin{eqnarray}\label{local_integrals}\fl
 \hat{I}_{j} & = & \frac{1}{L}\sum_{k} \int\!\!\!\int\!dxdy\,  k^{j} \mathrm{e}^{ik(x-y)} \hat{\Psi}^{\dag}(x) \hat{\Psi}(y)
=  \int\!\!\!\int\!dxdy\, \hat{\Psi}^{\dag}(x)\!\left[\int\!\frac{dk}{2\pi}\,k^{j}\mathrm{e}^{ik(x-y)}\right]\!\hat{\Psi}(y)\nonumber \\ \fl
 & = &  \int\!\!\!\int\!dxdy\,  (-i)^{j}\hat{\Psi}^{\dag}(x)\delta^{(j)}(x-y)  \hat{\Psi}(y)
 =\int\!dx \, \hat{\Psi}^{\dag}(x)(-i)^{j}\frac{\partial^{j}}{\partial x^{j}}  \hat{\Psi}(x).
\end{eqnarray}
This means that all powers of the 
momentum operator $\hat{P} \equiv -i\partial_{x}$ are conserved, as 
well known for free fermions. The Hamiltonian is $H=\hat I_2/2$.

\begin{figure}[t]
\includegraphics[width=0.33\textwidth]{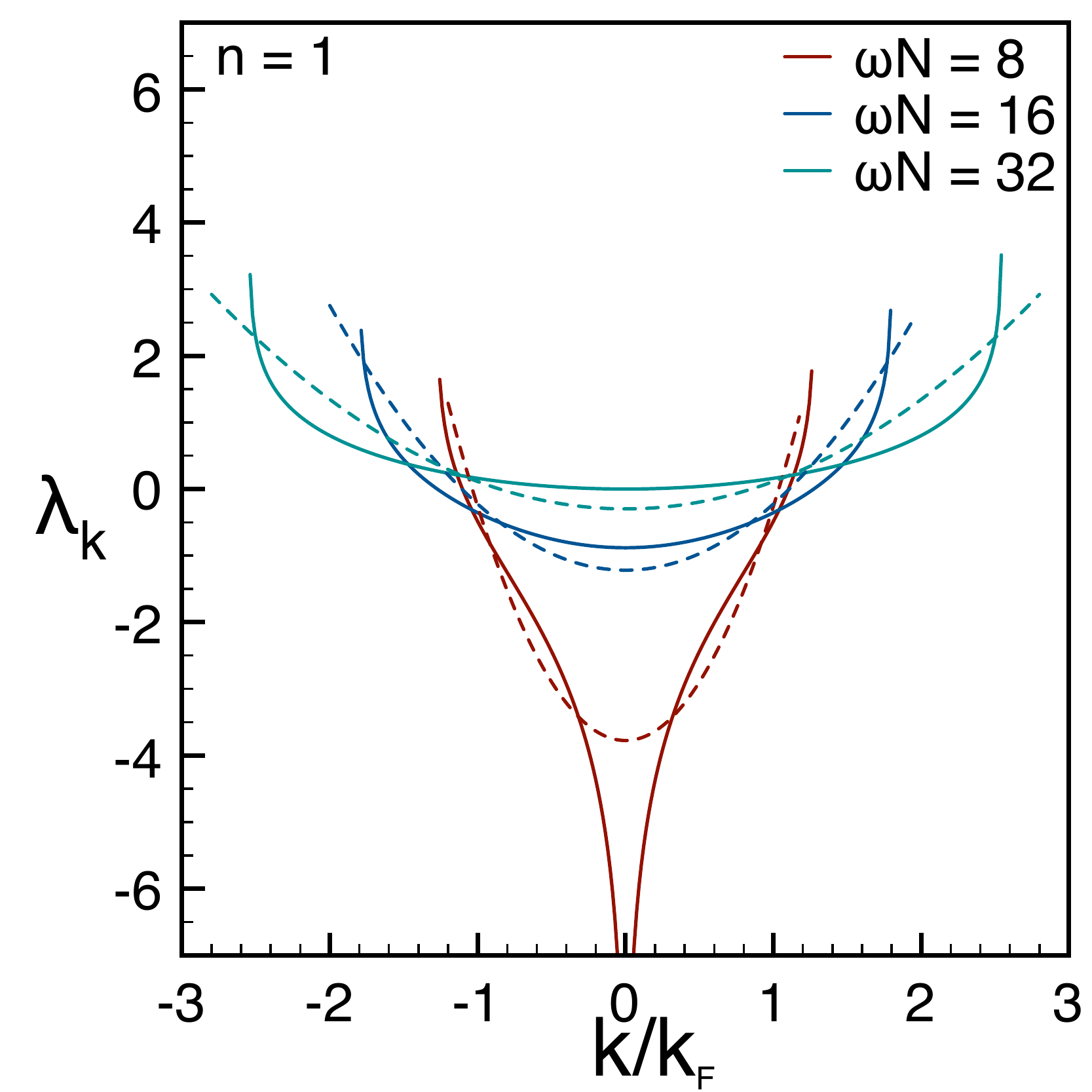}\includegraphics[width=0.33\textwidth]{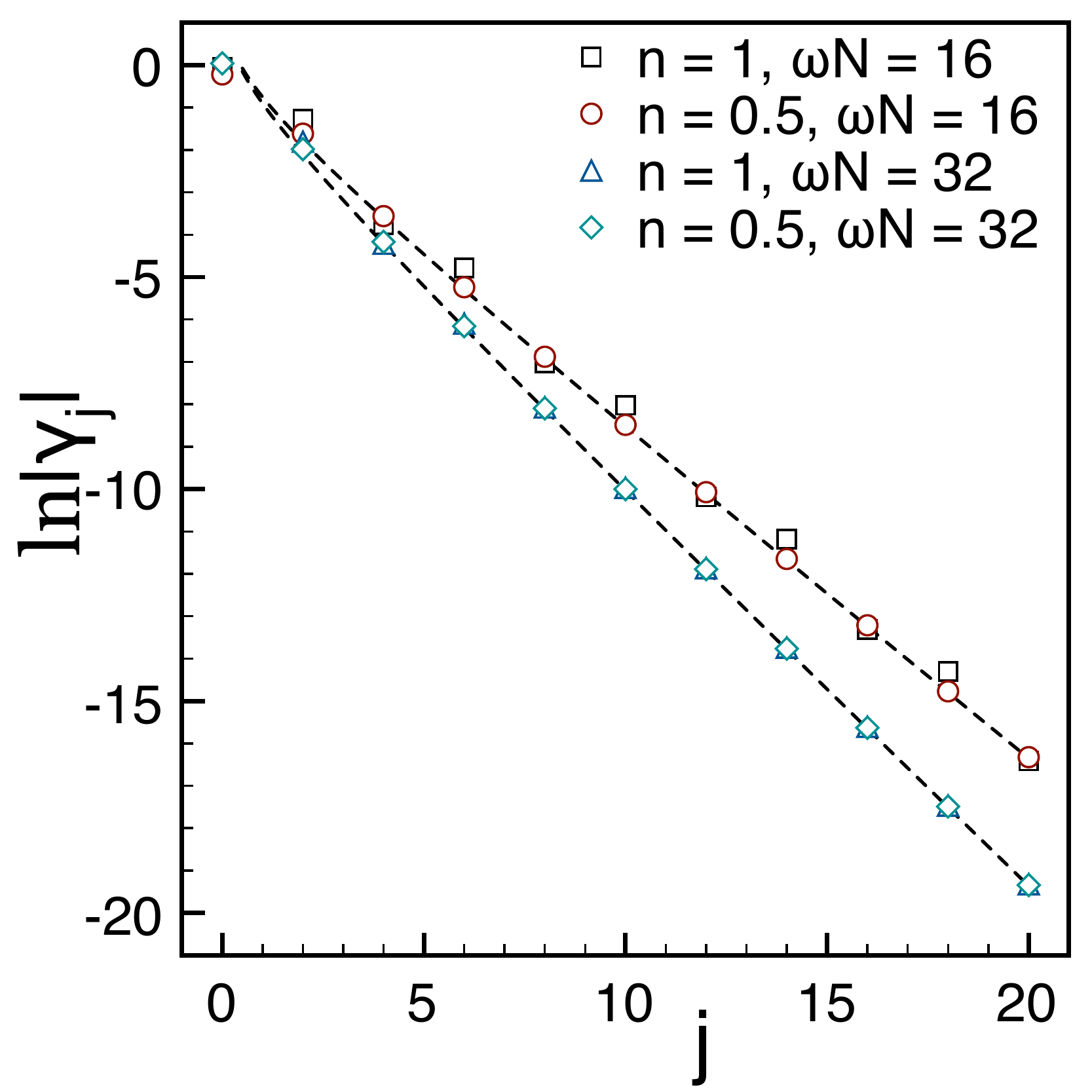}\includegraphics[width=0.33\textwidth]{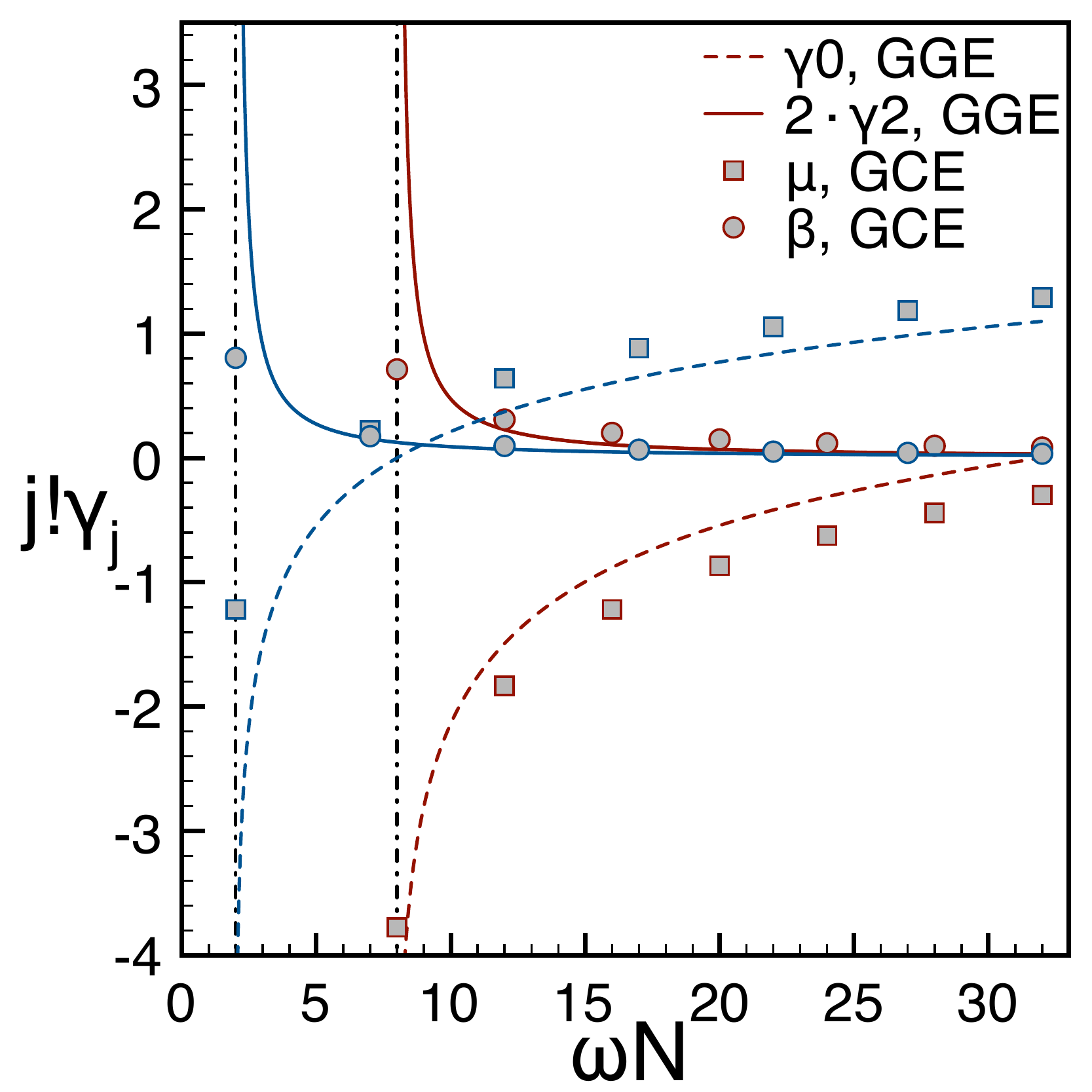}
\caption{(Left) The GGE Lagrange multipliers $\lambda_k$ (full lines) as a function of $k/k_F$ ($k_F=\pi n$) for different initial 
trapping potentials $\omega N$. Notice the singularity in zero for $\omega N = 8$ (i.e. $n_0=n$). 
For comparison  also the grancononical approximation (dashed line) is reported.
(Central) The even Lagrange multipliers $|\gamma_{j}|$  of the local GGE as a function of $j$ in logarithmic 
scale and for different initial conditions. 
The large-$j$ behavior (dashed lines)  depends only on $\omega N$. 
(Right) The Lagrange multipliers $\gamma_{0}$ and $\gamma_{2}$ (corresponding to the first two local conserved 
charges $\hat{N}$ and $2 \hat{H}$) as predicted by the GGE (full and dashed lines) 
and numerically evaluated in the case of GCE (symbols). 
Different colors represent different densities: $n=1/2$, blue; $n=1$, red. 
The vertical dot-dashed lines $\sqrt{2\omega N} = 4n$ delimit the region ($\sqrt{2\omega N} > 4n$, i.e. $n_0>n$) 
where  $\gamma_{0,2}$ are defined.
} 
\label{figLambdak}
\end{figure}

We want now to construct a GGE in terms of these local integrals of motion
\be
\rho_{\rm GGE}= Z_{\rm GGE}^{-1} \exp\Big(-\sum_{j=0}^{\infty} \gamma_i \hat{I}_j\Big), 
\ee
and understand its relation and equivalence with the one in momentum modes. 
To this aim, it is useful to  formally expand the $\lambda_{k}$ in powers of $k$ (in the TD limit $k$ is a continuous variable)
\begin{equation}   
\lambda_{k} 
= \sum_{j=0}^{\infty} \frac{\Gamma_{j}}{j!} k^{j},\quad \Gamma_{j}\equiv \left.\frac{d^{j}}{dk^{j}}\lambda_{k}\right|_{k=0},
\label{explam}
\end{equation}
where $\Gamma_{j}=0$ for all $j$ odd because $n_{GGE}(k)$ is an even function of $k$ 
due to the invariance of the initial ground state under  $k\to -k$ (which also implies $\langle \Psi_{0}| \hat{I}_{j} | \Psi_{0} \rangle = 0$, 
for all $j$ odd). 
This allows us to rearrange the sum over the  occupation number operators $\hat{n}_{k}$ in terms of the local charges $\hat{I}_{j}$
%
\begin{equation}
\sum_{k=-\infty}^{\infty}\lambda_{k}\hat{n}_{k} = \sum_{j=0}^{\infty} \frac{\Gamma_{j}}{j!} \sum_{k=-\infty}^{\infty}k^{j}\hat{n}_{k} 
= \sum_{j=0}^{\infty}  \frac{\Gamma_{j}}{j!} \hat{I}_{j} = \gamma_{0}\hat{N} + 2\gamma_{2} \hat{H} + \cdots,
\end{equation}
from which we conclude that the Lagrange multipliers of the local charges $\hat I_j$ are just $\gamma_j=\Gamma_j/j!$, i.e. the 
$\gamma_j$ are proportional to the derivatives of $\lambda_k$ in Eq. (\ref{lamkgge}) evaluated at $k=0$.
The Lagrange multipliers are written in a more compact form in terms of the initial average density in Eq. (\ref{n00}), i.e. 
$n_0=\sqrt{\omega N/8}$.
The first two Lagrange multipliers are  
\be
\gamma_{0} = \ln \Big[  \frac{n_0}n-1 \Big],\qquad  {\rm and} \qquad 
\gamma_{2} = \frac1{4\omega N(1 - n/n_0)}. 
\ee
The explicit analytic form of the higher order multipliers is more cumbersome to be written here, but  
they all diverge for $n_0\to n$ as 
\be
\gamma_{2j}\sim \frac1{(2j)!(2\omega N)^{j/2} (1-n/n_0)^{j}},
\label{gamth}
\ee
with the $j=0$ term becoming a logarithm. 
This divergence coincides with 
the trap release condition $\ell < L/2$, i.e. $n_0 > n$. 
In the opposite limit $\sqrt{2\omega N}\to\infty$ (i.e. $n_0\gg n$) all the Lagrange multipliers are vanishing except the first one, i.e.
\be
\gamma_{2j} \sim (\omega N)^{-j}, \qquad {\rm and} \qquad \gamma_{0}\sim\ln(\omega N).
\label{lamla}
\ee

It is also possible to extract the behavior of the Lagrange multipliers $\gamma_{j}$ for large $j$, which gives information about 
the weight that each local charge $\hat{I}_{j}$ has in the GGE expansion. 
After some algebra, one finds the leading contribution 
\be \gamma_{2j} \sim (2\omega N)^{-j}/(2j), \qquad {\rm for}\; j\gg 1,
\ee 
which does not depend on the density $n$ (see Fig. \ref{figLambdak}).
In Figure \ref{figLambdak} we plot both the coefficients $\lambda_{k}$ and $\gamma_{j}$ for different initial trapping potentials.

\subsection{Comparing the GGE  with the canonical  and grand canonical ensemble}

For a generic non-integrable system, the only local conserved charge is the post-quench Hamiltonian $\hat{H}$ and 
if a large-time stationary state exists 
it should be described by the Canonical Ensemble (CE) 
\begin{equation}
\rho_{CE} = Z_{CE}^{-1} \mathrm{e}^{-\beta_C \hat{H}},
\end{equation}
where, once again, the inverse temperature $\beta_C$ is fixed by the 
condition $\langle\Psi_{0}|\hat{H}|\Psi_{0}\rangle=\Tr[\rho_{CE}\hat{H}]$. 

It is worth investigating the qualitative and quantitative differences between the canonical ensemble  and the GGE 
to provide testable predictions for experiments and numerical analyses. 
In our case, in order to estimate the differences in the expectation values of local observables, 
we compare the results previously obtained in the GGE with the canonical ensemble. 
In the diagonal basis the post-quench Hamiltonian is
\begin{equation}
\hat{H}=\sum_{k} \frac{k^{2}}{2}\hat{n}_{k},\quad k=\frac{2\pi m}{L},
\end{equation}
so that the condition  fixing $\beta_C$ (the only multiplier here) is 
\begin{equation}\label{beta_eq1}
\sum_{k}\frac{k^2}{2}\frac{1}{1+\mathrm{e}^{\beta_C k^2 / 2}}=\sum_{k}\frac{k^2}{2} \langle \Psi_0| \hat n_k |\Psi_0\rangle .
\end{equation}
In the TD limit, 
Eq. (\ref{beta_eq1}) becomes
\begin{equation}
\int_{-\infty}^{\infty} \frac{k^2\,dk}{1+\mathrm{e}^{\beta_C k^2 / 2}}=\frac{2n}{\omega N} \int_{-\sqrt{2\omega N}}^{\sqrt{2\omega N}}dk\,k^{2} \sqrt{2\omega N - k^{2}}, 
\end{equation}
which gives
\begin{equation}
\beta_{C} = \left[\frac{(\sqrt{2}-1)\zeta(3/2)}{n\,\omega N\sqrt{\pi}}\right]^{2/3}= 0.719653\dots \,(n\,\omega N)^{-2/3}.
\end{equation}
The canonical approximation is qualitatively incorrect for this quench, indeed the momentum distribution
\be
n_{CE}(k) \equiv \Tr[\rho_{CE}\hat{n}_{k}] = \frac{1}{1+\mathrm{e}^{\beta_{C}k^2 /2}},
\label{nCE}
\ee
has infinite support (but decays for large $k$ as $e^{-\beta_c k^2}$). 
A comparison between the canonical and GGE $n(k)$ is reported in the left panel of Fig. \ref{figSk}. 
Furthermore we have $n_{CE}(0)=1/2$ independently of the initial conditions, contrarily to the infinite time 
limit (which is also GGE) which retains information about the initial state.  

We can improve the canonical approximation by considering also the number of particles operator $\hat{N}$ as a conserved charge, i.e. 
using the grand canonical ensemble (GCE)
\begin{equation}
\rho_{GCE} = Z_{GCE}^{-1} \mathrm{e}^{-\beta_{GC} \hat H-\mu_{GC} \hat{N}}.
\end{equation}
The two Lagrange multipliers $\beta_{GC}$ and $\mu_{GC}$ are fixed by the set of equations (in the TD limit)
\begin{eqnarray}
\int_{-\infty}^{\infty} \frac{k^2\,dk}{1+\mathrm{e}^{\beta_{GC} k^2 / 2 + \mu_{GC}}} & = &
\frac{2n}{\omega N} \int_{-\sqrt{2\omega N}}^{\sqrt{2\omega N}}dk\,k^{2} \sqrt{2\omega N - k^{2}}, \\
\int_{-\infty}^{\infty} \frac{dk}{1+\mathrm{e}^{\beta_{GC} k^2 / 2 + \mu_{GC}}} & = &
\frac{2n}{\omega N} \int_{-\sqrt{2\omega N}}^{\sqrt{2\omega N}}dk\, \sqrt{2\omega N - k^{2}},\nonumber
\end{eqnarray}
which can be recast
in terms of polylogarithm functions 
$\mathrm{Li}_{n}(z)=\sum_{k=1}^{\infty}z^{k}/k^{n}$ as
\begin{equation}\label{beta_mu_eq}
 \frac{\sqrt{2\pi}\,\mathrm{Li}_{3/2}(-\mathrm{e}^{-\mu_{GC}})}{\beta_{GC}^{3/2}}  =- \pi\,\omega N,\quad 
   \frac{\sqrt{2\pi}\,\mathrm{Li}_{1/2}(-\mathrm{e}^{-\mu_{GC}})}{\beta_{GC}^{1/2}}  =  - 2\pi n,
\end{equation}
from which  $\mu_{GC}$ is given by the solution of 
\begin{equation}
\frac{{\rm Li}_{3/2}(-{\rm e}^{-\mu_{GC}})}{\left[ {\rm Li}_{1/2}(-{\rm e}^{-\mu_{GC}}) \right]^3} = \frac{\omega N}{4\pi n^3}=
\frac{2 n_0^2}{\pi n^3},
\end{equation}
and then $\beta_{GC}$ plugging the numerically found $\mu_{GC}$ in one of the two equations in Eq. (\ref{beta_mu_eq}).

We can compare the GGE values of $\gamma_{0,2}$ with the GCE ones. 
The values of $\mu_{GC}$ and $\beta_{GC}$ in the GCE are reported in Fig. \ref{figLambdak} (right) and show a qualitative 
behavior similar to the GGE multipliers $\gamma_{0,2}$, but they are quantitatively rather different. 
For the multipliers $\lambda_k$ of the momentum occupation numbers, using the GCE ensemble amounts to considering 
only a second order expansion in $k$ of Eq. (\ref{lamkgge}). 
The resulting parabolic form with $\mu_{GC}$ and $\beta_{GC}$ is plotted in Fig. \ref{figLambdak} (left), 
showing a behavior rather different from the GGE one.

We can also compare $n_{GCE}(k)$
\be
n_{GCE}(k) \equiv \Tr[\rho_{GCE}\hat{n}_{k}] = \frac{1}{1+\mathrm{e}^{\beta_{GC}k^2 /2+ \mu_{GC} N }},
\label{nGCE}
\ee
with the GGE as shown in Fig. \ref{figSk} (left panel). 
One may note a relatively good match of the GCE and GGE curves for $\omega N = 8$ for density $n=1$. 
This is easily understood: close to the threshold value $\omega N=8 n^2$, among the GGE multipliers $\gamma_{2j}$
the ones with higher weight are those with smaller $j$ ($\gamma_{2j}/\gamma_{0,2}\rightarrow 0,\,\forall j>1$, cf. 
Eq. (\ref{gamth})) which are exactly those kept in the GCE approximation. 
Furthermore, in the opposite limit $\omega N\to\infty$, the GCE description exactly matches to the GGE description because only 
$\gamma_0$ is non-vanishing (cf. Eq. (\ref{lamla})), and indeed the 
momentum distribution flattens. 
This is not the case for the CE in which the dominant operator $\hat{N}$ is absent.

\section{The density-density correlator and the static structure factor}
\label{sec5}

An important experimentally measurable quantity is the equal-time density-density correlation function
\be
G(x,y;t)\equiv \langle \hat{\Psi}^{\dag}(x)\hat{\Psi}(x) \hat{\Psi}^{\dag}(y)\hat{\Psi}(y) \rangle=
 \langle \hat{\Phi}^{\dag}(x)\hat{\Phi}(x) \hat{\Phi}^{\dag}(y)\hat{\Phi}(y) \rangle,
\ee
which is the same both for fermions and bosons because the Jordan-Wigner string contributions trivially cancel.
Using Wick theorem, we can rewrite $G(x,y;t)$ as
\bea
G(x,y;t) & = & \delta(x-y)\langle\hat{\Psi}^{\dag}(x)\hat{\Psi}(y) \rangle +  \langle \hat{\Psi}^{\dag}(x)  \hat{\Psi}^{\dag}(y)\hat{\Psi}(y)\hat{\Psi}(x) \rangle\\
& = & \delta(x-y)\langle\hat{\Psi}^{\dag}(x)\hat{\Psi}(y) \rangle + \langle \hat{n}(x) \rangle \langle \hat{n}(y) \rangle - |\langle\hat{\Psi}^{\dag}(x)\hat{\Psi}(y) \rangle| ^{2}\nonumber \\
& = & n(x,t)n(y,t) + C(x,y;t)[\delta(x-y)-C(y,x;t)],
\eea
which, for free fermions, depends only on the fermionic correlation function. 
In the TD and large-times limits we know from the previous section that the fermionic correlation function becomes 
translationally invariant $C^{\infty}(x-y)\equiv C(x,y;t\rightarrow\infty)$ and therefore we can define a stationary structure 
factor $S(k)$ as the Fourier transform of the connected density-density correlators $G_{c}^{\infty}(x-y)\equiv G^{\infty}(x-y)-n^2$
\bea\label{Skdef}
S(k) & \equiv & \frac{1}{N} \int\!\!\int\!dx\,dy\,\mathrm{e}^{ik(x-y)} G_{c}^{\infty}(x-y)=  
\frac{1}{n} \int\!dz \, \mathrm{e}^{ikz} G_{c}^{\infty}(z)\\
 & = & 1 - \frac{1}{n} \int\! dz \, \mathrm{e}^{ikz} |C^{\infty}(z)|^2 =  1 -\frac{1}{n} \int\! \frac{dq}{2\pi}\,n(q)n(k-q),\nonumber
\eea
where $n(k)$ is the fermionic momentum distribution. 
Before calculating $S(k)$ in the GGE, let us give for 
comparison the structure factor for the free-fermionic ground-state (GS) with  
$n_{GS}(k)=\theta(k_{F}-|k|)$: 
\bea\fl
S_{GS}(k) & = &  1 - \frac{1}{2\pi\,n}\int_{-k_{F}}^{k_{F}} dq\, \theta(k_{F}-|k-q|) 
=  \left\{\begin{array}{cc}|k|/2k_{F} & |k|<2k_{F} \\1 & |k|>2k_{F}\end{array}\right. ,
\eea
which is also reported in Fig. \ref{figSk}.

\subsection{The structure factor in the GGE.}

\begin{figure}[t]
\includegraphics[width=0.5\textwidth]{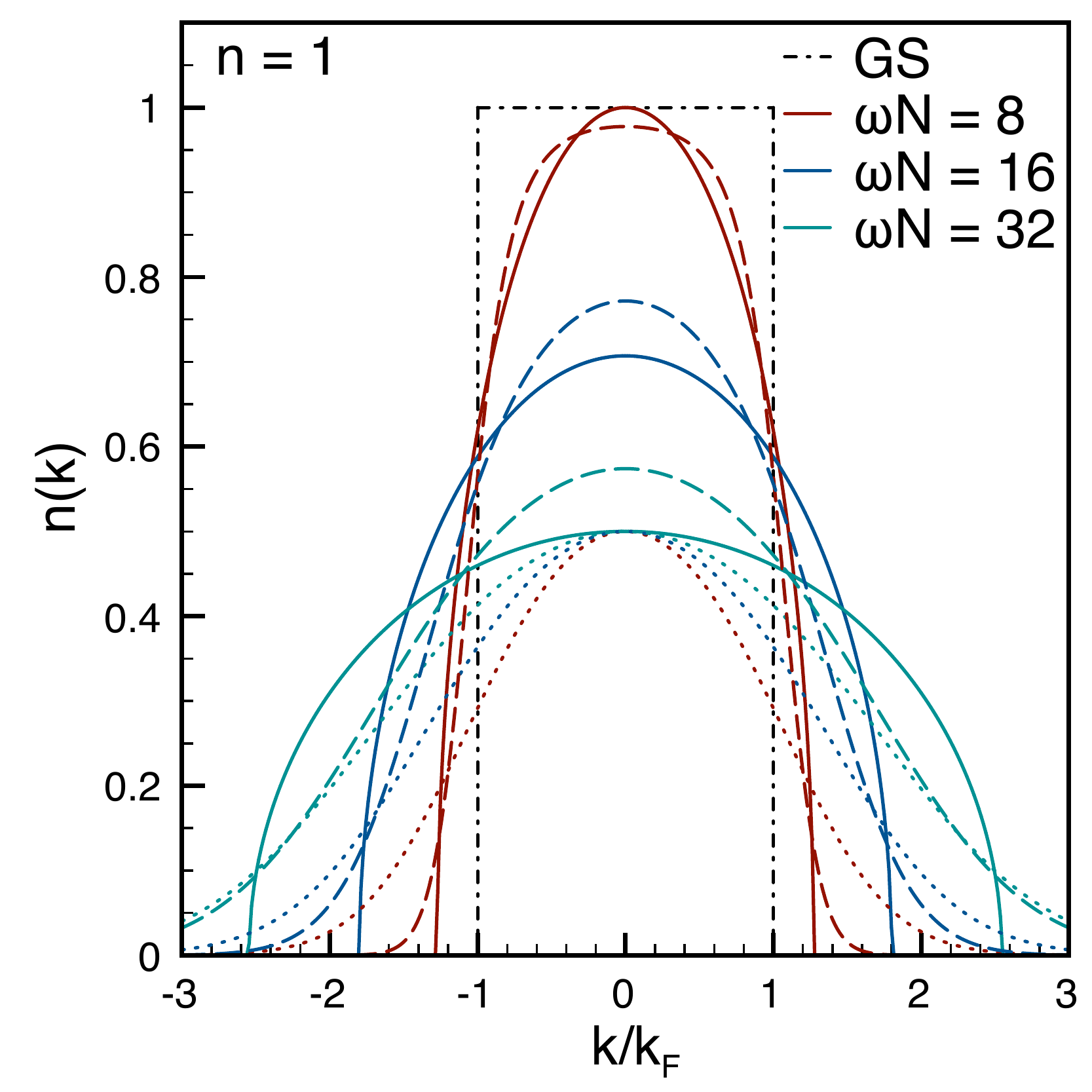}\includegraphics[width=0.5\textwidth]{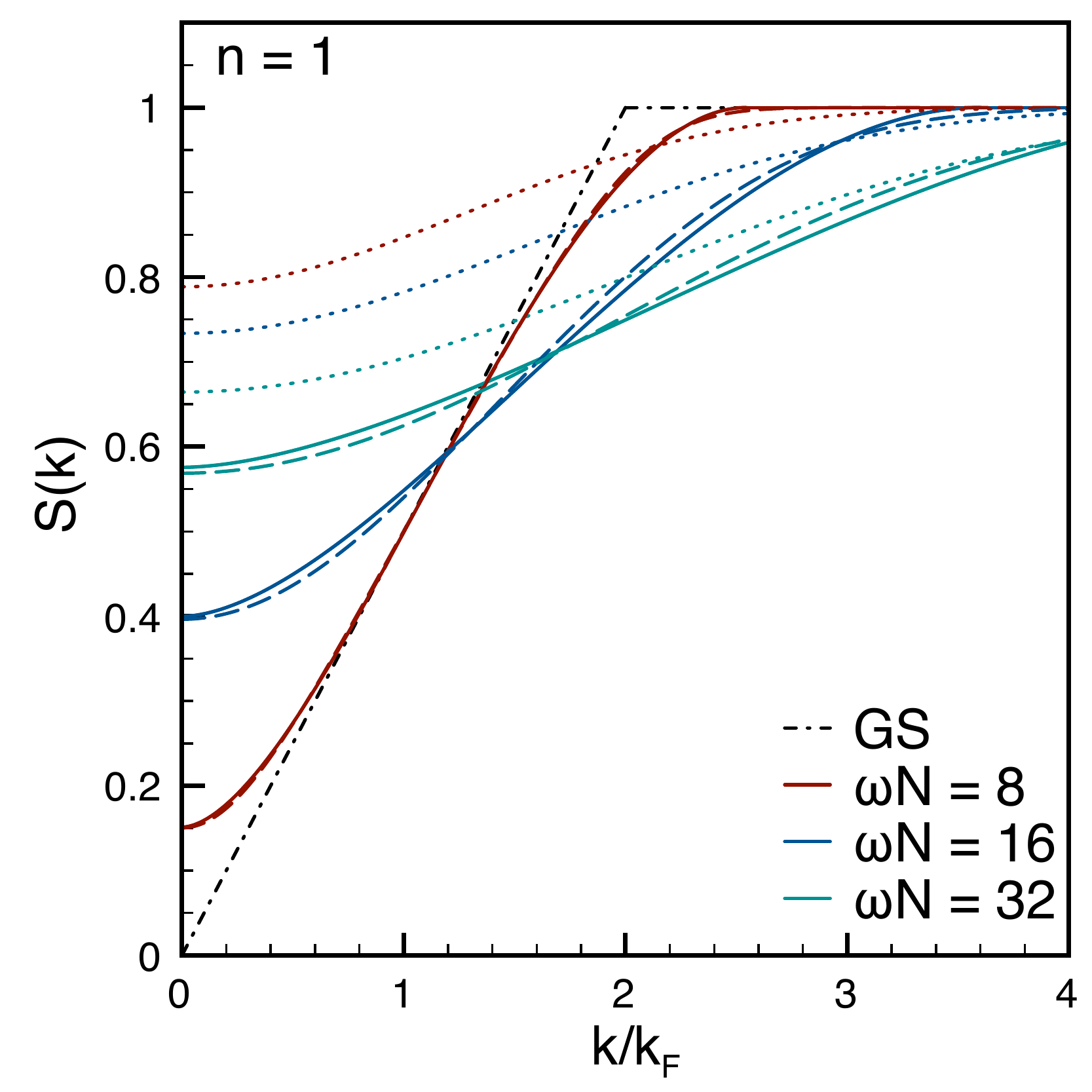}
\caption{The GGE momentum distribution $n(k)$ (on the left) and structure factor $S(k)$ (on the right) as a 
function of $k/k_F$ ($k_F=\pi n$) for different initial trap potentials $\omega N$  compared with the CE (dotted lines), 
the GCE (dashed lines) and the ground-state (GS) result (dot-dashed line).
} 
\label{figSk}
\end{figure}

The structure factor $S(k)$ in the GGE is obtained plugging the GGE $n_{GGE}(k)$ (cf. Eq. (\ref{nkGGE})) 
in Eq. (\ref{Skdef}), obtaining
\be
S(k)= 1 -\frac{1}{n} \int\! \frac{dq}{2\pi}\,n_{GGE}(q)n_{GGE}(k-q)
=1- \frac{4 \sqrt2 n}{\pi \sqrt{\omega N}} f\Big(\frac{k}{\sqrt{2\omega N}}\Big),
\ee
where the explicit result of the integration is 
\be
f(x) =
\left\{
\begin{array}{ll}\displaystyle
 \left[(4 + x^2) E\Big(1 - \frac{4}{x^2}\Big) - 8 K\Big(1 - \frac{4}{x^2}\Big)\right] \frac{|x|}6  &  {\rm if}\;  |x|<2  \\
 &\\
 0 & {\rm if}\;  |x|>2,
 \end{array}
\right.
\ee
and $E(z)$ and $K(z)$ are standard elliptic functions.
Notice that for $x>0$, $f(x)$ is a monotonous function with maximum  $f(0)=4/3$. 
Thus $S(k)$ turns out to be an even function of $k$ and monotonous for $k>0$. 
The plot of $S(k)$ for fixed density $n=N/L=1$ and 
for different  initial trapping potentials  $\omega N$ is reported in Fig. \ref{figSk}. 
$S(k)$ qualitatively resembles the one found numerically in Ref. \cite{ck-12} for the Lieb-Liniger gas. 
Because of the trap release constraint $\sqrt{\omega N} > 2\sqrt2 n$, we have
$S(k)>S(0)\geq 1-8/3\pi=0.151174\dots$.

We can also compare the GGE structure factor with the canonical and grand canonical ones by plugging in 
Eq. (\ref{Skdef}) the corresponding mode occupation functions. 
For the canonical ensemble, using $n_{CE}(k)$ in Eq. (\ref{nCE}), we have
\begin{equation}
S_{CE}(k) = 1 -\frac{1}{n}\int_{-\infty}^{\infty}\frac{dq}{(1+\mathrm{e}^{\beta_C q^2 / 2})(1+\mathrm{e}^{\beta_C (k-q)^2 / 2})},
\end{equation}
In Fig. \ref{figSk} this is compared to the GGE results and it is clear that they differ substantially. 
For the grand canonical ensemble, using the mode occupation in Eq. (\ref{nGCE}) and the numerically calculated 
Lagrange multipliers $\beta_{GC}$ and $\mu_{GC}$ we obtain the results reported in Fig. \ref{figSk}. 
Oppositely to the canonical ensemble, 
there is a relatively good match of the GCE and GGE data for all considered values of $\omega N$.
As already discussed in the previous section, this is due to the fact that both close to the lower
threshold $\omega N=8n$ and for large $\omega N$ the GGE gets a higher weight from the two lowest 
charges $\hat{N}$ and $\hat{H}$ which are the ones considered in the grand canonical ensemble.

\section{The two-point bosonic correlation function}
\label{sec6}

In this section we consider the equal time bosonic two-point correlation
\be
C_{B}(x,y;t)\equiv \langle\hat{\Phi}(x,t)\hat{\Phi}(y,t) \rangle,
\ee
also known as one-particle density matrix whose Fourier transform is the (bosonic) 
momentum distribution function commonly measured in cold atoms experiments. 
This can be expressed in terms of the fermionic correlations using the Jordan-Wigner mapping 
in Eq. (\ref{JordanWigner}) and  Wick theorem.
Indeed, for $y>x$ and suppressing for simplicity the time dependence of operators, we have
\begin{equation}
C_{B}(x,y;t) = \left\langle \hat{\Psi}^{\dag}(x)\,\exp\left\{-i\pi\int_{x}^{y}dz\,\hat{\Psi}^{\dag}(z) \hat{\Psi}(z)\right\}  \hat{\Psi}(y) \right\rangle.
\end{equation}
Taylor expanding the exponential this becomes
\begin{eqnarray}\fl
C_{B}(x,y;t) & = & \sum_{n=0}^{\infty}\frac{(-i\pi)^n}{n!}\int_{x}^{y}dz_{1}\cdots\int_{x}^{y}dz_{n}
\langle \hat{\Psi}^{\dag}(x) \hat{\Psi}^{\dag}(z_{1})\hat{\Psi}(z_1)\cdots  \hat{\Psi}^{\dag}(z_{n})\hat{\Psi}(z_n)  \hat{\Psi}(y)\rangle,
\nonumber
\end{eqnarray}
which can be rearranged in normal order\footnote{Using the anticommutation relations it is straightforward to show  
\bea\fl
\sum_{n}\frac{a^n}{n!}\int_{x}^{y}dz_{1}\cdots\int_{x}^{y}dz_{n} \hat{\Psi}^{\dag}(z_{1})\hat{\Psi}(z_1)\cdots  \hat{\Psi}^{\dag}(z_{n})\hat{\Psi}(z_n)  = 
&& \nonumber \\
\sum_{n}\frac{(\mathrm{e}^{a}-1)^n}{n!}\int_{x}^{y}dz_{1}\cdots\int_{x}^{y}dz_{n} \hat{\Psi}^{\dag}(z_{n})\cdots\hat{\Psi}^{\dag}(z_1)\hat{\Psi}(z_1)\cdots\hat{\Psi}(z_n), &&\nonumber
\eea
which for $a=-i\pi$ gives the desired result.
}
and, using Wick theorem, we finally have 
\begin{eqnarray}\label{C_B_fredholm}
C_{B}(x,y;t) & = & \sum_{n=0}^{\infty}\frac{(-2)^n}{n!}\int_{x}^{y}dz_{1}\cdots\int_{x}^{y}dz_{n} \, \det_{ij} \langle \hat{\Psi}^{\dag}(x_{i})\hat{\Psi}(y_j) \rangle \nonumber\\
& = & \sum_{n=0}^{\infty}\frac{(-2)^n}{n!}\int_{x}^{y}dz_{1}\cdots\int_{x}^{y}dz_{n} \, \det_{ij} C(x_i,y_j;t),
\end{eqnarray}
where the indices $i,j$ run from $0$ to $n$, and we used the convention 
$x_{i}=y_{i}\equiv z_{i},\,\forall i>0$, and $x_{0}\equiv x,\,y_{0}\equiv y$. 
Eq. (\ref{C_B_fredholm})  is a Fredholm's minor of the first order \cite{fredholm, feinberg}. 
Following Ref. \cite{feinberg},  $C_{B}(x,y;t)$ can be rewritten in an operatorial form as 
\begin{equation}\label{C_B_operatorial}
C_{B}(x,y;t) =\left.  \mathcal{D}_{[x,y;t]}(\lambda) \langle y | \frac{C}{1 - \lambda C} | x \rangle\right|_{\lambda=2},
\end{equation}
where we introduced the Fredholm's determinant 
\be\label{DFD}
\mathcal{D}_{[x,y;t]}(\lambda) = {\rm Det} \left[\delta(z-z') -\lambda C(z,z';t) \right],
\ee 
in which the kernel $C(z,z';t)$ and the identity $\delta(z-z')$ are restricted to the interval $[x,y]$.
We stress that in Eq. (\ref{C_B_operatorial}) the fraction stands for the multiplication by inverse operator 
of the denominator and is not the simple numerical ratio.

Although Eq. (\ref{C_B_operatorial}) is compact and elegant, its direct evaluation is not straightforward. 
It is convenient to rewrite $C_{B}(x,y;t)$ in terms of the time-evolved single particle wave functions $\phi_{j}(z,t)$ in 
Eq. (\ref{phi_L_t}) following Ref. \cite{pezer}, which in practice is just a change of basis.
To this aim we introduce the $N\times N$ overlap matrix $\mathbb{A}(x,y;t)$ with elements \cite{cmv-11}
\begin{equation}
\mathbb{A}_{ij}(x,y;t) = \int_{x}^{y}dz\,\phi^{*}_{i}(z,t)\phi_{j}(z,t),\quad i,j\in[0,\ldots,N-1],
\label{ovA}
\end{equation}
in terms of which we have \cite{pezer}
\begin{equation}\label{C_B_overlap}
C_{B}(x,y;t) = \sum_{i,j=0}^{N-1}\phi^{*}_{i}(x,t) \mathbb{B}_{ij}(x,y;t) \phi_{j}(y,t),
\end{equation}
where the $N\times N$ matrix $\mathbb{B}(x,y;t)$ is
\begin{equation}
\mathbb{B}(x,y;t) = \det[\mathbb{P}](\mathbb{P}^{-1})^{T},
\end{equation}
with the $N\times N$ matrix $\mathbb{P}$ related to the overlap matrix via 
\be 
\mathbb{P}(x,y;t) = \mathbb{I} -2\,\mathrm{sgn}(y-x)\mathbb{A}(x,y;t), 
\ee
where $\mathbb{I}$ is the $N\times N$ identity matrix. 

\begin{figure}[t]
\center\includegraphics[width=0.485\textwidth]{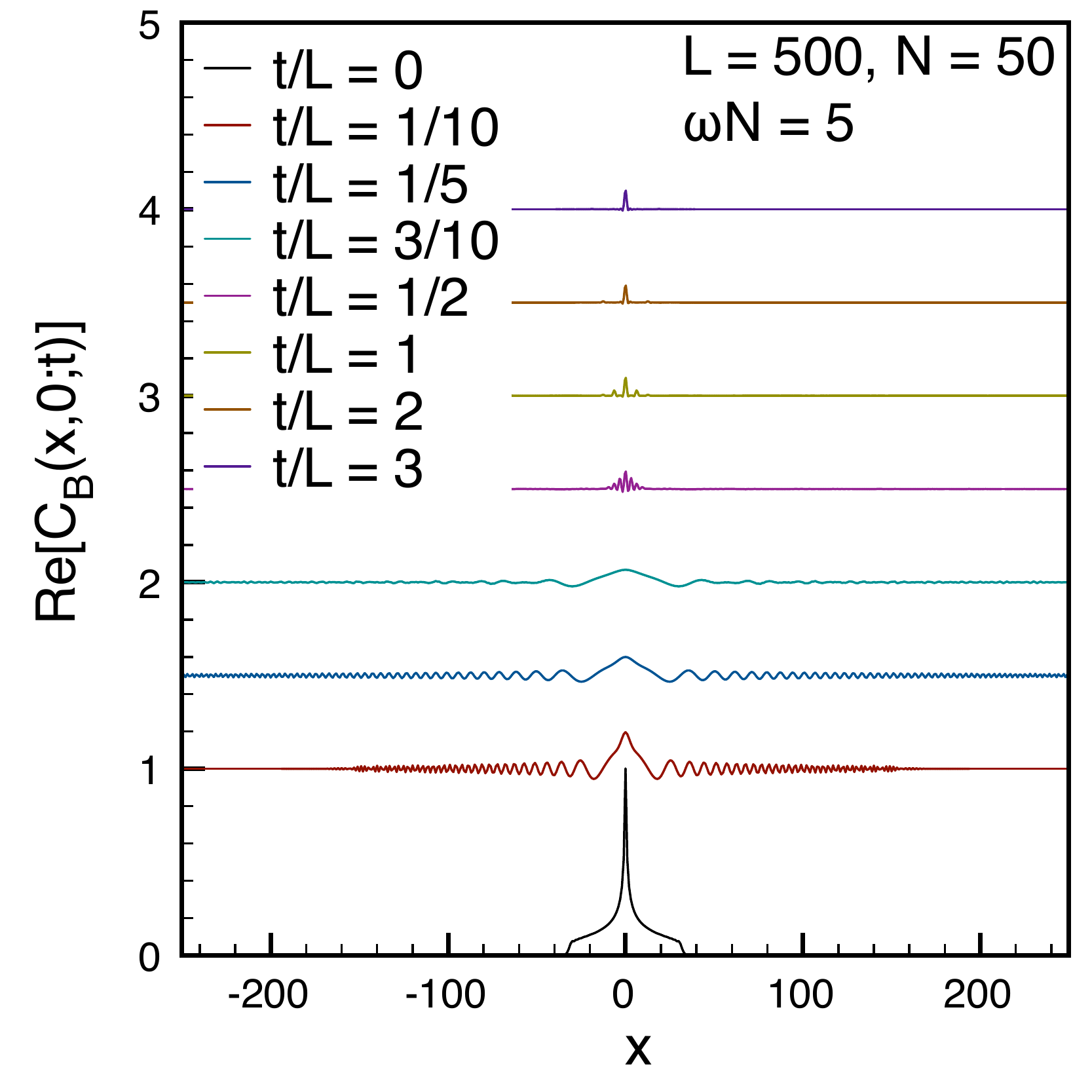}\includegraphics[width=0.515\textwidth]{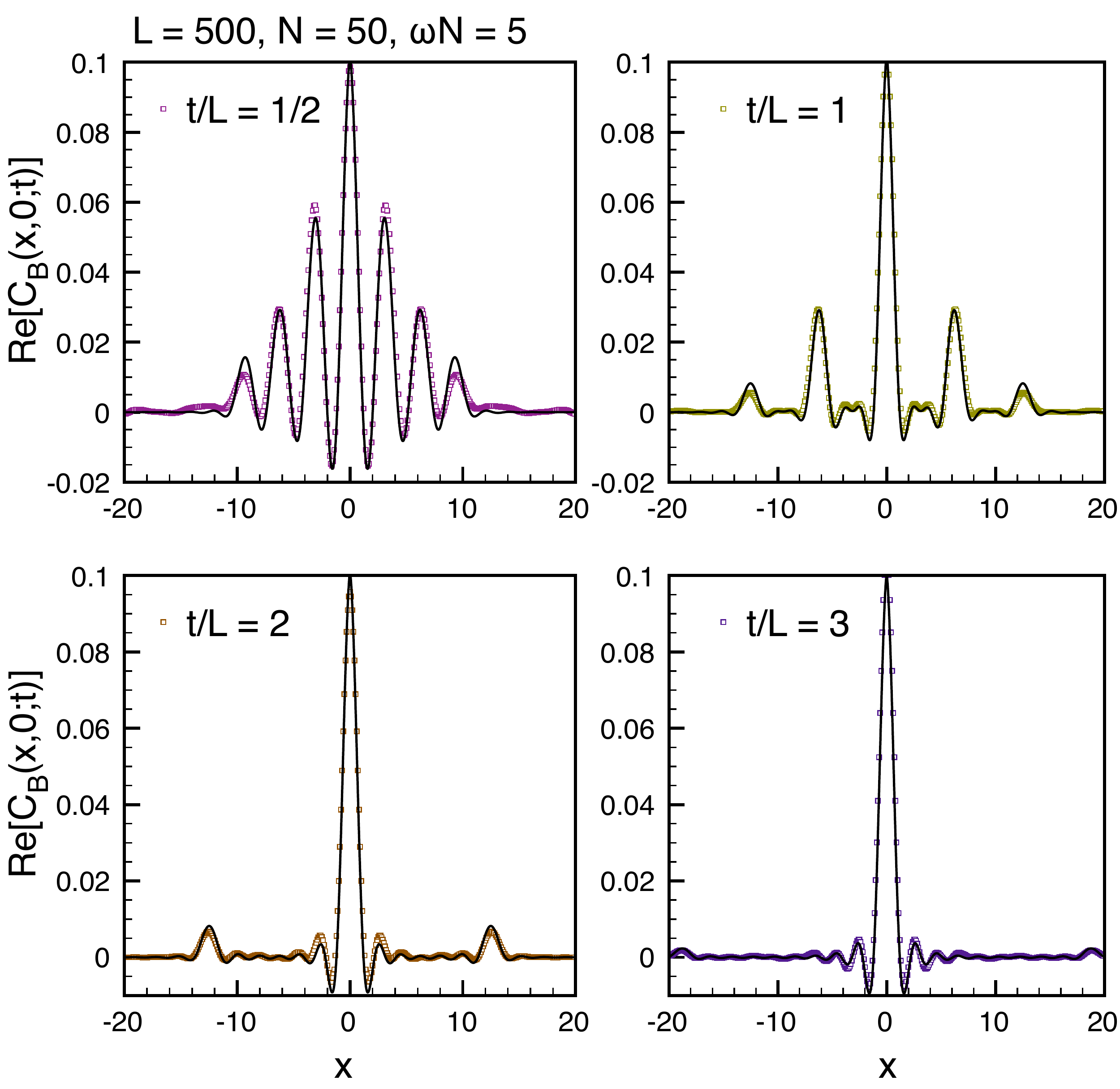}
\caption{(Left) Snapshot of the bosonic correlation $\mathrm{Re}[C_{B}(x,0;t)]$ at different rescaled times $t/L$. 
The profiles have been shifted along the vertical axis for clarity. The time $t/L=1/v\simeq 3/10$ approximatively separates 
the expansion in full space from the equilibration regime. 
(Right) Zoom close to the origin for the profiles at $t/L=1/2,1,2,3$. 
The full black lines correspond to the large-time behavior including corrections as in Eq. (\ref{CBcorr}) 
which match the data perfectly, including the moving peaks. 
In particular, for $t/L=1/2$ and $t/L=1$, in order to quantitatively describe all peaks, 
it has been necessary to include large-time corrections respectively up to the $3^{\rm rd}$ and $2^{\rm nd}$ order in Eq. (\ref{CBcorr}).} 
\label{fig_boson_corr}
\end{figure}

\subsection{The large-time limit of the bosonic correlators}
Eq. (\ref{C_B_overlap}) is not only a practical formula to calculate the bosonic correlation function at any time, but it is also 
the ideal starting point to evaluate its large-time limit. 
Indeed, from the single particle wave functions (\ref{phi_L_t}) and using the fact that only diagonal replica terms survive 
in the product, in the large-time regime $t/L\gg 1$ with $L \to \infty$ (but with $t/L^2 \ll 1$) 
one can approximate $\phi^{*}_{a}(z,t)\phi_{b}(z,t)$ as
\begin{eqnarray}
\phi^{*}_{a}(z,t)\phi_{b}(z,t) & \sim & \frac{\mathrm{e}^{i\pi(a-b)/2}}{\omega t} \sum_{p=-\infty}^{\infty} \chi_{a}^{*}\Big(\frac{z + pL}{\omega t}\Big)\chi_{b}\Big(\frac{z + pL}{\omega t}\Big) \nonumber \\
& \sim & \frac{i^{a-b}}{L}\int_{-\infty}^{\infty} dx\, \chi_{a}^{*}\left(\frac{z}{\omega t}+x\right) \chi_{b}\left(\frac{z}{\omega t}+x\right) = \frac{1}{L}\delta_{ab},
\end{eqnarray}
where in the last equality we used the orthonormality of the eigenfunctions $\chi_{a}(x)$. 
Consequently, the large-time behavior of the $\mathbb{A}$ and $\mathbb{P}$ matrices is
\begin{equation}\fl
\mathbb{A}_{ab}\equiv  \int_{x}^{y}dz\,\phi^{*}_{a}(z,t)\phi_{b}(z,t)= \frac{y-x}L \delta_{ab},\qquad
\mathbb{P}(x,y;t\rightarrow \infty) = \left(1-2\frac{|x-y|}{L}\right)\mathbb{I}.
\label{AandP}
\end{equation}
Clearly the above equations are valid as long as the rhs' are finite, i.e. when $|x-y|/L\sim O(1)$.
For $|x-y|\ll L$ different approaches must be used, as e.g. expanding the determinant in Eq. (\ref{C_B_fredholm}).

From Eq. (\ref{AandP}), the $\mathbb B$ matrix is 
\begin{equation}\label{Bmatrix_larget}\fl 
\mathbb{B} (x,y;t\rightarrow\infty) =  \left(1-2\frac{|x-y|}{L}\right)^{N-1} \mathbb{I},\quad {\rm and}
\quad \lim_{N\to\infty}  \mathbb{B} (x,y;t\rightarrow\infty)= \mathbb{I}\, \mathrm{e}^{-2n|x-y|} ,
\end{equation}
where the large $N$ limit has been taken keeping, as usual, $n=N/L$ constant.
Substituting Eq. (\ref{Bmatrix_larget}) in Eq. (\ref{C_B_overlap}) one finally finds
\begin{eqnarray}\label{bosonic_corr_eq}\fl
C_{B}(x,y;t\rightarrow\infty) & = & C(x,y;t\rightarrow\infty)\mathrm{e}^{-2n|x-y|} 
=2n\frac{J_{1}[ \sqrt{2\omega N}(x-y)]}{\sqrt{2\omega N}(x-y)}\mathrm{e}^{-2n|x-y|}. 
\end{eqnarray}
This result analytically explains the exponential decay of the bosonic correlation function that was already found for a different quench
in the GGE \cite{kormos}, providing also the exact prefactor $2n$ in the exponent. 

We can evaluate the approach to the stationary value of the bosonic correlator as we have already done for the fermionic one. 
The corrections to the factor $\mathrm{e}^{ -2n|x-y|}$ are exponentially small for $t/L\to\infty$ and, therefore, 
the leading corrections to the Eq. (\ref{bosonic_corr_eq}) come from corrections to the fermionic factor (cf. Eq. (\ref{CF_corr})), 
i.e. 
\be
C_{B}(x,y;t \gg 1) \simeq 2n\,\mathrm{e}^{-2n|x-y|} 
\sum_{m=-\infty}^{\infty}\frac{J_{1}[ \sqrt{2\omega N}(x-y + 2\pi m t /L) ]}{\sqrt{2\omega N}(x-y + 2\pi m t /L) }.
\label{CBcorr}
\ee
In Fig. \ref{fig_boson_corr}, not only the location, but also the amplitudes of the secondary peaks is well 
described by the above equation, although the fermionic correlation is not (see Fig. \ref{figCxy}). 
This is simply due to the exponential reduction of the peaks' amplitude in Eq. (\ref{CBcorr}).

In order to check the validity of these asymptotic predictions we numerically 
evaluate  $C_{B}(x,0;t)$ using Eq. (\ref{C_B_overlap}). 
In Fig. \ref{fig_boson_corr} we report the bosonic correlation function at different times for a system of size $L=500$ with 
$N=50$ and $\omega N = 5$. 
Notice how,  the bosonic correlation function 
reduces its amplitude and acquires a long-range oscillating behavior reminiscent of the fermionic correlation function
as long as the expanding gas does not feel the boundary (i.e. for $t/L\lesssim 1/v$). 
However, for larger times, the periodic boundary conditions destroy these oscillations and give rise to the exponential tail 
predicted by Eq. (\ref{CBcorr}) which perfectly describes the data for large enough time.

\begin{figure}[t]
\includegraphics[width=0.5\textwidth]{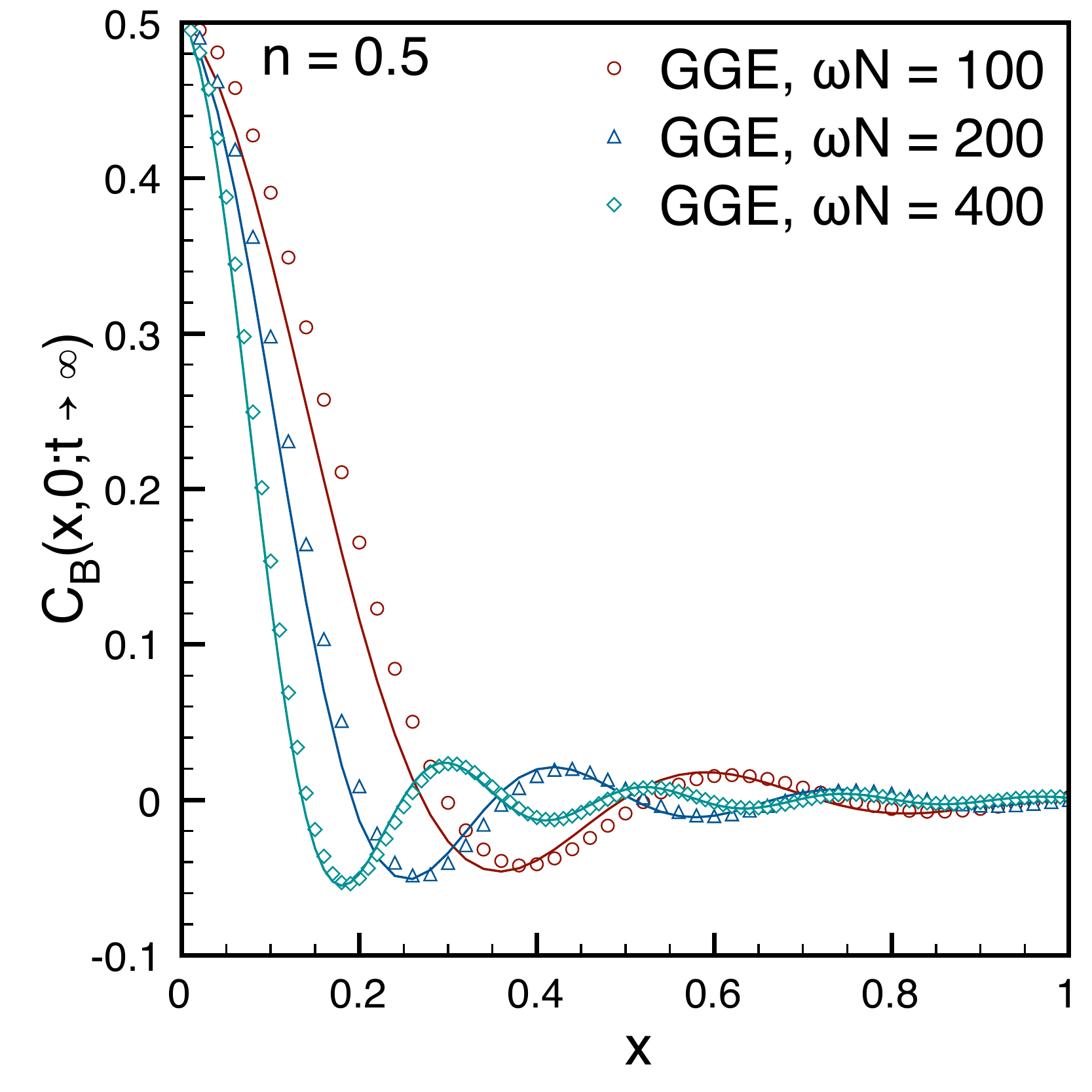}
\includegraphics[width=0.5\textwidth]{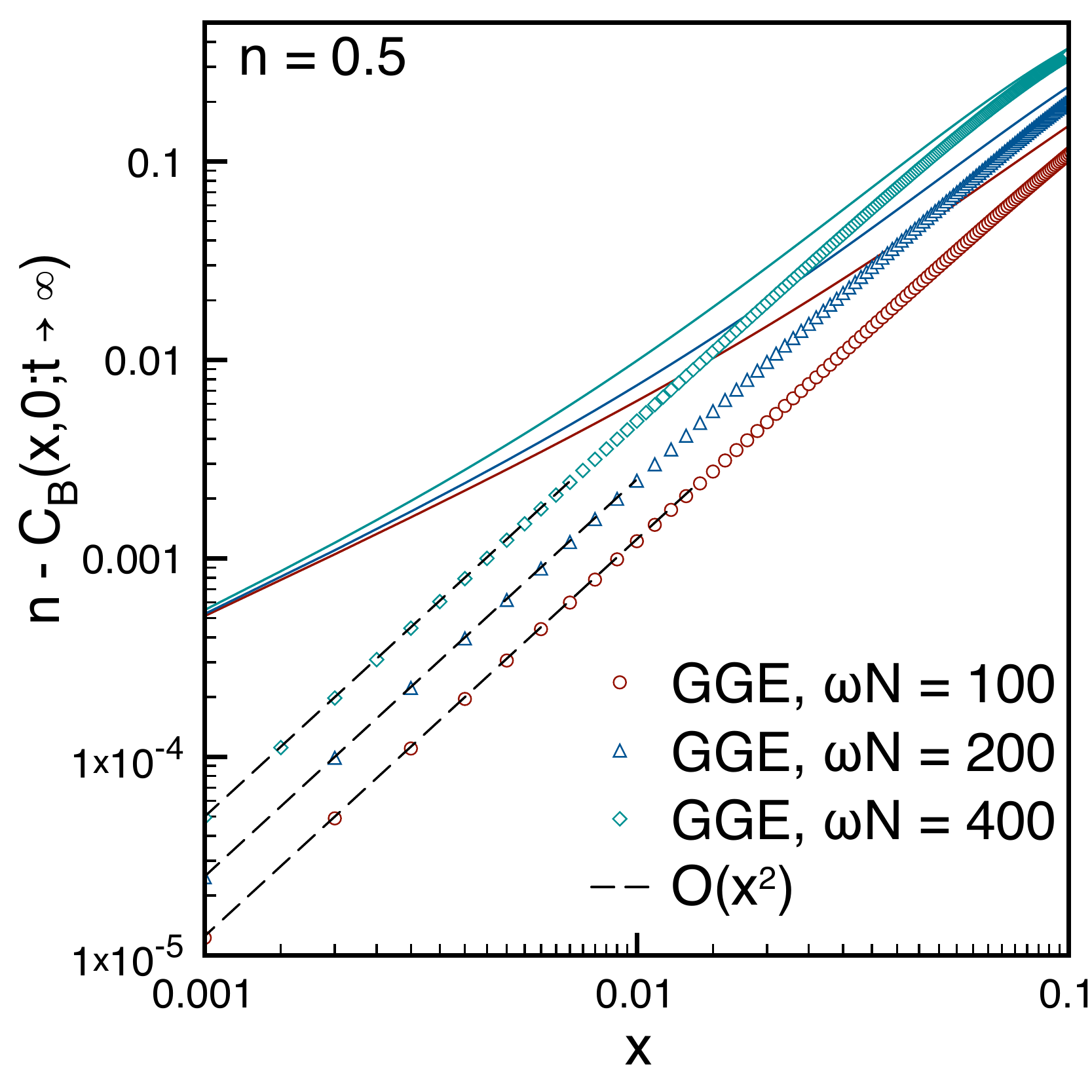}
\caption{(Left): Exact bosonic correlation function $C_{B}(x,y;t\rightarrow\infty)$ calculated by discretizing the Fredholm's 
minor in Eq. (\ref{C_B_fredholm}). For large enough $x$, the data always agree with the prediction in Eq. (\ref{bosonic_corr_eq}) 
(full lines), 
while for smaller $x$, the data approach it only for large enough $\omega N$.
(Right): Zoom for very small $x$ and in double logarithmic scale.
For small enough $x$ the behavior is always quadratic (dashed line), but increasing $x$ 
the bosonic correlation always crosses over to Eq. (\ref{bosonic_corr_eq}) (full lines).
} 
\label{newfig}
\end{figure}

\subsection{Bosonic correlators: the short-distance behavior}
The asymptotic formula (\ref{bosonic_corr_eq}) is strictly valid only in the TD limit because for any finite $N$, 
at very small distances $C_{B}(x,y;t\rightarrow\infty)$ crosses over to $|x-y|^3$
as expected from general arguments \cite{oldtan,tan1D,vm-13}. 
Consequently, the momentum distribution function has a large momentum tail of the form $k^{-2}$
which crosses over to the standard $k^{-4}$ for even larger $k$. 
This large-momentum crossover should be a measurable signature of the GGE. 

Eq. (\ref{C_B_overlap}) is not easy to manipulate for large enough number of particles needed to show this crossover in the GGE.
For this reason, we proceed by discretizing the Fredholm's minor in Eq. (\ref{C_B_fredholm}) 
as explained in Ref. \cite{adi,kormos}.
In order to do so, we proceed as follows: 
(i) we discretize the space interval $[0,x]$ in $M+1$ points, introducing the lattice spacing $a= x/(M+1)$; 
(ii) we define the $(M+1)\times (M+1)$ matrices (indices run form $1$ to $M+1$):
\bea 
\mathbb{R}_{nm}& = & \delta_{nm} - \delta_{n1}\delta_{1m}\\
\mathbb{S}_{nm}& = & C[(n-m)a,0;t\to\infty)\;{\rm for}\;n>1,\quad \mathbb{S}_{1m}  =   C(x-ma,0;t\to\infty),\nonumber
\eea
where $C(x,0;t\to\infty)$ is the GGE fermionic correlation in Eq. (\ref{Cxyt_larget_leading}).
Therefore, the bosonic correlator is given by the limit
\be
C_{B}(x,0;t\to\infty) = \lim_{a\to 0} \frac{\det(2a\, \mathbb{S}-\mathbb{R})}{2a}.
\ee
(In practice we evaluate the ratio in the rhs of the above equation for small enough spacing $a$
and check that it does not vary to the required precision by making it smaller.)
In this way, we numerically calculate $C_{B}(x,0;t\to\infty)$ as a function of $x$ for different values of $\omega N$
(we recall $n=N/L$ is constant) and the results are reported in Fig. \ref{newfig}. 
It is clear that increasing $N$, the numerical data approach the asymptotic result in Eq. (\ref{bosonic_corr_eq}).
However, if we zoom in the region of very small distances, as done in the left panel of Fig. \ref{newfig}, 
the $|x-y|$ singularity is absent, and the main singularity is of the form  $|x-y|^3$ while the leading behavior is non-singular $(x-y)^2$.
This small distance behavior can be worked out analytically in a very simple way. 
Indeed, for small $|x-y|$, the only terms contributing to the expansion $C_{B}(x,y;t\to\infty)$
up the $3^{rd}$ order are those with $n=0,1$  in Eq. (\ref{C_B_fredholm}) (analogously to a similar expansion for a different 
physical problem in Ref. \cite{vm-13}). 
The sum of these two contributions is 
\be\fl
C_{B}(x,y;t\to\infty) \sim n -\frac{n\omega N}{4}(x-y)^{2} - \frac{n^2 \omega N}{6}|x-y|^{3} + O((x-y)^4).
\ee
The right panel of Fig. \ref{newfig} reports the numerically evaluated $C_{B}(x,y;t\to\infty)$ for small $x$ which perfectly agrees
with the expansion above, and it crosses over to the TD value (\ref{bosonic_corr_eq}) for larger values of $x$.

\begin{figure}[t]
\includegraphics[width=0.5\textwidth]{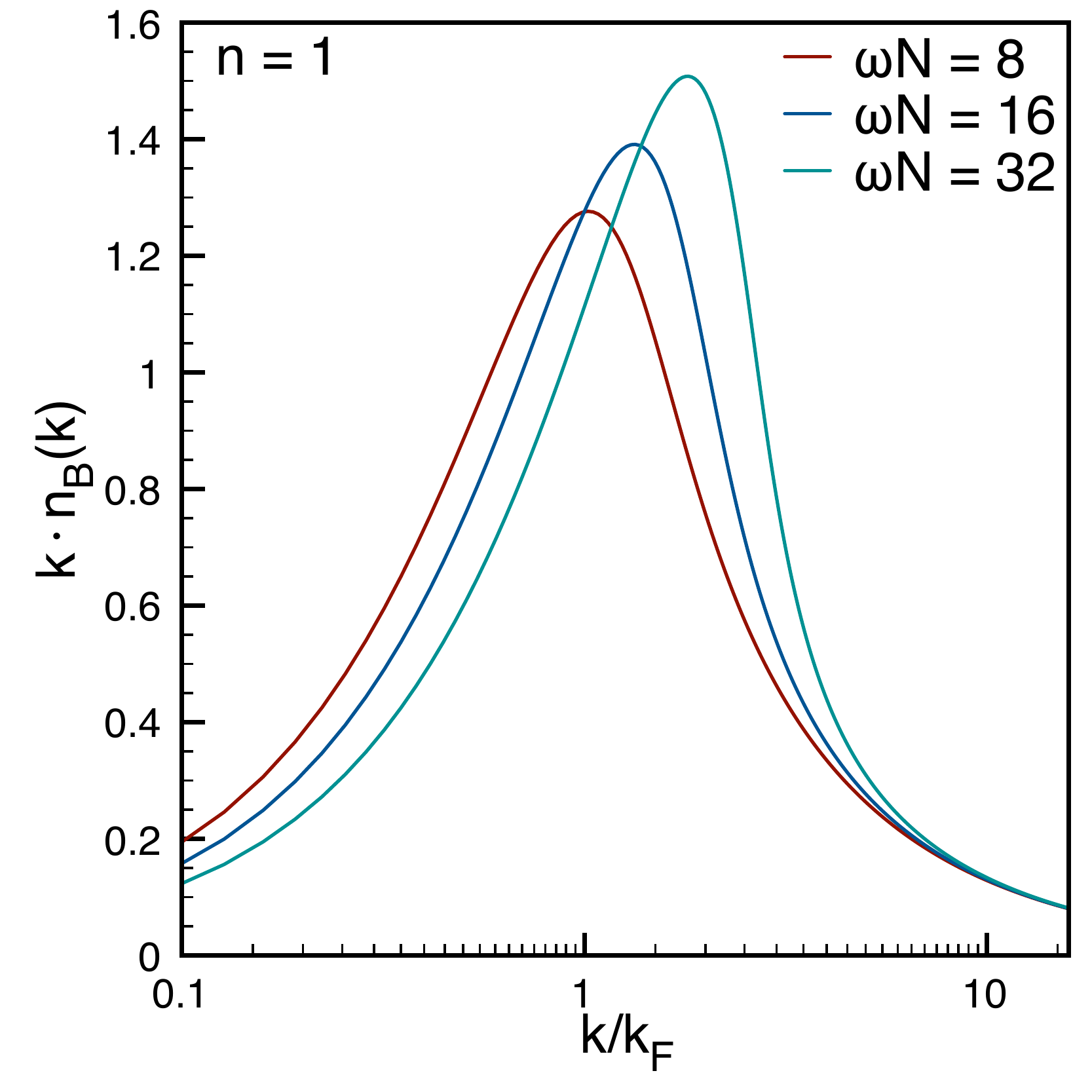}
\includegraphics[width=0.5\textwidth]{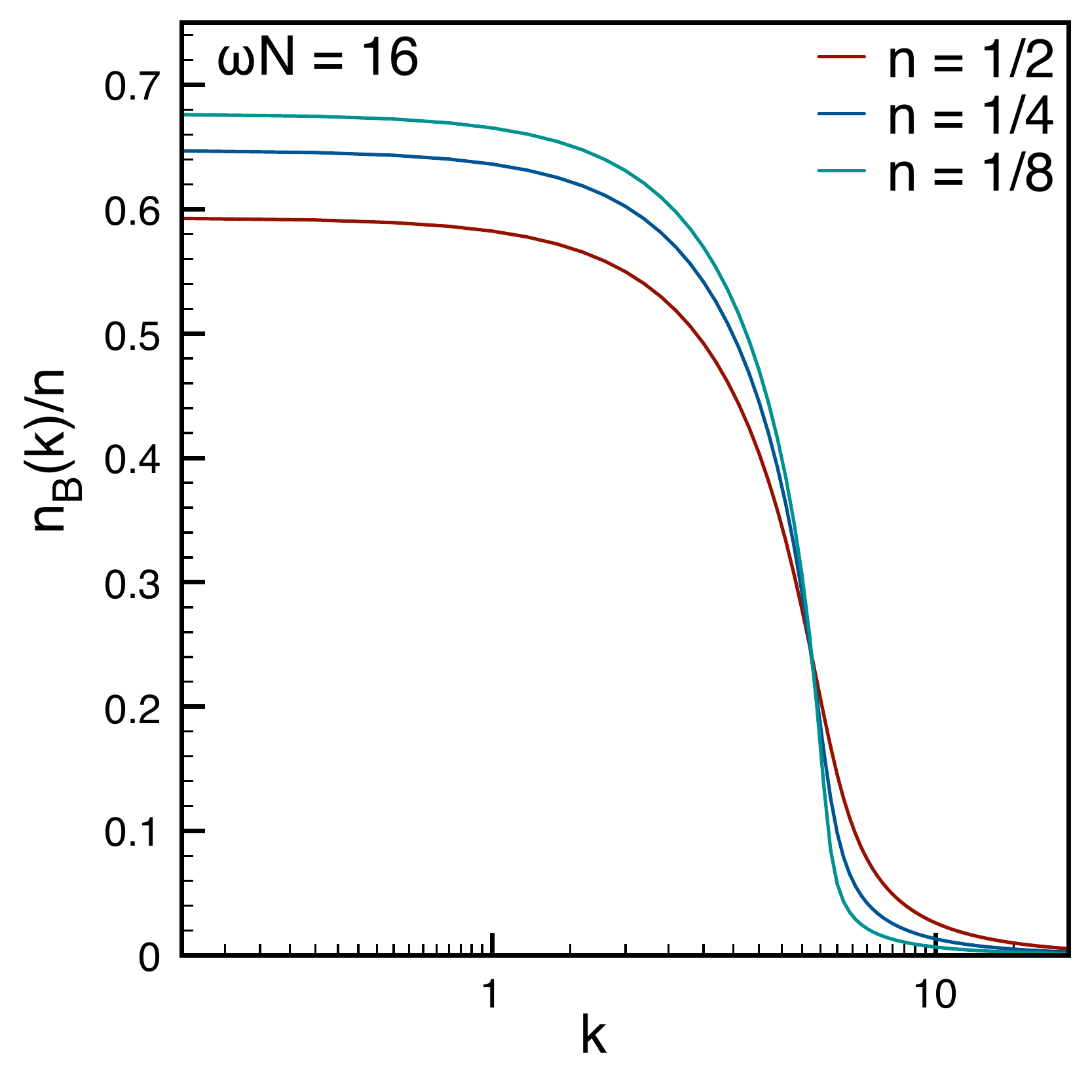}
\caption{
(Left): The GGE bosonic momentum distribution $n_{B}(k)$ (actually $k n_{B}(k)$ in order to have 
a plot resembling the one in Ref. \cite{ck-12})
for different initial conditions $\omega N$ and for fixed density $n=1$, as a function of $k/k_{F}$ ($k_{F}=n\pi$).
(Right): $n_{B}(k)/n$ for fixed initial condition $\omega N=16$ and several final densities $n$ as function of $k$. 
Notice as $n$ decreases the emergence of a singularity at $k=\sqrt{2\omega N}$.
As stressed in the text, for any finite $\omega N$, there will be a crossover from the plotted $k^{-2}$ tail for large $k$
to a standard $k^{-4}$. The location of the crossover depends on the value of $\omega N$ and it is not encoded in Eq. (\ref{nbGGE}).
} 
\label{fig_nB_k}
\end{figure}

\subsection{The bosonic momentum distribution}
The bosonic correlation function $C_{B}(x,y;t)$ in the large-time limit is translationally invariant, as it should. 
Thus for the bosonic occupation number operator 
\be
\hat{n}_{B}(k) \equiv \frac{1}{L}\int\!\!\int\! dx dy\,\mathrm{e}^{ik(x-y)} \hat{\Phi}^{\dag}(x)\hat{\Phi}(y),
\ee
one obtains in the large-time limit expectation value
\be
n_{B}(k) = \lim_{t\to\infty}\langle\hat{n}_{B}(k,t)\rangle =  \int dx\, \mathrm{e}^{ikx} C_{B}(x,0;t\to \infty).
\ee
Since the real space correlator (\ref{bosonic_corr_eq}) is a product, its Fourier transform is the convolution 
\be
n_{B}(k) = \int_{-\sqrt{2\omega N}}^{\sqrt{2\omega N}} \frac{dq}{2\pi} \, n_{GGE}(q)  \frac{1/n}{1+(k-q)^2/4n^2},
\label{nbGGE}
\ee
with $n_{GGE}(k)$ given in Eq. (\ref{nkGGE}).
In Fig. \ref{fig_nB_k} we plot the stationary bosonic momentum distribution as a function of $k/2\pi$ for different initial conditions. 
Notice that our result differs from the Gangardt and Pustilnik result in \cite{gp-08}, where using the stationary phase 
approximation it was found $n_{B}(k) = n_{GGE}(k)$ for $t\to \infty$. 
Indeed, Ref. \cite{gp-08} analyzes the expansion of a bosonic gas in the full space, for which the density of particle $n$ goes to zero.
However, their result is encoded in our solution if one considers the na\"ive limit $n\to 0$. 

The large momentum behavior of Eq. (\ref{nbGGE}), for any finite $\omega N$,  is ($k_F^0=\sqrt{2\omega N}$) 
\be
n_{B}(k\gg k^0_F) \simeq\frac{4n^2}{k^2}.
\ee
However, as previously discussed, this result does not reproduce the correct $k^{-4}$ large momentum tail for $n_{B}(k)$. 
Indeed, there will be a crossover at large enough $k$, whose location depends on $\omega N$,
from this $k^{-2}$ tail to the expected $k^{-4}$ as a consequence of the crossover for small $x$ in Fig. \ref{newfig}.

\section{Entanglement entropies of a subsystem}
\label{sec7}

Up to know we have only considered the correlation functions of local observables either bosonic or fermionic.
Another extremely important quantity for a full description of the out of equilibrium dynamics of 
quantum systems is the bipartite 
entanglement entropy. 
Indeed the amount of entanglement contained in a quantum system is the main limitation \cite{swvc-08,cv-09}
to simulate on a classical computer a quantum system (at least for 
numerical methods based on tensor network states). 
This observation motivated an intense study on the evolution of the entanglement
entropy and in particular of its growth with time. 
Based on results from conformal field theory \cite{cc-05,cc-07b,ds-11}
and on analytical \cite{cc-05,fc-08} or numerical calculations \cite{dmcf-06,lk-08,ep-08,ep-08b,isl-09,va-12,cc-13,sled-13,hgf-09} 
for specific models, it is known that the entanglement entropy 
grows linearly with time for a  global quench, while at most logarithmically for a local one. 
As a consequence a local quench is effectively simulable on a classical computer up to large times, while for a global
quench one can access only a relatively short time dynamics.
It is then natural to wonder whether the trap release dynamics studied here, displays an asymptotically logarithmic or 
extensive behavior for large times. To this goal, let us first introduce the basic definition and then move to the actual calculations.

For a general bipartition of a pure state $|\Psi\rangle$ of a quantum system (i.e. writing the whole Hilbert space of the system 
 as a direct product of two parts $\HH=\HH_A\otimes\HH_B$),
the R\'enyi entropy of the reduced density matrix $\rho_{A}={\rm Tr}_B |\Psi\rangle\langle\Psi|$ 
of the subsystem $A$ \cite{rev}
\begin{equation}
S^{(\alpha)}_{A} = \frac{1}{1-\alpha}\ln\mathrm{Tr}\rho^{\alpha}_{A},
\end{equation}
is a measurement of the entanglement between the two parts. 
In the limit $\alpha\to 1$, $S^{(1)}_{A}$ reduces to the more studied von Neumann entanglement entropy
but, the knowledge of the R\'enyi entropies for any $\alpha$ gives far more 
information than the $\alpha=1$ case because it provides  
the full spectrum of the reduced density matrix \cite{cl-08}.
In the ground-state of a one-dimensional conformal critical system (which include the quantum gases studied here)
in the case when $A$ is an interval of length $\ell$ 
embedded in a finite system of length $L$, the asymptotic behavior of the R\'enyi
entropies is  given by \cite{Holzhey,cc-04}
\begin{equation}
S^{(\alpha)}_A = \frac{c}{6}\left( 1 + \frac{1}{\alpha}\right)\ln \left[\frac{L}\pi \sin\Big(\frac{\pi\ell}L\Big)\right] + cst,
\end{equation}
where $c$ is the central charge of the underlying conformal field theory \cite{c-lec}. 
Conformal invariance can also be exploited to predict the behavior of the entanglement entropy for global \cite{cc-05} 
and local \cite{cc-07b,ds-11} quantum quenches. 

The bipartite entanglement entropies of a spatial subsystem of a one-dimensional quantum gas 
can be obtained exploiting the systematic framework of Refs. \cite{cmv-11,cmv-11b}
whenever the model can be mapped to a noninteracting fermion system. 
Indeed, the bipartite entanglement entropies of an interval $[x,y]$ can be always expressed in terms 
of the Fredholm's determinant in Eq. (\ref{DFD}) 
as the integral \cite{cmv-11b,jk-04}
\be\label{entropies_contour}
S^{(\alpha)}_{[x,y]}
= \oint\frac{d\lambda}{2\pi i}e_{\alpha}(\lambda)\frac{d}{d\lambda}\ln \mathcal{D}_{[x,y]}(\lambda^{-1}),
\ee
over a contour which encircles the segment $[0,1]$ and where $e_{\alpha}(\lambda)=\ln[\lambda^\alpha+(1-\lambda)^\alpha]/(1-\alpha)$. 
As a crucial point, the Fredholm's determinant can be evaluated in terms of the 
overlap matrix $\mathbb{A}$  because \cite{cmv-11,cmv-11b}
\bea\fl 
\Tr C^n\equiv \int_{x}^y dz_1\dots dz_n C(z_1,z_2) C(z_2,z_3)&& \dots C(z_n,z_1)= \nonumber \\ &&
=\sum_{i_1,i_2,\dots i_n} \mathbb{A}_{i_1i_2}\mathbb{A}_{i_2i_3}\dots \mathbb{A}_{i_ni_1}=\Tr \mathbb{A}^n,
\eea
where $C(x,y)$ is the fermionic correlation function. Using this trace identity 
 one has
\bea\fl 
\ln \mathcal{D}_{[x,y]}(\lambda^{-1}) & = & -\sum_{n=1}^{\infty}\frac{\Tr C^{n}}{n\lambda^{n}} = -\sum_{n=1}^{\infty}\frac{\Tr \mathbb{A}^{n}}{n\lambda^{n}} = -\sum_{n=1}^{\infty}\sum_{m=0}^{N-1}\frac{a_{m}^{n}}{n\lambda^{n}}
= \sum_{m=0}^{N-1}\ln(\lambda-a_{m}),
\eea
where $\{a_0,\ldots a_{N-1}\}$ are the eigenvalues of $\mathbb{A}$ and we have neglected the term proportional to $\ln\lambda$ because it does not contribute to the integral in Eq. (\ref{entropies_contour}). Finally, inserting this last result into the contour integral, one obtains
\be
S^{(\alpha)}_{[x,y]} 
= \oint\frac{d\lambda}{2\pi i} \sum_{m=0}^{N-1}\frac{e_{\alpha}(\lambda)}{\lambda-a_{m}} = \sum_{m=0}^{N-1}e_{\alpha}(a_{m}).
\label{SvsA}
\ee
This result is extremely  useful because it reduces the calculation of a problem initially defined on a continuum interval to the diagonalization of a $N\times N$ matrix that can be easily handled at least numerically. 
This approach has been already applied to a variety of equilibrium and non-equilibrium situations \cite{cmv-11,cmv-11b,cmv-12w,cmv-12h,v-12b,nv-13,v-12}. 
The result for the  interval $[x,y]$ in ground state of a gas of $N$ impenetrable bosons
 in a ring of length $L$  is \cite{cmv-11,cmv-11b}
\be
S^{(\alpha)}_{[x,y]}= \frac{1+\alpha^{-1}}{6}\ln\left[2N\sin \frac{\pi|y-x|}{L}\right]+cst.
\label{SaGS}
\ee

\begin{figure}[t]
\includegraphics[width=0.5\textwidth]{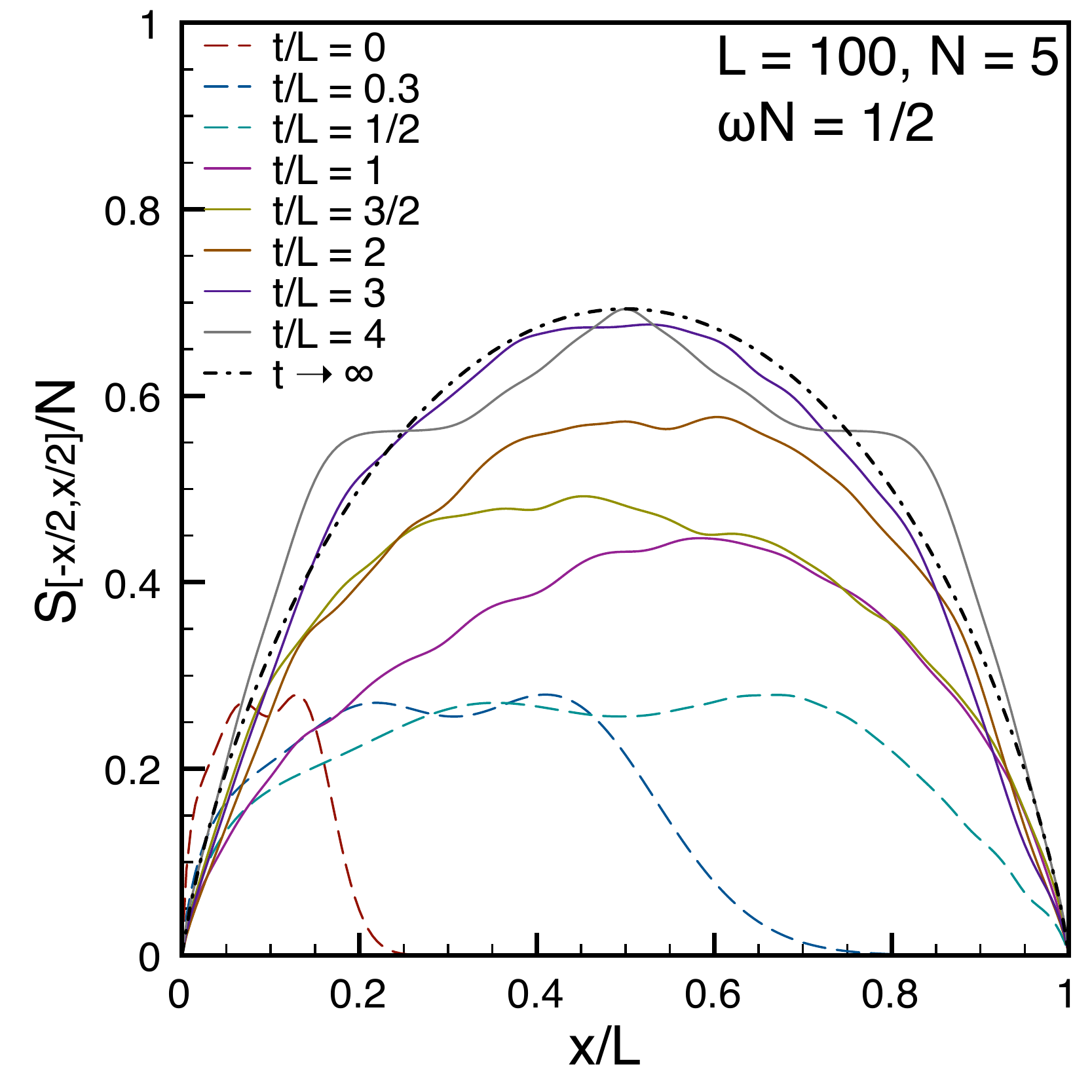}
\includegraphics[width=0.5\textwidth]{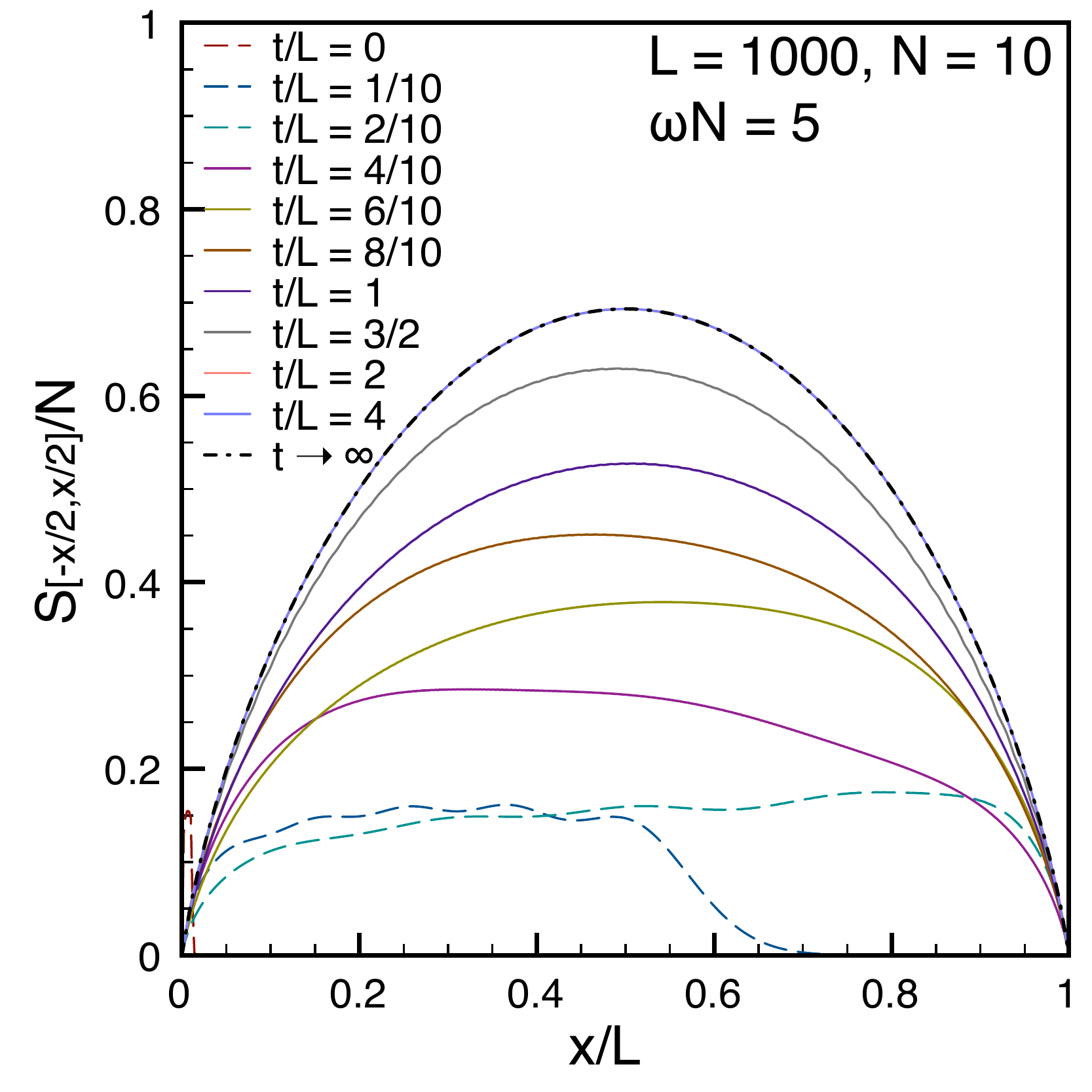}
\caption{Entanglement entropy profiles $S_{[-x/2,x/2]}(t)$, i.e. around the center of the system, as a function of the rescaled 
subsystem length $x/L$ for different times and initial conditions ($\omega N=1/2$ on the left and $\omega N=5$ on the right). 
The initial profile at time $t=0$ extends up to $x/L\simeq2\ell/L=4n/v$, where $v\equiv\sqrt{2\omega N}$ is the expansion 
velocity of the gas. 
This allows us to identify two different time regimes: for $t/L<1/2v$ (dashed lines) the entropy profiles expand 
following the expansion of the gas; for $t/L>1/2v$ (full lines) the entropy profiles grow approaching the equilibration profile 
(dot-dashed line). 
For sound velocity $v=1$ (left), the revival period $\tau/L \sim L /2\pi \simeq 16$ is small and the 
revival effects appear before the TD equilibration profile is reached. 
See, e.g., the curve for $t/L=4$ for which the revival effects are clearly present. 
For $v \simeq 3.16$ (right), the equilibration time is such that for $t/L=4$ the gas has already travelled around 
the ring about $12$ times allowing a perfect equilibration long before the revival effects take place ($\tau/L\sim 160$).
} 
\label{figS1}
\end{figure}

All numerical data in this section have been obtained by calculating explicitly the time dependent overlap matrix and 
plugging its eigenvalues in Eq. (\ref{SvsA}). 
As a first example, we report in Fig. \ref{figS1} the von Neumann entropy of the interval  $A=[-x/2,x/2]$
centered around the origin as a function of the length of the subsystem $x$ for different times and initial conditions.

\begin{figure}[t]
\includegraphics[width=0.5\textwidth]{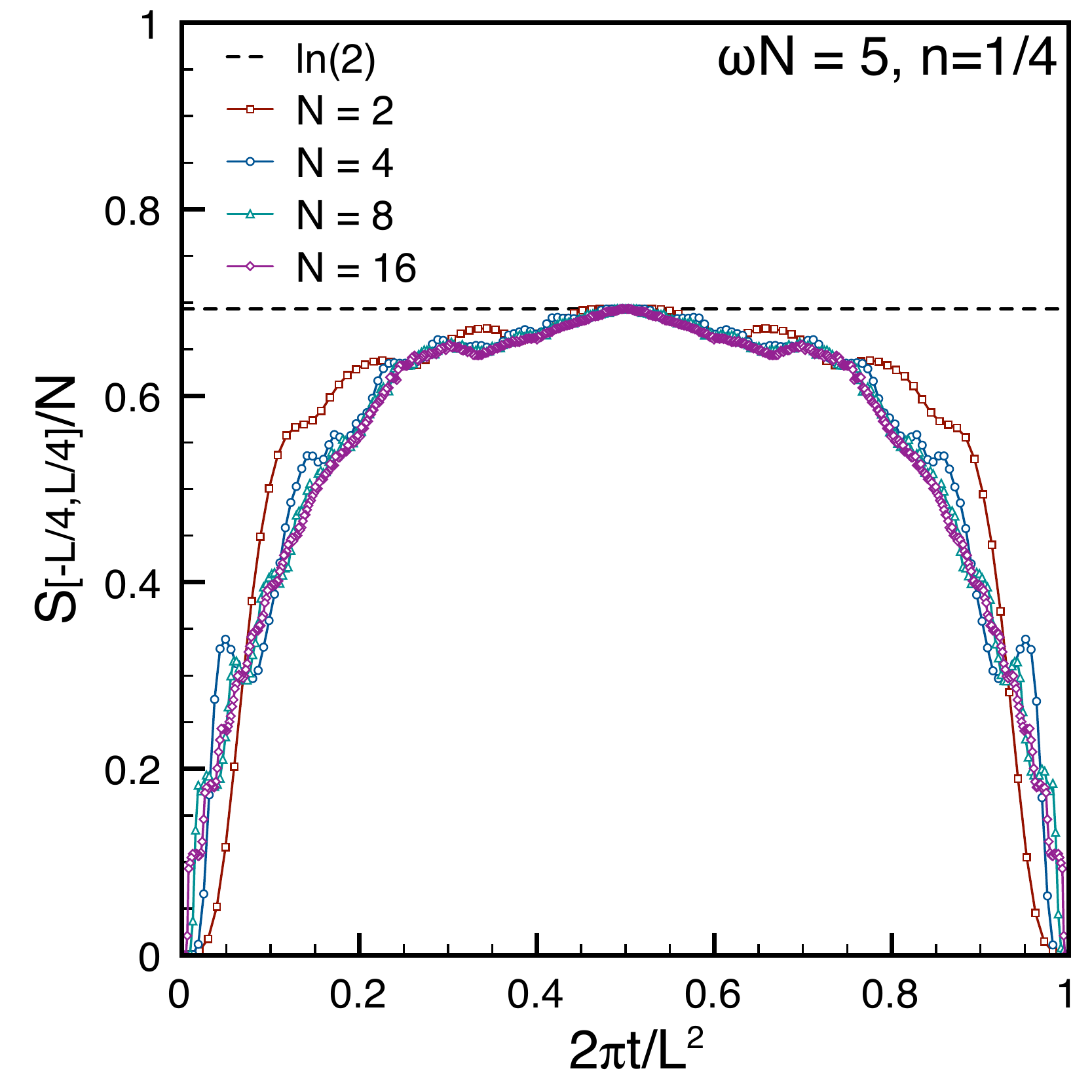}\includegraphics[width=0.5\textwidth]{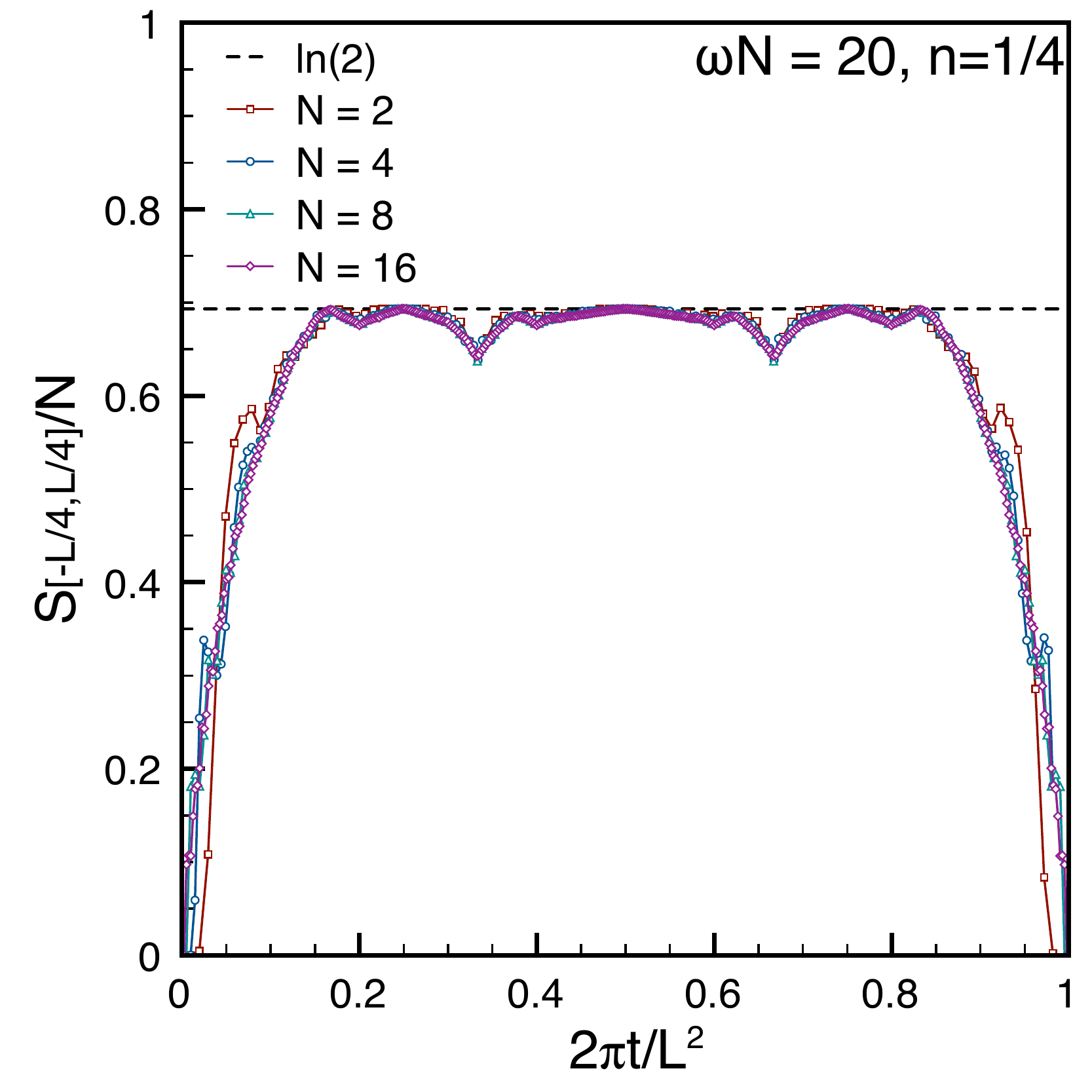}
\caption{The evolution of the entanglement entropy $S_{[-L/4,L/4]}(t)$ for a subsystem of length $L/2$ around the center 
of the initial trap, for system sizes $L=8,16,32,64$ and constant density $n=1/4$. 
The time is rescaled in terms of the revival period $\tau=L^2/2\pi$ in order to show the revival effects. 
The entanglement entropy stays for most of the time close to the stationary plateaux $\ln2$ which is approached 
before for larger $\omega N$, i.e. larger expansion velocity.
} 
\label{figS1_scaling}
\end{figure}

\subsection{The large-time limit of the entanglement entropies}
While for arbitrary finite time, the analytical diagonalization of  the overlap matrix $\mathbb{A}$  for an interval $[x,y]$ 
is a hard task,  in the TD and large-time limit the overlap matrix reduces
to the diagonal form in Eq. (\ref{AandP}), i.e. 
to $\mathbb{A}(t\to\infty) = \frac{|y-x|}{L}\mathbb{I}$. 
The R\`enyi entropies after the equilibration are then straightforwardly obtained as 
\be\label{Salpha_eq}
S^{(\alpha)}_{[x,y]}(t\to\infty) =N e_{\alpha}\left(\frac{|y-x|}{L}\right).
\ee
For $\alpha\to 1$ the previous formula gives the von Neumann entropy (we define $S_A\equiv S^{(1)}_A$)
\be\label{S1_eq}\fl
\frac{S_{[x,y]}(t\to\infty)}{N} = -\frac{|y-x|}{L}\ln\left(\frac{|y-x|}{L}\right) -\left(1-\frac{|y-x|}{L}\right)\ln\left(1-\frac{|y-x|}{L}\right),
\ee
and for $\alpha\to\infty$ the single copy entanglement 
\be\label{Sinf_eq}
\frac{S^{(\infty)}_{[x,y]}(t\to\infty)}{N} = -\ln\left(\frac{1}{2}+\left| \frac{|y-x|}{L} - \frac{1}{2} \right| \right).
\ee
We could have achieved the same result using directly the limit of the Fredholm's determinant 
$$
\ln\mathcal{D}_{[x,y;t]}(\lambda^{-1}) \to N \ln \left[ \frac{1}{\lambda} \left( \lambda- \frac{|y-x|}{L} \right) \right],
$$
that plugged in Eq. (\ref{entropies_contour}) 
gives again
$$
S^{(\alpha)}_{[x,y]}(t) = N  \oint\frac{d\lambda}{2\pi i} \frac{e_{\alpha(\lambda)}}{\lambda-|y-x|/L} = N e_{\alpha}\left(\frac{|y-x|}{L}\right).
$$

The extensive behavior of the entanglement entropies for large time in Eq. (\ref{Salpha_eq}) 
is a direct consequence 
of the fact that the overlap matrix is proportional to the identity for any $N$.
This contrasts the result for the expansion in full space \cite{cmv-11,v-12} for which, as a consequence of the zero-density 
limit for large time, for any finite subsystem $A$ one has $\mathbb{A}\to 0$ and so also the entanglement entropies
$S_A^{(\alpha)}(t\to\infty)\to 0$.
Indeed, using the exact solution of the one-particle dynamics for the release from a harmonic potential to the full line, 
it has been shown that the time-dependent entanglement entropy is \cite{cmv-11,v-12,nv-13} 
\be
S^{(\alpha)}_{[x,y]}(t)=S^{(\alpha)}_{[x/s(t),y/s(t)]}(0), \qquad {\rm with}\quad  s(t)=\sqrt{1+\omega^2 t^2}.
\label{Sfullsp}
\ee

\subsection{The time evolution of the entanglement entropy}

\begin{figure}[t]
\includegraphics[width=0.33\textwidth]{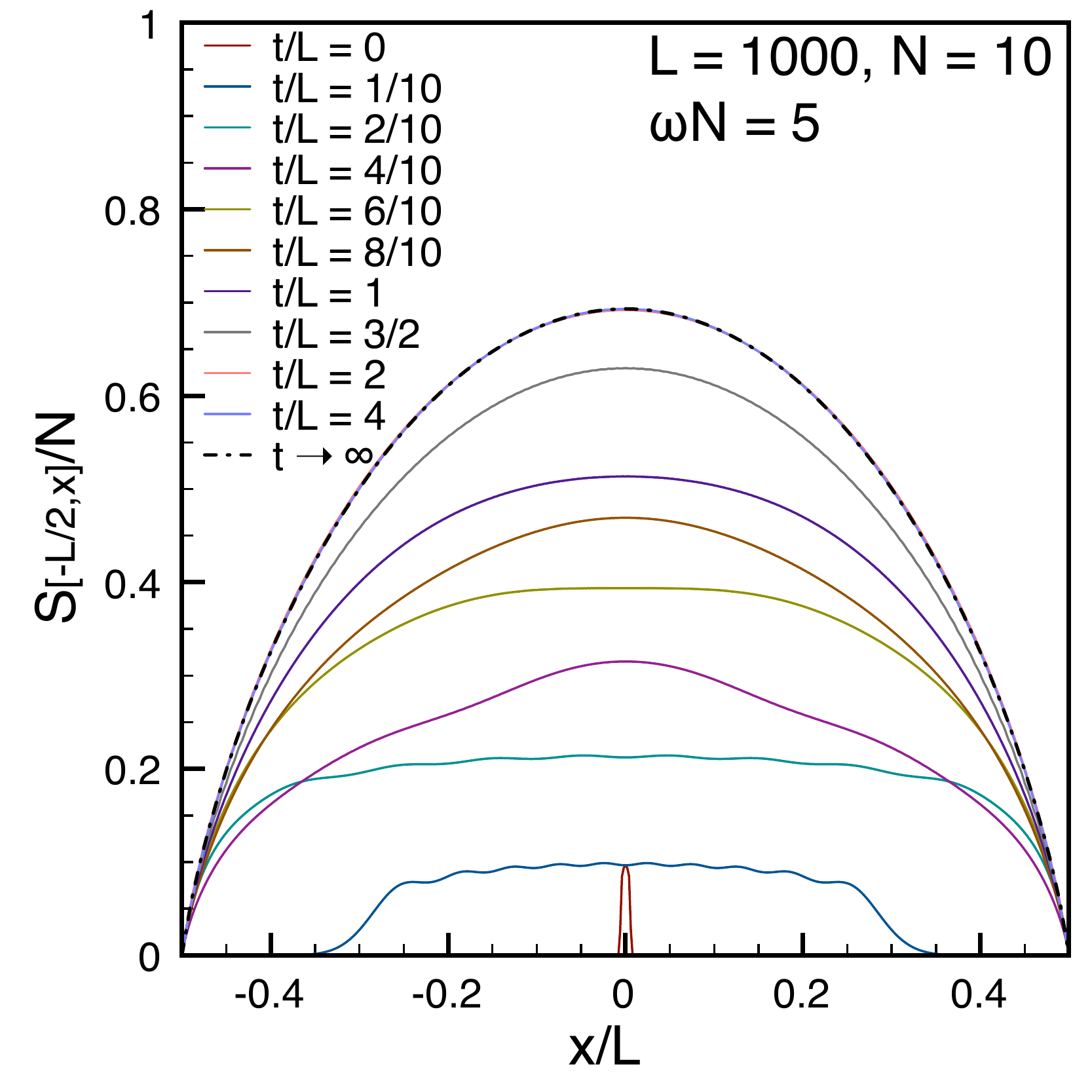}
\includegraphics[width=0.33\textwidth]{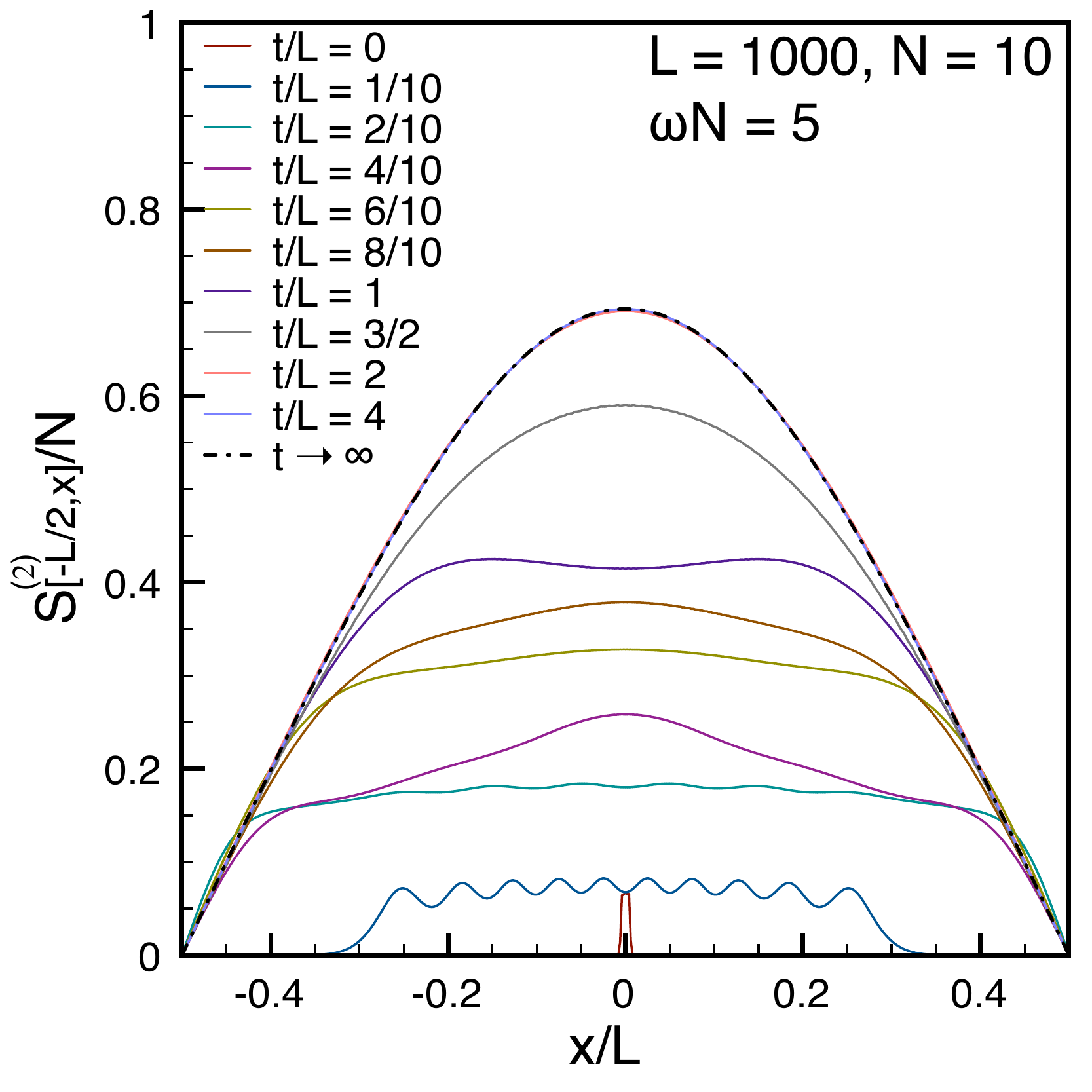}
\includegraphics[width=0.33\textwidth]{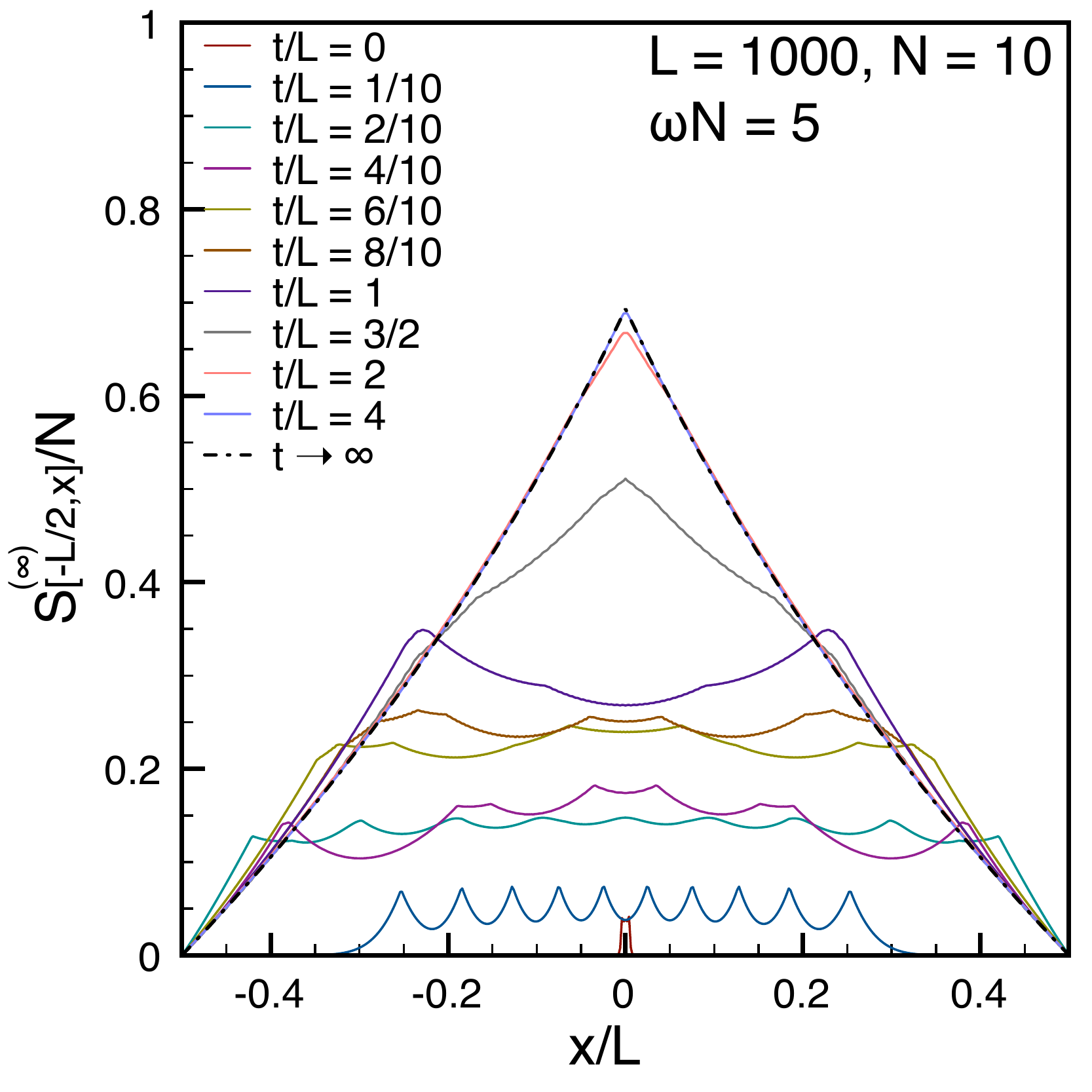}
\caption{Profiles of the entanglement entropies  $S^{(\alpha)}_{[-L/2,x]}(t)$ of the subsystem $[-L/2,x]$ 
for $\alpha=1,2,\infty$ as a function of the rescaled distance $x/L$ for different times. 
As expected, with increasing time the entropies tend to their equilibrium profiles.  
} 
\label{figSa_border}
\end{figure}

We now consider the time evolution of the entanglement entropies as obtained from the exact numerical 
diagonalization of the overlap matrix. 
Figs. \ref{figS1}, \ref{figS1_scaling}, and \ref{figSa_border} show different aspects of this time evolution. 
First it is evident from all figures that there are two clearly separated time regimes:
(i)
for $t/L< 1/2v$ (we recall $v=\sqrt{2\omega N}$) the gas expands without feeling the PBC and follows the dynamics as 
in full space in Eq. (\ref{Sfullsp});
(ii) once the particles go around the circle, they begin to mix with each other and therefore produce an increase of the entropy
that, after many turns, approaches the asymptotic value in Eq. (\ref{Salpha_eq}).
There are many other interesting details in this time evolution.
In the initial regime (i), as long as also $\omega t\ll 1$, in Eq. (\ref{Sfullsp}) $s(t)\sim 1$ and the support of the 
entanglement entropy just expands without changing considerably from the initial value in Eq. (\ref{SaGS}) as is clear from 
Figs. \ref{figS1} and  \ref{figSa_border}.
In the regime (ii) the gas starts turning around the circle and the entanglement entropies grow from the initial 
logarithmic dependence in $N$ to the extensive asymptotic one in Eq. (\ref{Salpha_eq}).
However, in order to ensure that the equilibration will be fully achieved before the revival effects take place, 
the inequality $L/2\pi \gg t/L \gg 1/v$ should be fulfilled  to ensure that the gas goes around the ring a sufficient number of 
times to equilibrate without ever reaching the revival time $L^2 /2\pi$. 
To further analyze the revival influence on the entanglement entropies, we report in Fig. \ref{figS1_scaling} the 
time-evolution of $S^{(1)}_{[-L/4,L/4]}(t)$ for a subsystem of length $L/2$ around the center of the initial trap. 
Comparing the two panels, we see that the sharper is the initial confinement, the faster the entropy reaches its equilibration value $\ln 2$. 
Moreover, the equilibrium value is held (apart from finite-size effects) for longer time for higher expansion velocity. 
In Fig. \ref{figSa_border} we report the different entropies $\alpha=1,2$ and $\infty$ for the subsystem $[-L/2,x]$ which 
starts from the left boundary: all the discussed effects are evident such as the initial expansion and after a few
turns around the circle a perfect match to the expected stationary behavior. 

While the quantitative aspects we have found are specific of the free fermionic model, the two time regimes
and their gross features are expected to be valid for a general trap release of any interacting 1D model.
This  allows us to conclude that it is possible to handle numerically by means of tensor network algorithms 
the gas expansion in the first time regime of expansion in full space, but not after the gas has reached the boundaries.
This observation explain a posteriori why it has been possible to obtain very accurate tDMRG simulation 
of interacting expanding gas on the full line \cite{hm-v,hm-v2,rsb-13}.

Finally we need to stress that the entanglement entropies show the expected large time and TD limit behavior also for very few 
particles (the maximum value of $N$ in the three Figures \ref{figS1}, \ref{figS1_scaling}, and \ref{figSa_border}  is $N=16$), 
as opposite to other observables discussed so far.
This is a quite standard fact both in and out of equilibrium, because, being the entanglement a more global
observable, all the small length and short time non-universal physics is averaged out giving at most subleading corrections in $N$.  
The main limitation for the applicability of the asymptotic result is that 
the length $L$ should be sufficiently large to guarantee the orthonormalization of all the initially occupied one-particle levels 
in order to ensure the validity of Eq. (\ref{AandP}).
Indeed, since we are working with the eigenfunctions $\chi_{j}(x)$ of the harmonic oscillator defined in the whole space,  the 
normalization condition in $[-L/2,L/2]$ is correct up to the correction in Eq. (\ref{mapeta}) of the order of 
$\sim j^{-1/4}(\omega L)^{j-3/4}\mathrm{e}^{-\omega L^2 / 8}$.

\subsection{Entanglement entropies and particle fluctuations}
The R\'enyi entropies 
characterize the non-trivial connections between different parts of an extended quantum system. 
For systems which can be mapped to free fermions as the present one, they are intimately related to the expectation values of the 
correlations of local operators. 
Indeed,  the entanglement entropies can be formally related to the even cumulant $V^{(2k)}_{[x,y]}$ 
of the particle-number distribution \cite{kl-09,srh-10,srf-12,cmv-12}
\be\fl
V^{(k)}_{[x,y]} = (-i\partial_{\lambda})^{k}\ln\langle \mathrm{e}^{i\lambda \hat{N}_{[x,y]}} \rangle |_{\lambda = 0},
\qquad {\rm where}\quad
\hat{N}_{[x,y]} = \int_{x}^{y}\! dz \, \hat{\Psi}^{\dag}(z)\hat{\Psi}(z),\label{genfu}
\ee
is the operator counting the number of particles in the interval $[x,y]$. 
Indeed, it has been shown that the following formal expansion holds \cite{srf-12}
\be\label{Salpha_cumulant}
S^{(\alpha)}_{[x,y]}  = \sum_{k=1}^{\infty} s^{(\alpha)}_{k} V^{(2k)}_{[x,y]},\quad s^{(\alpha)}_{k}  =  \frac{(-1)^{k}(2\pi)^{2k}2\zeta[-2k,(1+\alpha)/2]}{(\alpha-1)\alpha^{2k} (2k)!},
\ee
where $\zeta[n,x]\equiv \sum_{k=0}^{\infty}(k+x)^{-n}$ is the generalized Riemann zeta function. 
When the generating function (\ref{genfu}) is specialized to the trap release dynamics, it assumes a particularly simple form 
after taking the TD and large-time limits, as a consequence of the diagonal form of the overlap matrix. 
Indeed we have
\be
\frac{V^{(k)}_{[x,y]}}{N}  =  (-i\partial_{\lambda})^{k}\ln[1-(1-\mathrm{e}^{i\lambda})z] |_{\lambda = 0},
\ee
where $z=(y-x)/L$ is the rescaled length of the interval. In particular, using the expansion of the logarithm $\ln(1-a z) = -\sum_{p=1}^{\infty}p^{-1}a^{p}z^{p} $, one can rewrite the cumulants as
\be
\frac{V^{(k)}_{[x,y]}}{N} = \sum_{p=1}^{\infty} w^{(k)}_{p} z^{p},\quad w^{(k)}_{p} = -(-i)^{k}p^{-1}\partial_{\lambda}^{k}(1-\mathrm{e}^{i\lambda})^{p}|_{\lambda=0},
\ee 
where the coefficients $w^{(k)}_{p}$ can be evaluated using the binomial theorem, obtaining
\be
w^{(k)}_{p}  =  \frac{1}{p}\sum_{n=0}^{p}(-1)^{n+1}\frac{p!}{n!(p-n)!}n^{k},
\ee
which are zero for $p>k$. Plugging the last result into the expansion for the cumulants, we have
\bea
\frac{V^{(k)}_{[x,y]}}{N} & = & \sum_{p=1}^{\infty} (p-1)! \,z^{p} \sum_{n=0}^{p} \frac{(-1)^{n+1}\,n^{k}}{n!(p-n)!} \nonumber \\
 & = & \sum_{n=0}^{\infty} \frac{(-1)^{n+1}}{n!} n^{k} \sum_{p=n}^{\infty} \frac{(p-1)!}{(p-n)!} z^{p} - [(-1)^{n+1}n^{k-1}z^{n}]|_{n=0}\nonumber \\
 & = &  \sum_{n=0}^{\infty} \frac{(-1)^{n+1}}{n!} n^{k} z^{n} \sum_{q=0}^{\infty}\frac{(q+n-1)!}{q!}z^{q} +\delta_{k,1}\nonumber \\
&=& \delta_{k,1} + \sum_{n=0}^{\infty}(-1)^{n+1}n^{k-1} \left(\frac{z}{1-z}\right)^{n} = -\mathrm{Li}_{1-k}\left(\frac{z}{z-1}\right),
\label{Vext}
\eea
in terms of the polylogarithm functions $\mathrm{Li}_{k}(z)=\sum_{n=1}^{\infty}z^{n}/n^{k}$.

\begin{figure}[t]
\includegraphics[width=0.33\textwidth]{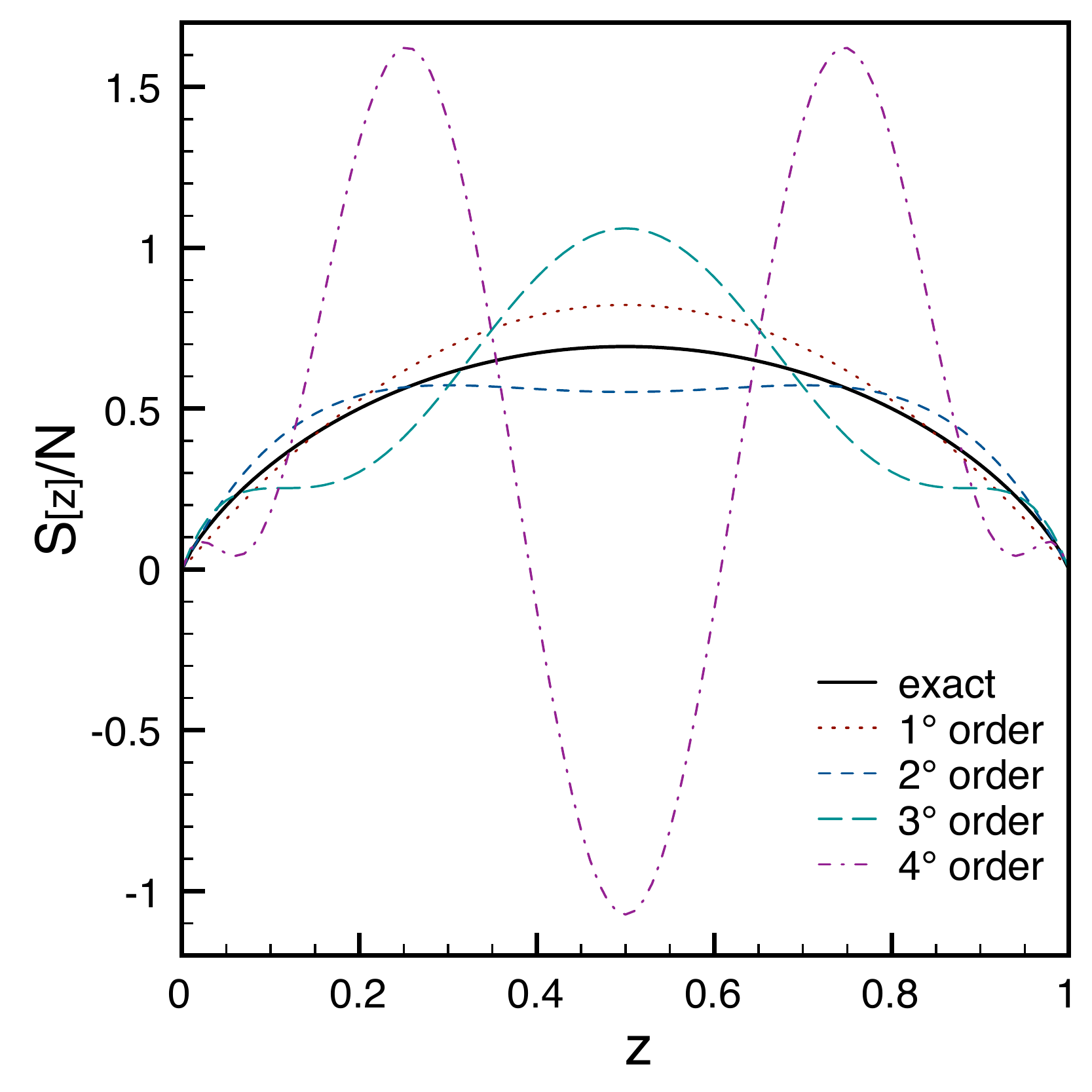}
\includegraphics[width=0.33\textwidth]{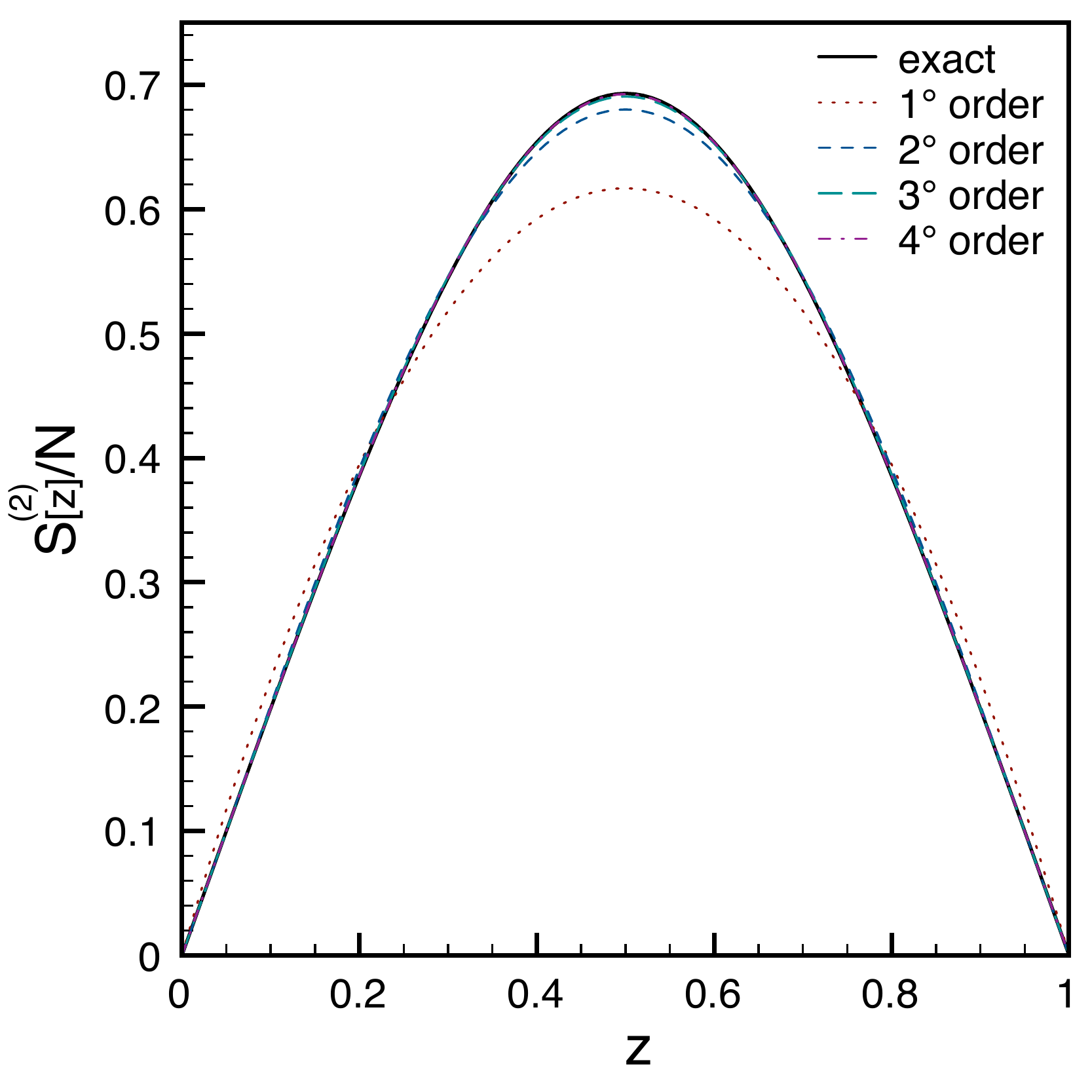}
\includegraphics[width=0.33\textwidth]{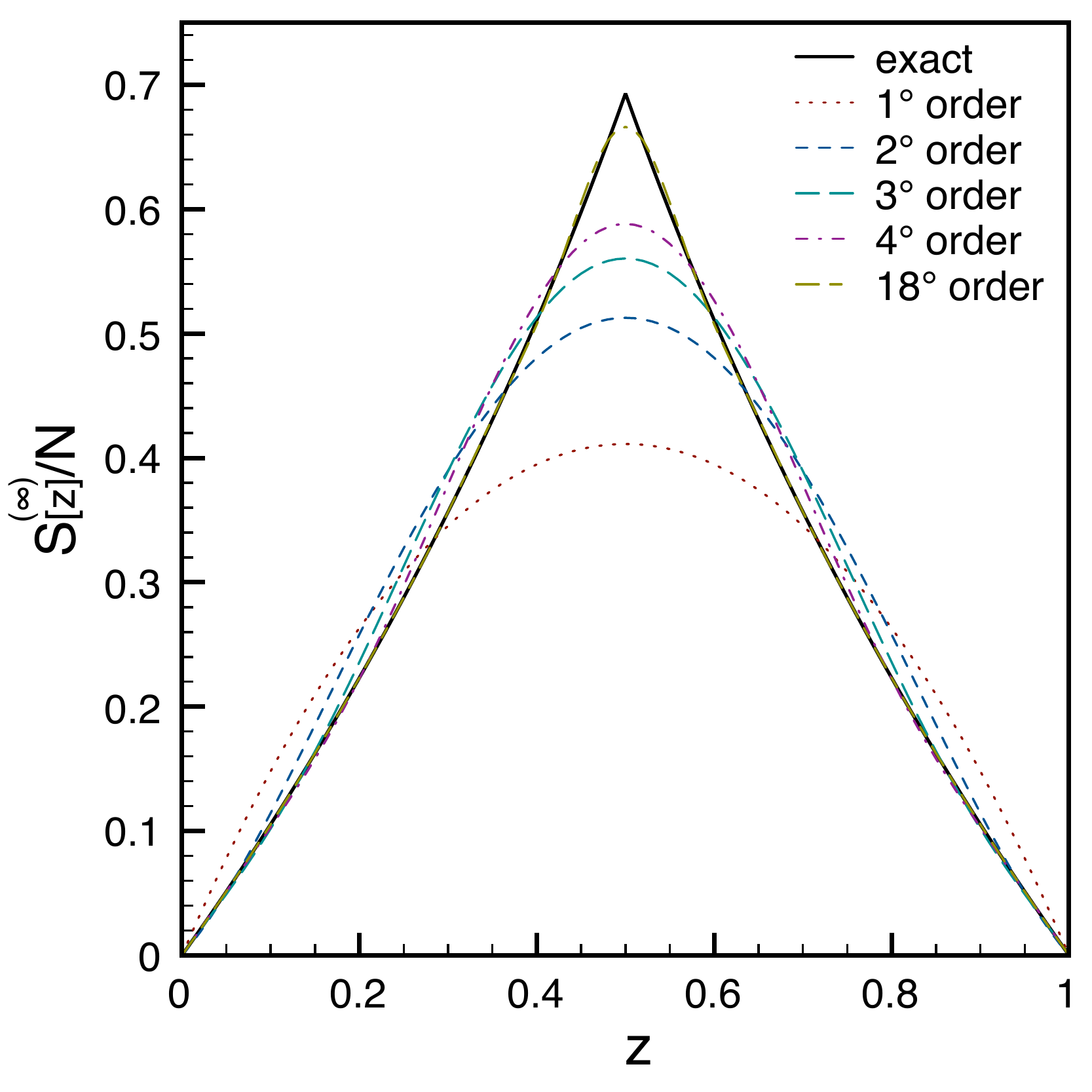}
\caption{Exact entanglement entropy profiles (full thick lines) in the stationary state for a subsystem of length $zL$
compared with the cumulant expansion given by Eq. (\ref{Salpha_cumulant}) and truncated up to a given finite order (dashed lines). 
While for $\alpha=2,\infty$ (and all other integer $\alpha$), the convergence of the cumulant expansion is very fast, 
for $\alpha=1$ (and all non-integer $\alpha$) the series is only asymptotic and the direct sum does not converge as clear
from the first panel.
}
\label{figS_cumulant}
\end{figure}

In Fig. \ref{figS_cumulant} we report the entanglement entropies $S^{(1)}_A, S^{(2)}_A$ and $S^{(\infty)}_A$ 
for a subsystem $A$ of length $zL$ ($0\leq z\leq 1$) embedded in the ring of length $L$.
The corresponding approximations given by the expansion (\ref{Salpha_cumulant}) calculated as a sum up to a finite order 
are also reported. Several comments are now in order. 
As opposed to ground-state results \cite{srh-10,srf-12,cmv-12,v-12b} and other non-equilibrium situations \cite{kl-09,v-12,nv-13}, 
all cumulants contribute to the leading behavior of the entanglement entropies and the expansion 
(\ref{Salpha_cumulant}) does not get effectively truncated at the second order. 
Indeed all even cumulants in Eq. (\ref{Vext}) are linear  in the particle number $N$ (i.e. extensive).
This also implies that the relation $S_A^{(\alpha)}= V_A^{(2)} (1+\alpha^{-1})\pi^2/6$ does {\it not} hold, as clear from the 
exact results.
Finally it is worth mentioning, as already noticed elsewhere \cite{cmv-11,cmv-12w}, that when all cumulants contribute to the 
expansion (\ref{Salpha_cumulant}), such series is well defined and convergent only for integer $\alpha>1$.
For all other values, and in particular the important one $\alpha=1$, the coefficients $s^{(\alpha)}_k$ grow 
too quickly with $k$ and the resulting series is only asymptotic and adequate resummation schemes should 
be used to extract quantitative information from it.
In fact, the left panel in Fig. \ref{figS_cumulant} shows how by adding more terms to the expansion (\ref{Salpha_cumulant})
for the von Neumann entropy, we have worse and worse results. 

\section{Trap to trap release}
\label{sec8}

So far we have considered the case of a gas release from a trap into a circle. We now consider the release from one trap into another larger one. The Hamiltonian before the quench $H_0$  is now given by (\ref{H_2}) with trap frequency $\omega_0$ and the Hamiltonian after the quench $H$ by the same equation but with frequency $\omega<\omega_0$. If we denote by $\hat\xi$ the operators that diagonalise the pre-quench Hamiltonian and by $\hat\zeta$ those that diagonalise the post-quench Hamiltonian, then from Eq. 
(\ref{psi-xi}) we find that the relation between them is
\begin{eqnarray}
\hat{\xi}_{i} & = & \int_{-\infty}^{+\infty} dx \, \chi^{*}_{i}(\omega_0;x) \hat{\Psi}(x) \nonumber \\
& = &  \sum_{j=0}^\infty \int_{-\infty}^{+\infty} dx \, \chi^{*}_{i}(\omega_0;x) \chi_{j}(\omega;x) \; \hat{\zeta}_j 
= \sum_{j=0}^\infty B_{i,j}(\omega_0,\omega) \hat{\zeta}_j ,
\end{eqnarray}
where we defined
\begin{equation} 
B_{i,j}(\omega_0,\omega) \equiv \int_{-\infty}^{+\infty} dx \, \chi^{*}_{i}(\omega_0;x) \chi_{j}(\omega;x) = \langle \chi_i(\omega_0) | \chi_j(\omega) \rangle.
\end{equation}
Note that $\chi(\omega_0;x)=({\omega/\omega_0})^{\frac14}\chi_j(\omega ; x\sqrt{\omega/\omega_0})$. Even though we will not need the explicit expressions for the overlaps $B_{i,j}$ since, as we will see below, the evolution of the system's wavefunction can be derived in a simpler way, we report their value for completeness
\begin{eqnarray}\fl 
B_{i,j}(\omega_0,\omega) & = & \frac{1}{\sqrt{2^{i}i!}} \frac{1}{\sqrt{2^{j}j!}}\left(\frac{ab}{\pi}\right)^{\frac{1}{2}} 
\Gamma\left(\frac{i+j+1}{2}\right)2^{i+j}(ab)^{i}(a^{2}-1)^{(i-j)/2} \nonumber \\
& \times & \, _{2}F_{1}\left[ -\frac{j}{2};\frac{1-j}{2}; \frac{1-i-j}{2}; \frac{(\omega_0+\omega)^2}{4 \omega_0 \omega}\right],
\end{eqnarray}
for $i+j$ even and zero otherwise. In the last equation $a=\sqrt{{2\omega}/({\omega_{0}+\omega})}$, $b=\sqrt{{2\omega_{0}}/({\omega_{0}+\omega})}$ and $_{2}F_{1}$ is the hypergeometric function. 

The inverse relation is simply given by interchange of the $\omega_0$ and $\omega$
\begin{equation}
\hat{\zeta}_{i} = \sum_{j=0}^\infty B_{i,j}(\omega,\omega_0) \hat{\xi}_j ,
\end{equation}
Unlike (\ref{mapeta}) the above inverse relation is exact without any further assumptions. 
The initial state is given, as in the previous case, by (\ref{GS0}).

We are now able to calculate the time evolution of physical observables, in particular of the two-point fermionic correlation 
function $C(x,y;t)=\langle\hat{\Psi}^{\dag}(x)\hat{\Psi}(y)\rangle_{t}$ from which, as we have seen, all other local observables 
can be derived 
\begin{eqnarray}\label{C_F2}
C(x,y;t) & = & \langle\Psi_{0}|\mathrm{e}^{iHt}\hat{\Psi}^{\dag}(x)\hat{\Psi}(y)\mathrm{e}^{-iHt}|\Psi_{0}\rangle
 =\sum_{j=0}^{N-1} \phi^{*}_{j}(x,t) \phi_{j}(y,t),
\end{eqnarray}
where $\phi_{j}(x,t)$ are, as before, the time evolved one-particle eigenfunctions
\begin{equation}\label{one_particle_01}
\phi_{j}(x,t) = \sum_{n=0}^{\infty}B_{n,j}(\omega,\omega_0) \chi_{n}(\omega;x)\mathrm{e}^{-i \omega \left(n+\frac12\right) t },
\end{equation}
i.e. the solution of the single particle Schr\"odinger equation
\begin{equation}
i\frac{\partial \phi_{j}(x,t) } {\partial t} = -\frac12 \frac{\partial^{2} \phi_{j}(x,t) }{\partial x^2} + \frac12 \omega^2 x^2 \phi_{j}(x,t) \label{Schr}
\end{equation}
with $\phi_{j}(x,0)=\chi_{j}(\omega_0;x)$. 
This differential equation belongs to a class of problems (harmonic oscillator with time dependent frequency 
\cite{mg-05,grit-pol}
that can be solved elegantly by means of a scaling ansatz
\begin{equation}\label{ansatz}
\phi_{j}(x,t) = a(t)  \phi_j\left( \frac{x}{b(t)},0 \right)  \exp\left[ \frac12 i\Omega(t) {x^2} + i \varphi(t)\right],
\end{equation}
where the functions $a(t),b(t)$ and $\varphi(t)$ are assumed real. 
By substituting the ansatz into (\ref{Schr}) and using the time-independent Schr\"odinger equation satisfied by $\phi_{j}(x,0)$ 
\begin{equation}
E_{0j}  \phi_{j}(x,0) = -\frac12 \frac{\partial^{2} \phi_{j}(x,0) }{\partial x^2} + \frac12 \omega_0^2 x^2 \phi_{j}(x,0)
\end{equation}
(with $E_{0j}=\omega_0 (j+1/2)$) 
we see that the ansatz (\ref{ansatz}) is correct, if we choose the auxiliary functions to satisfy the equations
\begin{eqnarray}
& \ddot b(t) + \omega^2 b(t)  = {\omega_0^2}/{b^3(t)} \label{b_eq} , \quad b(0)=1,\quad \dot b(0) = 0, \label{scaling} \\
& \Omega(t)  = {\dot b(t)}/{b(t)},\\
& a(t) = 1/\sqrt{b(t)}, \\
& \varphi(t) = - E_{0j} \int^t_0 dt/b^2(t).
\end{eqnarray}
With these definitions the evolved wave functions are given by
\begin{equation}
\phi_{j}(x,t) = \frac1{\sqrt{b(t)}} \; \phi_j\left( \frac x{b(t)},0 \right) \exp\left [ \frac12 i \frac{\dot b(t)}{b(t)} {x^2} - i E_{0j} \int^t_0 dt/b^2(t) \right],
\end{equation}
where the solution to the differential equation (\ref{scaling}) for $b(t)$ is
\begin{equation}
b(t) = \sqrt{1+(\omega_0^2-\omega^2) \sin^2\omega t / \omega^2} .
\end{equation}

Going back to the two-point fermionic correlation function, we obtain
\begin{eqnarray}\label{C_F2b}
C(x,y;t) & = \frac1{b(t)} \exp\left [ \frac12 i \frac{\dot b(t)}{b(t)} {(y^2 - x^2)} \right]  \sum_{j=0}^{N-1} \phi^{*}_{j}\left(\frac{x}{b(t)},0\right)  \phi_{j}\left(\frac{y}{b(t)},0\right) \nonumber \\
& = \frac1{b(t)} \exp\left [ \frac12 i \frac{\dot b(t)}{b(t)} {\left(y^2 - x^2\right)} \right] C\left(\frac{x}{b(t)},\frac{y}{b(t)};0\right)
\end{eqnarray}
In particular, for the particle density $n(x;t)$, i.e. the diagonal part of the fermionic correlation function, we obtain the simple expression
\begin{equation}\label{n2}
n(x;t) = \frac1{b(t)}  n\left(\frac{x}{b(t)};0\right),
\end{equation}
that is, the density profile is simply given by its initial form (the Wigner semicircle), rescaled periodically in time, 
a phenomenon known as ``breathing''. Similarly, the density-density correlation function is given 
by
\begin{equation}
G(x,y;t) 
= \frac1{b(t)^2} \; G\left(\frac{x}{b(t)},\frac{y}{b(t)};0\right) .
\end{equation}
Note that the evolution of the system is periodic with period $\pi/\omega$.

At this point we realise that there is a crucial difference in comparison with the case of trap-to-circle release and in fact with most typical quantum quench problems studied in the literature. In the previous case the evolution was characterised by two different time scales, well separated in the thermodynamic limit: the time needed in order for the particles to travel around the circumference of the circle (so that they overlap and interfere among each other leading to dephasing) which scales like $\sim L$, and the revival time determined by the inverse of the fundamental frequency of the evolution, which scales like $\sim L^2$. This fact allowed us to consider the regime of intermediate times where equilibration takes place, i.e. times such that $t/L\to\infty$ but $t/L^2\to0$. In the present case instead, there is only one characteristic time scale determined by $\omega$, meaning that both dephasing and revival have the same frequency. 
Therefore there is no regime in which the first effect can lead to equilibration before the second takes place. 
However a \emph{weak} version of the GGE conjecture may still be applicable, if it refers to the \emph{long time averages} of the expectation values of observables \cite{bch-11}, instead of the values themselves, which exhibit persistent oscillations. 

The GGE density matrix in this case can be written using the post-quench occupation number operators $\hat{n}_{i}= \hat{\zeta}^{\dag}_{i} \hat{\zeta}_{i}$ as conserved charges 
\begin{equation}
\hat{\rho}_{GGE} =\frac{1}{Z}\exp \left( -\sum_{i=0}^{\infty} \lambda_{i}\hat{n}_{i} \right ),
\end{equation}
with $Z=\mathrm{Tr}\left[ \exp\left(-\sum_{i} \lambda_{i}  \hat{n}_{i}\right) \right]$ and where the Lagrange multipliers 
are defined through the equations
\begin{equation}\label{cond2}
\mathrm{Tr}[\hat{n}_{i} \hat{\rho}_{GGE}] = \langle\Psi_{0}|\hat{n}_{i}|\Psi_{0}\rangle,
\end{equation}
which can be easily evaluated, giving
\begin{equation}
\frac{1}{1+\mathrm{e}^{\lambda_{i}}} = \sum_{k=0}^{N-1} B^{*}_{i,k}(\omega,\omega_0) B_{i,k}(\omega,\omega_0).
\end{equation}
We stress that these charges are clearly non-local and we do not know whether can be written in terms of linear 
combination of local charges as in the trap-to-circle quench. We use the above definition just as a prescription 
and check its correctness for two basic observables.

\begin{figure}[t]
\center\includegraphics[width=0.8\textwidth]{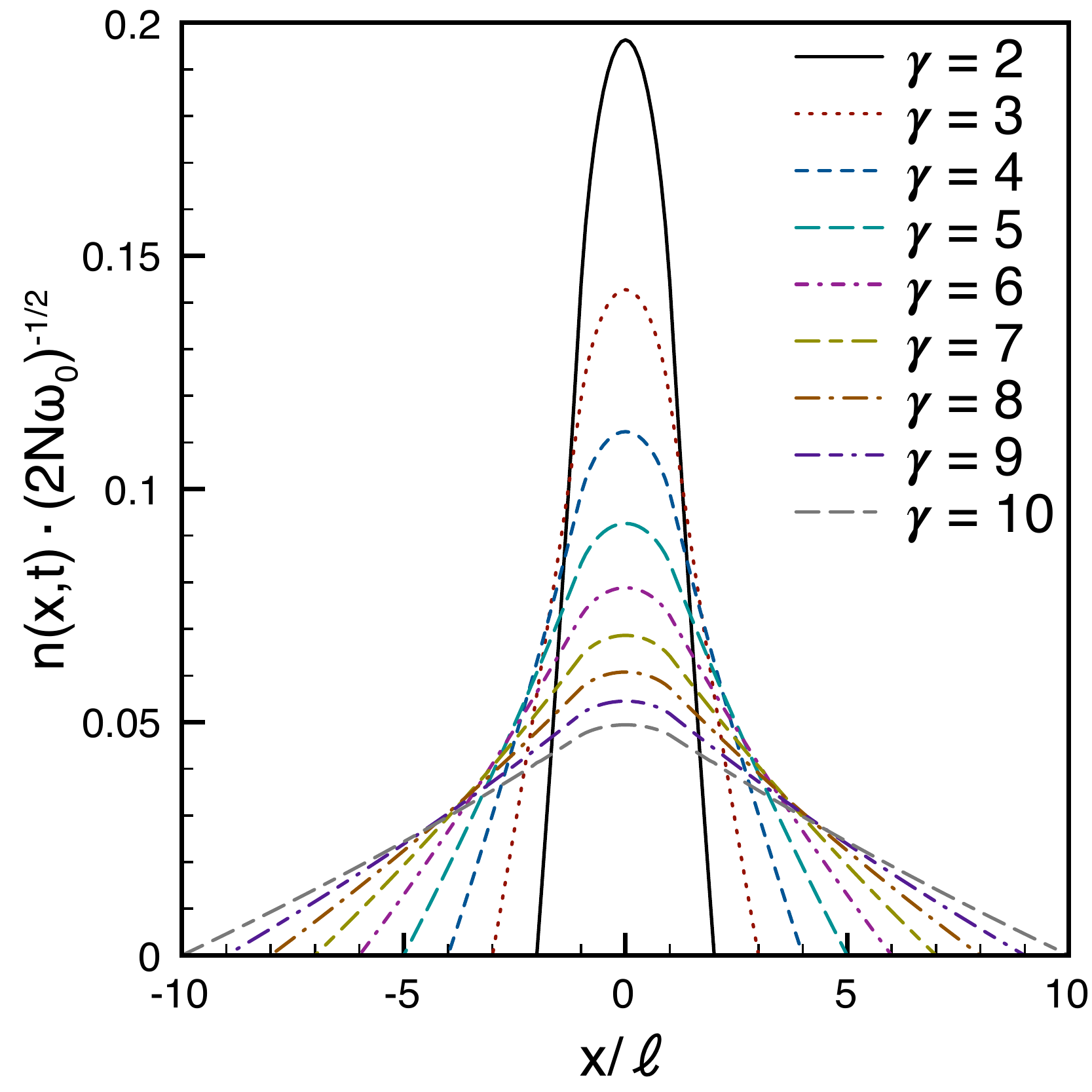}
\caption{Time averaged density profile for the trap-to-trap release for several values of the quench parameters 
parametrized as $\gamma=\omega_0/\omega$ as a function of $x/\ell$, where $\ell$ is the radius of the initial  
distribution.}
\label{rho_ta}
\end{figure}

It is straightforward to show that the time averaged observables are in agreement with the GGE. In fact one can show this using \emph{solely} the general properties of the model (essentially based on its non-interacting nature) rather than specific details. We will demonstrate this for the two point correlation function. The GGE prediction is given by
\begin{eqnarray}\label{Cgge2}\fl 
C_{GGE}(x,y) & = & \mathrm{Tr}[\hat{\Psi}^{\dag}(x)\hat{\Psi}(y)\hat{\rho}_{GGE} ] 
=\sum_{i,j=0}^{\infty} \chi^{*}_{i}(\omega;x)\chi_{j}(\omega;y) \mathrm{Tr}[\hat{\zeta}^{\dag}_i \hat{\zeta}_j \hat{\rho}_{GGE} ] 
\\\fl 
& = & \sum_{i,j=0}^{\infty} \chi^{*}_{i}(\omega;x)\chi_{j}(\omega;y) \delta_{ij} \mathrm{Tr}[\hat{n}_i  \hat{\rho}_{GGE} ] 
=\sum_{i=0}^{\infty} \chi^{*}_{i}(\omega;x)\chi_{i}(\omega;y)  \mathrm{Tr}[\hat{n}_i  \hat{\rho}_{GGE} ],\nonumber 
\end{eqnarray}
On the other hand, the exact time evolved expression is
\begin{eqnarray}\label{Cex2}\fl 
C(x,y;t) & = & \langle\Psi_{0}|\mathrm{e}^{iHt}\hat{\Psi}^{\dag}(x)\hat{\Psi}(y)\mathrm{e}^{-iHt}|\Psi_{0}\rangle
=\sum_{i,j=0}^{\infty} \chi^{*}_{i}(\omega;x)\chi_{j}(\omega;y) \langle\Psi_{0}|\mathrm{e}^{iHt}\hat{\zeta}^{\dag}_{i}\hat{\zeta}_{j}\mathrm{e}^{-iHt}|\Psi_{0}\rangle \nonumber \\ \fl 
& = &\sum_{i,j=0}^{\infty} \chi^{*}_{i}(\omega;x)\chi_{j}(\omega;y) \mathrm{e}^{i\omega(i-j)t} \langle\Psi_{0}|\hat{\zeta}^{\dag}_{i}\hat{\zeta}_{j}|\Psi_{0}\rangle,
\end{eqnarray}
and after time averaging, in which case only the diagonal terms of the sum survive,
\begin{eqnarray}\label{Cta2}
\overline{C(x,y;t)} & = &\sum_{i=0}^{\infty} \chi^{*}_{i}(\omega;x)\chi_{i}(\omega;y) \langle\Psi_{0}|\hat{n}_{i}|\Psi_{0}\rangle.
\end{eqnarray}
Now it is obvious that the two expressions (\ref{Cgge2}) and (\ref{Cta2}) are identical due to (\ref{cond2}). Since this holds for the two point correlation function, it is also true for any other observable that is a linear combination of values of the two point function. 
The key point here was that the field operators $\hat\Psi(x)$ are expressed as a linear combination of the diagonalization operators $\hat\zeta_i, \hat\zeta_i^\dagger$. Therefore the correlation function is just a linear combination of the expectation values of the conserved charges and the equality of its GGE value with its time averaged exact expression is a direct consequence of the defining condition of the GGE. 
For observables that are algebraic combinations of the two point function, one should check whether the time averaging commutes with the algebraic combination. In particular for the density-density correlation function $G(x,y;t)$ satisfying
\begin{equation}
G(x,y;t) = n(x,t)n(y,t) + n(x;t)\delta(x-y) -|C(x,y;t)|^2 ,
\end{equation}
even though $\overline{n(x,t)n(y,t)} \neq \overline{n(x,t)}\;\overline{n(y,t)}$ and $\overline{C(x,y;t)C^*(x,y;t)} \neq \overline{C(x,y;t)} \; \overline{C^*(x,y;t)}$, we can readily show that $\overline{n(x,t)n(y,t)} - \overline{C(x,y;t)C^*(x,y;t)} =\overline{n(x,t)}\;\overline{n(y,t)} - \overline{C(x,y;t)} \; \overline{C^*(x,y;t)}$ and therefore 
\begin{equation}
\overline{G(x,y;t)} = G_{GGE}(x,y;t) ,
\end{equation}
i.e. the time averaged density-density correlation function is also correctly predicted by the GGE.

Fig.~\ref{rho_ta} shows numerical plots of the time averaged density profile $\overline{n(x;t)}$ as a function of $x$ for several values of 
$\gamma\equiv\omega_0/\omega$. This is given by the following expression
\begin{equation}\label{n3}
\overline{n(x;t)} = \frac\omega\pi \int_0^{\pi/\omega} dt  \frac1{b(t)}  n\left(\frac{x}{b(t)};0\right) = 
\frac{2\omega}\pi \int_1^{\omega_0/\omega} db \; \frac1{\dot b(b) \, b}  n\left(\frac{x}{b};0\right) .
\end{equation}

\section{Summary and Discussion}
\label{concl}

In this paper we considered the non equilibrium dynamics of a gas of impenetrable bosons 
released from a harmonic trap to a periodic ring of length $L$ as sketched in Fig. \ref{sketch}. 
The results could be summarized as following.
\begin{itemize}
\item The TD limit should be handled with care. One should consider $N,L\to\infty$ with $n=N/L$ constant but also 
the initial trap frequency $\omega\to0$ with $\omega N$ constant, analogously to what done in Ref. \cite{ck-12}.
Among the other things this also implies that we have a finite average initial density $n_0$ and 
a finite expansion velocity $v=\sqrt{2\omega N}$.
\item Also the large time limit should be handled with care: we require $t\to\infty$, i.e $vt\gg L$, but $t/L^2\to 0$
in order to avoid the effect of the revivals.
\item The calculation significantly simplifies by physically assuming that the initial extension of the gas 
$2\ell=2\sqrt{2N/\omega}$ is smaller that the length of the ring $L$. 
This is equivalent to the requirement $n_0>n$.
Under this condition, the gas initially does not feel the presence of the periodic boundary conditions. 
\end{itemize}
While all these results were explicitly checked only for the specific model, they are expected to remain valid 
for any trap-to-ring release experiment, for integrable/non-integrable model.
Specifically for impenetrable bosons, we have shown the following results.
\begin{itemize}
\item We exactly calculate the time-dependent  fermionic two-point  correlation function (and hence, 
at equal points, the density).
We have shown that for large times it converges to a stationary value (equal to its time average). 
The approach to the stationary value is power-law of the form $t^{-3/2}$.
For intermediate times, the dynamics is much more complicated and it is described in Sec. \ref{sec3}.

\item We prove that for long time and in the TD limit,  any  subsystem becomes stationary and its behavior 
is described by a GGE. This provides the first analytic proof of a GGE for an inhomogeneous initial state. 
The GGE built with the fermionic momentum occupation and with the local integrals of motion are proved to be 
equivalent and all Lagrange multipliers are explicitly calculated. 
We also compared the GGE with the canonical and the grand canonical ensembles. 

\item The density-density correlation is analytically worked out in the large time limit.

\item In the stationary state, the bosonic two-point correlation function turns out to decay exponentially with the distance, 
contrarily to the fermionic one which is a power-law. 
We also find a very peculiar short distance behavior giving rise to a $k^{-2}$ tail in the bosonic momentum distribution
which crosses over to standard $k^{-4}$ only for very large momenta. 

\item We calculate the entanglement entropies of a compact subsystem which in the infinite time limit turned out to 
assume an extremely simple form. The finite time results are however rather complicated as reported in Sec. \ref{sec7}.

\item Finally we also considered the trap-to-trap release in which local observables oscillate forever. 
However, the time-average expectations of several calculated observables are still described by a proper GGE.

\end{itemize}

There are a few points which need some discussion.
We found that the mechanism responsible for the equilibration is the interference of the particles going around the circle 
many times, requiring $vt\gg L$ in order to observe a stationary behavior. 
This is very different from the one in a standard global quantum quench where instead 
the opposite requirement $vt\ll L$ should be satisfied to avoid revival effects.
While we have shown this equilibration mechanism only for a gas of impenetrable bosons, it is 
natural to expect the same for any one dimensional gas released into a circle, independently from the
fact that the stationary state is GGE or thermal (for integrable and non-integrable models respectively). 
It would be  interesting to check this statement for the time evolution of truly interacting models
such as the Lieb-Liniger or Gaudin-Yang fermionic gases, on the lines of Refs. \cite{ck-12,bck-13}, or 
by means of purely numerical methods.
Furthermore, the results derived here are also the starting point for the analytic study of the trap release dynamics
of the Lieb-Liniger model (\ref{HLL}) in a $1/c$ expansion.  

Finally there are several easy generalizations that are worth investigating 
such as the dynamics from different inhomogeneous initial states 
(e.g. due to non-harmonic trapping potentials) and the effect of different boundary 
conditions in the post-quench Hamiltonians (such as Dirichlet ones).

\section*{Acknowledgments} 
We are extremely grateful to Marton Kormos for very useful discussions.  
All authors  acknowledge the ERC  for financial  support under  Starting Grant 279391 EDEQS. 


\appendix

\section{From Fourier transform to Fourier series}\label{appA}
Let us suppose that we need to evaluate a function $f(x):[-L/2,L/2]\to\mathbb{C}$ which is given by the Fourier series
\begin{equation}
f(x) = \frac{1}{L}\sum_{m=-\infty}^{\infty}g(k_{m})\mathrm{e}^{-ik_{m}x},\quad k_{m}=\frac{2\pi m}{L},
\end{equation}
and we only know the Fourier transform of $g(k)$
\begin{equation}
\tilde{f}(x) \equiv \frac{1}{2\pi}\int_{-\infty}^{\infty}dk\,g(k)\mathrm{e}^{-ikx}\quad
\Leftrightarrow\quad
g(k) = \int_{-\infty}^{\infty} dx\, \tilde{f}(x)\mathrm{e}^{ikx}.
\end{equation}
This gives for the original series 
\begin{equation}\label{fourier}
f(x) = \frac{1}{L}\sum_{m=-\infty}^{\infty}\int_{-\infty}^{\infty} dy\, \tilde{f}(y)\mathrm{e}^{ik_{m}(y-x)} =  \sum_{j=-\infty}^{\infty}\tilde{f}(x+jL),
\end{equation}
where we used the Fourier series representation of the Dirac $\delta$-comb
\begin{equation}
\Delta_{L}(x)\equiv \sum_{j=-\infty}^{\infty}\delta(x-jL) = \frac{1}{L}\sum_{m=-\infty}^{\infty}\mathrm{e}^{2\pi i m x /L}.
\end{equation}
In simpler words, this tells us that whenever we know the solution of a Fourier problem defined in the whole 
space $[-\infty,\infty]$, the solution of the same problem restricted to a finite interval of length $L$ 
is given by the sum of the infinite-space solution shifted in space by integer multiples of $L$.

Moreover, we can use Eq. (\ref{fourier}) to evaluate a sum of a series in terms of an integral plus corrections. Indeed, let $F(\{x_{i}\},\delta p)$ be a function of a set of continuum variables $\{x_{i}\}$ and a discrete variable $\delta p$ evaluated in a set of point ($p\in\mathbb{N}$), then if $\delta\rightarrow 0$, one has
\begin{eqnarray}\label{sumtoint}
\fl \lim_{\delta\to0}\,\delta \sum_{p=-\infty}^{\infty}F(\{x_{i}\},\delta p)  =  \int_{-\infty}^{\infty} dz\, F(\{x_{i}\},z) \\ \fl\qquad\qquad
 +   2 \int_{-\infty}^{\infty} dz\, F(\{x_{i}\},z) \cos\left(\frac{2\pi z}{\delta}\right) 
 +   2 \int_{-\infty}^{\infty} dz\, F(\{x_{i}\},z) \cos\left(\frac{4\pi z}{\delta}\right)  \; + \;  \ldots \nonumber ,
\end{eqnarray}
where the integrals weighted over the $\cos$ functions are vanishing for $\delta\rightarrow 0$ due to the strong oscillations and give the next corrections to the leading term.


\Bibliography{99}
\addcontentsline{toc}{section}{References}


\bibitem{uc}
M.~Greiner, O.~Mandel, T.~W.~H\"ansch, and I.~Bloch,
Nature {\bf 419} 51 (2002).


\bibitem{kww-06}
T. Kinoshita, T. Wenger,  D. S. Weiss, 
 Nature {\bf 440}, 900 (2006).

\bibitem{tc-07}
S. Hofferberth, I. Lesanovsky, B. Fischer, T. Schumm, and J. Schmiedmayer,
Nature {\bf 449}, 324 (2007).

\bibitem{hgm-09}
E. Haller, M. Gustavsson, M.J. Mark, J.G. Danzl, R. Hart, G. Pupillo, and H.-C. Naegerl, Science {\bf 325}, 1224 (2009).

\bibitem{tetal-11}
S. Trotzky, Y.-A. Chen, A. Flesch, I. P. McCulloch, U. Schollw\"ock,
J. Eisert, and I. Bloch, 
Nature Phys. {\bf 8}, 325 (2012). 

\bibitem{cetal-12}
M. Cheneau, P. Barmettler, D. Poletti, M. Endres, P. Schauss, T. Fukuhara, C. Gross, I. Bloch, C. Kollath, and S. Kuhr,
Nature {\bf 481}, 484 (2012).

\bibitem{getal-11}
M. Gring, M. Kuhnert, T. Langen, T. Kitagawa, B. Rauer, M. Schreitl, I. Mazets, D. A. Smith, E. Demler, and J. Schmiedmayer,
Science {\bf 337}, 1318 (2012).


\bibitem{shr-12}
U. Schneider, L. Hackerm\"uller, J. P. Ronzheimer, S. Will, S. Braun, T. Best, I. Bloch, E. Demler, S. Mandt, D. Rasch, and A. Rosch,
Nature Phys. {\bf 8}, 213 (2012).


\bibitem{rsb-13}
J. P. Ronzheimer, M. Schreiber, S. Braun, S. S. Hodgman, S. Langer, I. P. McCulloch, F. Heidrich-Meisner, I. Bloch, and U. Schneider,
Phys. Rev. Lett. {\bf 110}, 205301 (2013) 

\bibitem{mmk-13}
F. Meinert, M. J. Mark, E. Kirilov, K. Lauber, P. Weinmann, A. J. Daley, H.-C. N\"agerl
 arXiv:1304.2628. 


\bibitem{fse-13}
T. Fukuhara, P. Schauss, M. Endres, S. Hild, M. Cheneau, I. Bloch, and C. Gross,
 arXiv:1305.6598
 
\bibitem{revq}
A. Polkovnikov, K. Sengupta, A. Silva, and M. Vengalattore, 
Rev. Mod. Phys. {\bf 83}, 863 (2011).


\bibitem{cc-06} P. Calabrese and  J. Cardy, 
Phys. Rev. Lett. {\bf 96}, 136801 (2006).

\bibitem{cc-06b} 
 P. Calabrese and  J. Cardy,  
J. Stat. Mech. P06008  (2007).


\bibitem{c-06}
M. A. Cazalilla, Phys. Rev. Lett. {\bf 97}, 156403 (2006).


\bibitem{ir-00}
F. Igloi and H. Rieger,  
Phys. Rev. Lett. {\bf 85}, 3233 (2000);\\
K. Sengupta, S. Powell, S. Sachdev,  
Phys. Rev. A {\bf 69} 053616 (2004). 

\bibitem{dmcf-06}
G. De Chiara, S. Montangero, P. Calabrese, and R. Fazio, 
J. Stat. Mech. L03001 (2006).

\bibitem{lk-08}
A. Laeuchli and C. Kollath, J. Stat. Mech. (2008) P05018.

\bibitem{ic-09}
A. Iucci, and M. A. Cazalilla, Phys. Rev. A {\bf 80}, 063619 (2009); \\
A. Iucci, and M. A. Cazalilla, New J. Phys. {\bf 12}, 055019 (2010).

\bibitem{fcc-09}
A. Faribault, P. Calabrese, and J.-S. Caux, J. Stat. Mech. P03018 (2009);\\
A. Faribault, P. Calabrese, and J.-S. Caux, J. Math. Phys. {\bf 50}, 095212 (2009).


\bibitem{ir-10}
F. Igloi and H. Rieger,   
Phys. Rev. Lett. {\bf 106}, 035701 (2011).

\bibitem{sc-10}
S. Sotiriadis and J. Cardy,
Phys. Rev. B {\bf 81}, 134305 (2010).

\bibitem{bpk-12}
P. Barmettler, D. Poletti, M. Cheneau, and C. Kollath,
Phys. Rev. A {\bf 85}, 053625 (2012).

\bibitem{CEF}
P. Calabrese, F. H. L. Essler and M. Fagotti, 
Phys. Rev. Lett. {\bf 106}, 227203 (2011).  

\bibitem{CEFI}
P. Calabrese, F. H. L. Essler and M. Fagotti,  J. Stat. Mech. (2012) P07016.


\bibitem{gcg-11} 
L. Foini, L. F. Cugliandolo, and A. Gambassi, Phys. Rev. B {\bf 84}, 212404 (2011); \\
L. Foini, L. F. Cugliandolo, and A. Gambassi, J. Stat. Mech. (2012) P09011.

\bibitem{ri-11}
H. Rieger and F. Igl\'oi,
Phys. Rev. B {\bf 84}, 165117 (2011);\\
B. Blass, H. Rieger, and F. Igl\'oi, EPL {\bf 99}, 30004 (2012);\\
S. Evangelisti, J. Stat. Mech. (2013) P04003. 

\bibitem{ors-12}
C. Karrasch, J. Rentrop, D. Schuricht, and V. Meden, Phys. Rev. Lett. {\bf 109}, 126406 (2012);\\
 J. Rentrop, D. Schuricht, and V. Meden, New J. Phys. {\bf 14}, 075001 (2012). 


\bibitem{se-12}
D. Schuricht and F. H. L. Essler, J. Stat. Mech. (2012) P04017.



\bibitem{gg} 
M. Rigol, V. Dunjko, V. Yurovsky,  and M. Olshanii,
Phys. Rev. Lett. {\bf 98}, 50405 (2007).


\bibitem{gg2}
M. Rigol, V. Dunjko,  and M. Olshanii,
Nature {\bf 452}, 854 (2008). 




\bibitem{msnm-07} 
S.R. Manmana, S. Wessel, R.M. Noack, and A. Muramatsu, Phys. Rev. Lett. {\bf 98}, 210405 (2007).
 

\bibitem{wk-07}
 M. Eckstein and M. Kollar, Phys. Rev. Lett. {\bf 100}, 120404 (2007); \\
 M. Kollar and M. Eckstein, Phys. Rev. A {\bf 78}, 013626 (2008); \\
 M. Kollar, F.A. Wolf, and M. Eckstein, Phys. Rev. B {\bf 84}, 054304 (2011).

\bibitem{cdeo-08}
M. Cramer, C. M. Dawson, J. Eisert, and T. J. Osborne, 
Phys. Rev. Lett. {\bf 100}, 030602 (2008);\\
M. Cramer and J. Eisert,
New J. Phys. 12, 055020 (2010).

\bibitem{bs-08}
T. Barthel and U. Schollw\"ock, 
Phys. Rev. Lett. {\bf 100}, 100601 (2008).

\bibitem{scc-09}
S. Sotiriadis, P. Calabrese,  and J. Cardy, 
EPL {\bf 87}, 20002, (2009).


\bibitem{r-09}
M. Rigol, Phys. Rev. Lett. {\bf 103}, 100403 (2009), Phys. Rev. A {\bf 80}, 053607 (2009)

\bibitem{dams-09}
 D. Rossini, A. Silva, G. Mussardo, and G. E. Santoro, Phys. Rev. Lett. {\bf 102}, 127204 (2009); \\
 D. Rossini, S. Suzuki, G. Mussardo, G. E. Santoro, and A. Silva, Phys. Rev. B {\bf 82}, 144302 (2010).

\bibitem{CEFII}
P. Calabrese, F. H. L. Essler and M. Fagotti, 
J. Stat. Mech. (2012) P07022.

\bibitem{f-13}
M. Fagotti,  Rev. B {\bf 87}, 165106 (2013).


\bibitem{eef-12}
F. H. L. Essler, S. Evangelisti, and M. Fagotti,
Phys. Rev. Lett. {\bf 109}, 247206 (2012).


\bibitem{ccss-11}
T. Caneva, E. Canovi, D. Rossini, G. E. Santoro, and A. Silva,  J. Stat. Mech. (2011) P07015.

\bibitem{rs-12}
M. Rigol and M. Srednicki, Phys. Rev. Lett. {\bf 108}, 110601 (2012).

\bibitem{bkl-10}
G. Roux,  Phys. Rev. A {\bf 79}, 021608 (2009);\\
G. Biroli, C. Kollath, and A. Laeuchli,
Phys. Rev. Lett. {\bf 105}, 250401 (2010).

\bibitem{bdkm-11}
G. P. Brandino, A. De Luca, R.M. Konik, and G. Mussardo, 
Phys. Rev. B {\bf 85}, 214435 (2012).


\bibitem{fm-10}
D. Fioretto and G. Mussardo,
New J. Phys. {\bf 12}, 055015 (2010).



\bibitem{can}
E. Canovi, D. Rossini, R. Fazio, G. E. Santoro, and A. Silva, 
Phys. Rev. B {\bf 83}, 094431 (2011).


 \bibitem{cic-12}
 M. A. Cazalilla, A. Iucci, and M.-C. Chung, Phys. Rev. E {\bf 85}, 011133 (2012).

\bibitem{sfm-12}
S. Sotiriadis, D. Fioretto, and G. Mussardo,
J. Stat. Mech. (2012) P02017.

\bibitem{mc-12} 
J. Mossel and J.-S. Caux, J. Phys. A {\bf 45}, 255001 (2012);\\
  E. Demler and A. M. Tsvelik, Phys. Rev. B {\bf 86}, 115448 (2012). 
  

\bibitem{ms-12}
J. Marino and A. Silva, Phys. Rev. B {\bf 86}, 060408 (2012).  

\bibitem{o-12}
M. Olshanii, arXiv:1208.0582. 


\bibitem{srgs-13}
N. Sedlmayr, J. Ren, F. Gebhard, and J. Sirker,
Phys. Rev. Lett. {\bf 110}, 100406 (2013);\\
J. Sirker, N.P. Konstantinidis, N. Sedlmayr; arXiv:1303.3064. 

\bibitem{ce-13}
J.-S. Caux and F. H.L. Essler, arXiv:1301.3806.

\bibitem{fe-13}
M.  Fagotti and  F. H.L. Essler, arXiv:1302.6944. 

\bibitem{m-13}
G. Mussardo, arxiv:1304.7599.

\bibitem{p-13}
B. Pozsgay, arxiv:1304.5374. 

\bibitem{fe-13b}
M. Fagotti and F.H.L. Essler, arxiv:1305.0468.



\bibitem{kla-07}
C. Kollath, A. Laeuchli, and E. Altman, Phys. Rev. Lett. {\bf 98}, 180601 (2007). 

\bibitem{rf-11}
A.C. Cassidy, C.W. Clark, and M. Rigol, Phys. Rev. Lett. {\bf 106}, 140405 (2011);\\ 
M. Rigol and M. Fitzpatrick, Phys. Rev. A {\bf 84}, 033640 (2011);\\
K. He and M. Rigol, Phys. Rev. A {\bf 85}, 063609 (2012);\\
 C. Gramsch and M. Rigol, Phys. Rev. A {\bf 86}, 053615 (2012).

\bibitem{bch-11}
M. C. Banuls, J. I. Cirac, and M. B. Hastings,
Phys. Rev. Lett. {\bf 106}, 050405 (2011). 

\bibitem{gme-11}
C. Gogolin, M. P. Mueller, and J. Eisert, Phys. Rev. Lett. {\bf 106}, 040401 (2011).

\bibitem{gm-11} 
P. Grisins and I. E. Mazets,  Phys. Rev. A {\bf 84}, 053635 (2011).

\bibitem{gp-08}
D. M. Gangardt and M. Pustilnik, Phys. Rev. A {\bf 77}, 041604 (2008).


\bibitem{g-13}  
V. Gurarie, J. Stat. Mech. P02014 (2013). 


\bibitem{cl-13}
A. Chandran, A. Nanduri, S. S. Gubser, and S. L.Sondhi,
arXiv:1304.2402. 


\bibitem{chl-08}
P. Calabrese, C. Hagendorf, and P. Le Doussal, 
J. Stat. Mech. P07013  (2008).

\bibitem{sc-08}
S. Sotiriadis and J. Cardy, 
J. Stat. Mech. P11003  (2008).

\bibitem{eip-09}
V. Eisler, F. Igloi, and I. Peschel, 
J. Stat. Mech.  P02011 (2009).


\bibitem{predra}
T. Antal, Z. Racz, A. Rakos, and G. M. Schutz, 
Phys. Rev. E {\bf 59}, 4912 (1999);\\ 
Y. Ogata, 
Phys. Rev. E {\bf 66}, 066123 (2002);\\
T. Antal, P. L. Krapivsky, and A. Rakos, 
Phys. Rev. E {\bf 78}, 061115 (2008).

\bibitem{dra}
D. Karevski, 
Eur. Phys. J. B {\bf 27}, 147 (2001);\\
T. Platini and D. Karevski, 
Eur. Phys. J. B {\bf 48}, 225 (2005);\\
T. Platini and D. Karevski, 
J. Phys. A {\bf 40}, 1711 (2007).

\bibitem{inh2}
J. Lancaster and A. Mitra, 
Phys. Rev. E  {\bf 81}, 061134 (2010);\\
S. Langer, M. Heyl, I. P. McCulloch, and F. Heidrich-Meisner,
Phys. Rev. B {\bf 84}, 205115 (2011);\\
M. Collura, H. Aufderheide, G. Roux, and D. Karevski;
Phys. Rev. A {\bf 86}, 013615 (2012);\\
V. Eisler and Z. Racz, 
Phys. Rev. Lett. {\bf 110}, 060602 (2013).


\bibitem{mpc-10}
J. Mossel, G. Palacios, and J.-S.  Caux, 
J. Stat. Mech.  L09001 (2010).

\bibitem{pv-13}
S. Peotta and M. Di Ventra, arXiv:1303.6916.



\bibitem{mg-05}
A. Minguzzi and D.M. Gangardt, Phys. Rev. Lett. {\bf 94}, 240404 (2005).

\bibitem{dm-06}
A. del Campo and J. G. Muga, Europhys. Lett. {\bf 74}, 965 (2006).\\
M. Collura and d. Karevski,  Phys. Rev. Lett. 104, 200601 (2010).

\bibitem{cro}
H. Buljan, R. Pezer, and T. Gasenzer, Phys. Rev. Lett. {\bf 100}, 080406 (2008);\\
D. Jukic, B. Klajn, T. Gasenzer, and H. Buljan, Phys. Rev. A {\bf 78}, 053602 (2008);\\
D. Jukic, B. Klajn, and H. Buljan, Phys. Rev. A {\bf 79}, 033612 (2009).

\bibitem{dc-08}
A. del Campo, Phys. Rev. A {\bf 78}, 045602 (2008).

\bibitem{a-12}
D. Iyer and N. Andrei, Phys. Rev. Lett. {\bf 109}, 115304 (2012);\\
D. Iyer, H. Guan, and N. Andrei, arXiv:1304.0506.


\bibitem{cv-10}
M. Campostrini and  E. Vicari, Phys. Rev. A {\bf 82}, 063636 (2010);

\bibitem{v-12}
E. Vicari, Phys.  Rev. A  {\bf 85},  062324  (2012).

\bibitem{hm-v}
F. Heidrich-Meisner, M. Rigol, A. Muramatsu, A.E. Feiguin, E. Dagotto, Phys. Rev. A {\bf 78}, 013620 (2008);\\
S. Langer, F. Heidrich-Meisner, J. Gemmer, I. P. McCulloch, U. Schollw\"ock, Phys. Rev. B {\bf 79}, 214409 (2009);

\bibitem{hm-v2}
C. J. Bolech, F. Heidrich-Meisner, S. Langer, I. P. McCulloch, G. Orso, and M. Rigol, Phys. Rev. Lett. {\bf 109}, 110602 (2012);\\
L. Vidmar, S. Langer, I. P. McCulloch, U. Schneider, U. Schollw\"ock, and F. Heidrich-Meisner, arxiv:1305.5496.

\bibitem{lbb-12}
H. Lu, L. O. Baksmaty, C. J. Bolech, and H. Pu, 
Phys. Rev. Lett. {\bf 108}, 225302 (2012).

\bibitem{r-10}
G. Roux, Phys. Rev. A {\bf 81}, 053604 (2010).

\bibitem{db-12}
A. del Campo, Phys. Rev. A {\bf 84}, 031606 (2011);\\
A. del Campo and M. G. Boshier, Sci. Rep. {\bf 2}, 648 (2012).




\bibitem{ck-12}
J.-S. Caux and R. M. Konik, Phys. Rev. Lett. {\bf 109}, 175301 (2012).

\bibitem{US}
M. Collura, S. Sotiriadis, and P. Calabrese, Phys. Rev. Lett. {\bf 110}, 245301 (2013)


\bibitem{LiebPR130} E. H. Lieb and W. Liniger, Phys. Rev. {\bf 130}, 1605 (1963);  
E. H. Lieb, Phys. Rev. {\bf 130}, 1616 (1963).


\bibitem{bck-13}
G. Brandino, J.-S. Caux, and R. M. Konik,  arXiv:1301.0308.

\bibitem{TG} L. Tonks, Phys. Rev. {\bf 50}, 955 (1936);  M. Girardeau, J. Math. Phys. {\bf 1}, 516 (1960).

\bibitem{deg-13}
E. Kaminishi, J. Sato, and T. Deguchi,
arXiv:1305.3412. 

\bibitem{mg-63}
J. B. McGuire, J. Math. Phys. {\bf 5}, 622 (1964).

\bibitem{cc-07}
P. Calabrese and J.-S. Caux, Phys. Rev. Lett. {\bf 98}, 150403 (2007);\\ 
P. Calabrese and J.-S. Caux, J. Stat. Mech. (2007) P08032.


\bibitem{ksc-13}
M. Kormos, A. Shashi, Y.-Z. Chou, J.-S. Caux, and A. Imambekov,
arXiv:1305.7202.  

\bibitem{grd-10}
V. Gritsev, T. Rostunov, and E. Demler, J. Stat. Mech. (2010) P05012.  

\bibitem{cmv-11}
P. Calabrese, M. Mintchev, and E. Vicari, Phys. Rev. Lett. {\bf 107}, 020601 (2011).

\bibitem{cmv-11b}
P. Calabrese, M. Mintchev, and E. Vicari, J. Stat. Mech. P09028 (2011).

\bibitem{pes}
I. Peschel, J. Phys. A {\bf 36}, L205 (2003); \\
I. Peschel, J. Stat. Mech. (2004) P06004;\\
I. Peschel and V. Eisler,  J. Phys. A {\bf 42}, 504003 (2009);\\
I. Peschel, Braz. J. Phys. {\bf 42}, 267 (2012).


\bibitem{dk-90-11}
B. Davies, Physica A {\bf 167}, 433-456 (1990); B. Davies and V. E. Korepin, arXiv:1109.6604; 







\bibitem{fredholm}
A. Jerri, \textit{Introduction to Integral Equations with Applications}, John Wiley \& Sons (1999);

\bibitem{feinberg}
J. Feinberg, J. Phys. A: Math. Gen. {\bf 37}, 6299 (2004).

\bibitem{pezer}
R. Pezer and H. Buljan, Phys. Rev. Lett. {\bf 98}, 240403 (2007).

\bibitem{kormos}
M. Kormos, A. Shashi, Y.-Z. Chou, A. Imambekov, arXiv:1204.3889.


\bibitem{adi}
A. Imambekov, I. E. Mazets, D. S. Petrov, V. Gritsev, S. Manz, S.Hofferberth, T. Schumm, E. Demler, and J. Schmiedmayer,
Phys. Rev. A {\bf 80}, 033604 (2009).

\bibitem{oldtan}
A. Minguzzi, P. Vignolo and M. P. Tosi, Phys. Lett. A {\bf 294}, 222 (2002);\\ 
M. Olshanii and V. Dunjko, Phys. Rev. Lett. {\bf 91}, 090401 (2003).

\bibitem{tan1D}
M. Barth and W. Zwerger, Ann. Phys. {\bf 326}, 2544 (2011).

\bibitem{vm-13}
P. Vignolo and A. Minguzzi, Phys. Rev. Lett. {\bf 110}, 020403 (2013).




\bibitem{swvc-08}
N. Schuch, M. M. Wolf, F. Verstraete, and J. I.Cirac 
Phys. Rev. Lett. {\bf 100}, 030504 (2008);\\
D. Perez-Garcia, F. Verstraete, M. M.  Wolf, J. I. and Cirac, 
Quantum Inf. Comput. {\bf 7}, 401 (2007).

\bibitem{cv-09}
J. I. Cirac and F. Verstraete 
J. Phys. {\bf 42}, 504004 (2009);\\
U. Schollw\"ok, Ann. Phys. {\bf 326}, 96 (2011);\\
P. Hauke, F. M. Cucchietti, L. Tagliacozzo, I. Deutsch, and M. Lewenstein, Rep. Progr. Phys. {\bf 75}, 082401 (2012).



\bibitem{cc-05} 
 P. Calabrese and  J. Cardy, J. Stat. Mech. P04010 (2005).

\bibitem{cc-07b}
P. Calabrese and J. Cardy, 
J. Stat. Mech. P10004 (2007).

\bibitem{ds-11}
J.-M. St\'ephan and J. Dubail,
J. Stat. Mech. (2011) P08019.


\bibitem{fc-08}
M. Fagotti and P. Calabrese,
Phys. Rev. A {\bf 78}, 010306 (2008). 



\bibitem{ep-08}
V. Eisler and  I. Peschel, J. Stat. Mech. (2007) P06005;\\
V. Eisler, D. Karevski, T. Platini, and I. Peschel, J. Stat. Mech. (2008) P01023.

\bibitem{ep-08b}
V. Eisler and  I. Peschel, Ann. Phys. (Berlin) {\bf 17}, 410 (2008).

\bibitem{isl-09}
F. Igloi, Z. Szatmari, and Y.-C. Lin, 
Phys. Rev.  B  {\bf  80},  024405 (2009);\\
F. Igloi, Z. Szatmari, and Y.-C. Lin, 
Phys.  Rev. B {\bf 85} 094417 (2012). 


\bibitem{va-12}
J. H. Bardarson, F. Pollmann, and J. E. Moore, Phys. Rev. Lett. {\bf 109}, 017202 (2012);\\
R. Vosk and E. Altman, arXiv:1205.0026. 


\bibitem{cc-13}
M. Collura and P. Calabrese, J. Phys. A {\bf 46}, 175001 (2013).


\bibitem{sled-13}
P. Hauke and L. Tagliacozzo,
arXiv:1304.7725;\\ 
J. Schachenmayer, B. P. Lanyon, C. F. Roos, A. J. Daley
 arXiv:1305.6880. 

\bibitem{hgf-09}
B. Hsu, E. Grosfeld, and E. Fradkin, 
Phys. Rev. B  {\bf 80}, 235412  (2009);\\
J. Cardy, 
Phys. Rev. Lett.  {\bf 106}, 150404 (2011);\\
D. A. Abanin and E. Demler,
Phys. Rev. Lett. {\bf 109}, 020504 (2012).


\bibitem{rev}
L. Amico, R. Fazio, A. Osterloh, and V. Vedral, 
Rev. Mod. Phys. {\bf 80}, 517 (2008);\\
J. Eisert, M. Cramer, and M. B. Plenio, 
Rev. Mod. Phys. {\bf 82}, 277 (2010);\\
P. Calabrese, J. Cardy, and B. Doyon Eds, J. Phys. A {\bf 42} 500301 (2009).

\bibitem{cl-08}
P. Calabrese and A. Lefevre,
Phys. Rev. A {\bf 78}, 032329 (2008).

\bibitem{Holzhey} C. Holzhey, F. Larsen, and F. Wilczek,
Nucl. Phys. B {\bf 424}, 443 (1994).

\bibitem{cc-04}
P.~Calabrese and J.~Cardy,
J. Stat. Mech. P06002 (2004);\\
%
P. Calabrese and J. Cardy,
J. Phys. A {\bf 42}, 504005 (2009).

\bibitem{c-lec}
J. Cardy,  
J. Stat. Mech. (2010) P10004.







\bibitem{jk-04}
B.-Q. Jin and V. E. Korepin,
J. Stat. Phys. {\bf 116}, 79 (2004);\\
P. Calabrese and F. H. L. Essler, 
J. Stat. Mech. (2010) P08029. 


\bibitem{cmv-12w}
P. Calabrese, M. Mintchev, and E. Vicari,
J. Phys. A {\bf 45} (2012) 105206

\bibitem{cmv-12h}
P. Calabrese, M. Mintchev and E. Vicari, EPL {\bf 97} (2012) 20009

\bibitem{v-12b}
E. Vicari, Phys Rev A {\bf 85}, 062104 (2012).

\bibitem{nv-13}
J. Nespolo and E. Vicari, Phys. Rev. A {\bf 87}, 032316 (2013).

\bibitem{kl-09}
I. Klich and L. Levitov,
Phys. Rev. Lett. {\bf 102}, 100502 (2009).


\bibitem{srh-10}
H. F. Song, S. Rachel, and K. Le Hur, Phys. Rev. B {\bf 82}, 012405 (2010);\\ 
H. F. Song, C. Flindt, S. Rachel, I. Klich, and K. Le Hur, Phys. Rev. B {\bf 83}, 161408 (2011). 

\bibitem{srf-12}
H. F. Song, S. Rachel, C. Flindt, I. Klich, N. Laflorencie and K. Le Hur, 
Phys. Rev. B {\bf 85}, 035409 (2012).



\bibitem{cmv-12}
P. Calabrese, M. Mintchev and E. Vicari, EPL {\bf 98}, 20003 (2012).


\bibitem{grit-pol}
A. Polkovnikov and V. Gritsev, Nature Phys. {\bf 4}, 477 (2008).



 
 
 




\end{thebibliography}
\end{document}